\documentclass[journal,11pt,onecolumn]{IEEEtran}

\ifCLASSINFOpdf
\else
   \usepackage[dvips]{graphicx}
\fi
\usepackage{url}
\usepackage{algorithmicx}
\usepackage{algpseudocode}
\usepackage{algorithm}
\algdef{SE}[DOWHILE]{Do}{doWhile}{\algorithmicdo}[1]{\algorithmicwhile\ #1}%
\usepackage{amsmath}
\usepackage{dsfont, cuted}
\usepackage{amsfonts}
\usepackage{amsthm}
\usepackage[utf8]{inputenc}
\usepackage{breqn}
\usepackage{subfigure}
\usepackage{multirow}
\usepackage{multicol}
\usepackage{booktabs}
\usepackage{fullwidth}
\usepackage[noadjust]{cite}
\usepackage{setspace}
\usepackage{flushend}
\usepackage{hyperref}
\usepackage{bmpsize}
\usepackage{graphicx}
\newtheorem{thm}{Theorem}

\usepackage{multirow,tabularx}
\newcolumntype{C}{>{\centering\arraybackslash}X}
\newcolumntype{L}{>{\raggedright \arraybackslash}X}
\newcolumntype{R}{>{\raggedleft \arraybackslash}X}
\begin{document}

\title{On Solving SAR Imaging Inverse Problems\\ Using Non-Convex Regularization with a\\ Cauchy-based Penalty}

\author{Oktay~Karakuş,~\IEEEmembership{Member,~IEEE,}
Alin~Achim,~\IEEEmembership{Senior Member,~IEEE}
        \thanks{This work was supported by the Engineering and Physical Sciences Research Council (EPSRC) under grant EP/R009260/1 (AssenSAR).}
        \thanks{Oktay Karakuş
        and Alin Achim are with the Visual Information Laboratory, University of Bristol, Bristol BS1 5DD, U.K. (e-mail: o.karakus@bristol.ac.uk; alin.achim@britol.ac.uk)}
}

\maketitle

\begin{abstract}
Synthetic aperture radar (SAR) imagery can provide useful information in a multitude of applications, including climate change, environmental monitoring, meteorology, high dimensional mapping, ship monitoring, or planetary exploration. In this paper, we investigate solutions to a number of inverse problems encountered in SAR imaging. We propose a convex proximal splitting method for the optimization of a cost function that includes a non-convex Cauchy-based penalty. The  convergence of the overall cost function optimization is ensured through careful selection of model parameters within a forward-backward (FB) algorithm. The performance of the proposed penalty function is evaluated by solving three standard SAR imaging inverse problems, including super-resolution, image formation, and despeckling, as well as ship wake detection for maritime applications. The proposed method is compared to several methods employing classical penalty functions such as total variation ($TV$) and $L_1$ norms, and to the generalized minimax-concave (GMC) penalty. We show that the proposed Cauchy-based penalty function leads to better image reconstruction results when compared to the reference penalty functions for all SAR imaging inverse problems in this paper.
\end{abstract}

\begin{IEEEkeywords}
Non-convex regularization; Convex optimization; Cauchy proximal operator; Inverse problems; Denoising; Image reconstruction.
\end{IEEEkeywords}

\IEEEpeerreviewmaketitle

\section{Introduction}
\IEEEPARstart{S}{ynthetic} aperture radar (SAR) is an important remote sensing technology capable of providing high-resolution images of the Earth, during day and night, for various terrains and in challenging conditions, for example due to adverse weather~\cite{moreira2013tutorial}. Thanks to recent technological developments, new generations of satellites have been launched and spatial resolutions that were previously unavailable are now offered by space-borne SAR remote sensing. Despite SAR images reaching resolutions as high as 1m, phenomena such as atmospheric delays \cite{bouman2016computational} 
or speckle noise, still impose limitations on SAR images quality.
It is therefore of significant importance to further enhance this in order to facilitate target detection and tracking, classification, security-related tasks, high dimensional mapping, maritime, or agricultural monitoring.

The problem of estimating an object of interest directly from the measurements (images) occurs in all imaging systems. Problems of this type are generically referred to as imaging inverse problems. The measurements and the forward-model connecting observations and sources are not enough to obtain solutions to these problems directly, due to their ill-posed nature. 
Unlike the forward-model which is well-posed almost every time \cite{hadamard2003lectures}, SAR imaging inverse problems are always ill-posed \cite{zhu2011inverse}. Therefore, having prior knowledge about the object of the interest plays a crucial role in reaching viable solutions in SAR imaging inverse problems. This leads to regularization based methods, which have received great attention in SAR applications including super-resolution \cite{babacan2008total, shkvarko2016solving}, despeckling \cite{achim2003sar,zhao2014adaptive, ozcan2015sparsity}, auto-focusing \cite{cetin2014sparsity, onhon2011sparsity}, ship wake detection \cite{karakucs2019ship, karakucs2019ship2}, or image formation/reconstruction \cite{baraniuk2007compressive, patel2010compressed, wei2019improved}. 


All these examples involve either a well-known regularization function, e.g. \texorpdfstring{$L_1$}{L1}, \texorpdfstring{$TV$}{TV}, or some combinations thereof. Despite its popularity, the $L_1$ norm penalty function tends to underestimate high intensity values, whilst $TV$ tends to over-smooth the data and may cause loss of details. Non-convex penalty functions can generally lead to better and more accurate estimations \cite{selesnick2, nikolova2005analysis, chen2014convergence} when compared to $L_1$, $TV$, or other convex penalty functions. Notwithstanding this, due to the non-convexity of the penalty functions, the overall cost function becomes non-convex, which leads to a multitude of sub-optimal local minima.
Convexity preserving non-convex penalty functions are essential in order to solve this problem, an idea that has been successfully exploited by Blake and Zimmerman \cite{blake1987visual}, and by Nikolova \cite{nikolova2010fast} by setting the penalty function in accordance with the data fidelity term. This was further investigated in \cite{parekh2015convex,lanza2016convex,selesnick2, malek2016class,selesnick2020non,anantrasirichai2020image,selesnick2017total}. Specifically, a convex denoising scheme is proposed with tight frame regularization in \cite{parekh2015convex}, whilst \cite{lanza2016convex} proposes the use of parameters non-convex regularizers to effectively induce sparsity of the gradient magnitudes. In \cite{selesnick2017total}, the Moreau envelope is used for TV denoising in order to preserve the convexity of the TV-based cost function. Finally, the non-convex generalized minimax concave (GMC) penalty function was proposed in \cite{selesnick2} for convex optimization problems. 

In this paper, we propose the use of the Cauchy distribution as a basis for the definition of a non-convex penalty function in a variational framework for solving SAR imaging inverse problems. The Cauchy distribution is a special member of the $\alpha$-stable distribution family ($\alpha=1$), which is known for its ability to model heavy-tailed data in various signal processing applications. It is a sparsity-enforcing  prior similar to the $L_1$ and $L_p$ norms \cite{mohammad2012bayesian} and it has generally been used in denoising applications by modelling sub-band coefficients in transform domains \cite{achim2003sar, bhuiyan2007spatially, chen2008wavelet, ranjani2010dual, gao2013directionlet}. The Cauchy distribution was also used as a noise model in image processing applications, in conjunction with quadratic \cite{sciacchitano2015variational} and TV norm \cite{mei2018cauchy} based penalty terms. Indeed, the approaches presented in \cite{sciacchitano2015variational,mei2018cauchy} preserve the convexity of the overall cost function while using the Cauchy distribution as data fidelity term rather than as penalty term as is the case in this paper.

Often, variational Bayesian methods are used to tackle Cauchy regularized inverse problems due to the lack of a closed-form analytical expression of the corresponding proximal operator. This prevented so far the Cauchy prior from being used in proximal splitting algorithms such as FB or ADMM. Moreover, having a proximal operator would make it applicable in advanced Bayesian image processing methodologies such as uncertainty quantification (UQ) via e.g., proximal Markov Chain Monte Carlo ($p$-MCMC) algorithms \cite{cai2018uncertainty, durmus2018efficient}.

In a recently submitted contributions~\cite{karakucs2019cauchy1}, we proposed such a proximal splitting algorithm employing a non-convex Cauchy-based penalty. 
Specifically, we developed a closed-form expression for the Cauchy proximal operator similar to the MAP estimator in \cite{wan2011segmentation}, and derived the necessary conditions (with appropriate proofs) for the Cauchy model parameters and splitting algorithm step size $\mu$ to ensure convergence. Furthermore, we showed the effect of the choice of the Cauchy scale parameter, $\gamma$, on performance, and presented results on generic 1D and 2D signal reconstruction examples.

In this paper, we extend the work in \cite{karakucs2019cauchy1} to advanced SAR imaging inverse problems. We investigate the performance of the proposed method in four different cases, including super-resolution, image-formation, despeckling, and ship wake detection. We follow the convexity conditions proposed in \cite{karakucs2019cauchy1} to guarantee the convergence for all the examples in this paper. The performance is then evaluated in comparison to methods based on several state-of-the-art penalty functions, such as $L_1$, $TV$ and GMC. All the minimization problems are solved via the FB algorithm proposed in~\cite{karakucs2019cauchy1} for all the penalty functions considered.

The rest of the paper is organized as follows: Section \ref{sec:SARIP} introduces the SAR imaging inverse problems considered in this study, whilst Section \ref{sec:Cauchyprox} presents the proximal splitting algorithm proposed for solving those problems. In Section \ref{sec:results}, the experimental validation and analysis of the proposed solutions are presented. We conclude our study by sharing remarks and future work directions in Section \ref{sec:conc}.

\section{SAR Imaging Inverse Problems}\label{sec:SARIP}
We begin this section by presenting the generic SAR image formation model, which can be expressed as
\begin{align}\label{equ:IP}
    Y = \mathcal{A}X + N,
\end{align}
where $Y$ denotes the observed SAR data, $X$ is the unknown SAR image, which can also be referred to as the target image (either an enhanced image or the raw data), $\mathcal{A}$ is the forward model operator and $N$ represents the noise.

Since recovering the object of interest $X$ from the observation $Y$ is an an ill-posed problem, we must consider prior information on $X$ to obtain a stable and unique reconstruction result. Under the assumption of an independent and identically distributed (iid) Gaussian noise case, we express the data fidelity term $\Psi(\cdot)$ (i.e. the likelihood) as
\begin{align}\label{equ:LHD}
   \Psi(Y, \mathcal{A}X) =  \frac{\|Y - \mathcal{A}X\|_2^2}{2\sigma^2}
\end{align}
where $\sigma$ refers to the standard deviation of the noise. Having the prior knowledge $p(X)$, the problem of estimating $X$ from the observed SAR image $Y$ by using the signal model in (\ref{equ:IP}) turns into a minimization problem 
\begin{align}\label{equ:mini1}
    \hat{X} &= \arg\min_X F(X),\\
    &= \arg\min_X \Bigg\{ \frac{\|Y - \mathcal{A}X\|_2^2}{2\sigma^2} - \log p(X) \Bigg\}
\end{align}
where we define $\psi(X)=-\log p(X)$ as the penalty function, and $F(X) = \Psi(Y, \mathcal{A}X) + \psi(X)$ is the cost function.
As discussed earlier, the selection of $\psi(X)$ (or equivalently $p(X)$) plays a crucial role in MAP estimation in order to overcome the ill-posedness of the problem and to obtain a stable solution. In the literature, depending on the SAR application, the penalty function $\psi(X)$ has various forms, among others the $L_1$, $L_2$, $TV$ or $L_p$ norms.


In the following sub-sections, we discuss the details of the four example SAR imaging inverse problem considered in this paper. Table \ref{tab:relation} presents the variable and operator relationships between the generic inverse problem in (\ref{equ:IP}) and the ones we introduce for each application in the following sub-sections.

\begin{table}[htbp]
\setstretch{0.7}\small
  \centering
  \caption{The relationship between the generic inverse problem in (\ref{equ:IP}) and applications. (\textit{$\mathcal{I}$ refers to the identity matrix.})}
    \resizebox{0.7\linewidth}{!}{\begin{tabular}{lccclll}
    \toprule
    Inverse Problem &    Equation &  & \multicolumn{3}{c}{Relation} \\
    \toprule
    \multirow{1}[0]{*}{Super-Resolution} & (\ref{equ:SRformation}) &       & $X \gets X $ & $Y \gets Y$&$\mathcal{A} \gets DH$\\
          \midrule
    \multirow{1}[0]{*}{Image Formation/} & (\ref{equ:SARRecons}) &       &  $X \gets f$ & $Y \gets y$ & $\mathcal{A} \gets \Phi$\\
    \multirow{1}[0]{*}{Reconstruction}  & &       &  &  & \\
          \midrule
    \multirow{1}[0]{*}{Despeckling} & (\ref{equ:wavelet}) &       & $X \gets \Psi_{(i)}$ & $Y \gets \Gamma_{(i)}$ & $\mathcal{A} \gets \mathcal{I}$ \\
          \midrule
    \multirow{1}[0]{*}{Ship Wake Detection} & (\ref{equ:wake}) &       &  $X \gets \Omega$ &  $Y \gets Y $ & $\mathcal{A} \gets \mathcal{C}$ \\
          \bottomrule
    \end{tabular}}%
  \label{tab:relation}%
\end{table}%

\subsection{Super-Resolution}
\label{sec:SR}
Super-resolution (SR) image reconstruction is a relatively common image processing technique, which seeks to reconstruct a high-resolution (HR) image through various approaches, starting with either a single or multiple low-resolution (LR) images~\cite{6732896}. Due to previously mentioned limitations of SAR imagery, various SR methodologies have been proposed specifically for SAR. These include among others an $L_1$ norm based Bayesian SR methodology proposed in \cite{villena2009bayesian} or a $TV$ regularization based approach based on a gradient profile prior in a MAP framework in \cite{liu2017texture}. In \cite{chavez2014super}, a SR method using wavelet domain interpolation with edge extraction was proposed whilst a dual-tree complex wavelet transform has been used in \cite{iqbal2012satellite}.

The SR image formation model employed here considers the HR SAR image $X$ being blurred and down-sampled, corresponding to an observed LR SAR image $Y$ \cite{liu2017texture} as
\begin{align}\label{equ:SRformation}
    Y = DHX + N,
\end{align}
where $H$ models a blurring filter, $D$ represents the down-sampling operator and $N$ is additive white Gaussian noise (AWGN). 

\subsection{Image Formation/Reconstruction}\label{sec:Formation}
A SAR system transmits a sequence of pulses and then receives echoes back-scattered from the interrogated surface and targets, which form the raw-data or so-called phase history data. A SAR image $f$ can be modelled as a linear system under the assumptions of (i) free space propagation, (ii) scalar wavefields, (iii) static targets. The corresponding linear image formation model is therefore \cite{kelly2011iterative}
\begin{align}\label{equ:SARRecons}
    y = \Phi f + n,
\end{align}
where $y$ represents the acquired complex phase history data, $\Phi$ refers to the measurement matrix  and $n$ is the additive system noise. 

The traditional reconstruction of a SAR scene can be performed using matched-filter based (MF) techniques by approximating the pseudo-inverse of $\Phi$ (filtered adjoint). Moreover, a least squares reconstruction approach can be used, which employs only a data fidelity term and ignores prior information on $f$, i.e. the function $\psi(\cdot)$ \cite{kelly2011iterative}. Despite their efficiency, traditional reconstruction techniques necessitate Nyquist rate samples of the echoes \cite{fang2013fast}.
Considering the ill-posedness of the problem, sparsity, and compressive signal cases with few samples, prior knowledge on $f$ should be taken into account in order to obtain a unique and stable solution. Hence, regularization based techniques are again suitable, as already proposed in the literature based on $L_1$ \cite{kelly2011iterative, alonso2010novel, fang2013fast}, $TV$ \cite{patel2010compressed} norms, or the GMC \cite{wei2019improved}. 

\subsection{Despeckling}\label{sec:despeckling}
A common and important problem hampering statistical inferences in SAR imagery is the presence of multiplicative speckle noise. In SAR systems, the received back-scattered signals sum up coherently and then undergo nonlinear transformations. This in turn causes a granular look in the resulting images, which is referred to as speckle noise. This may lead to loss of crucial details in SAR images and can cause problems in their analysis, e.g. for feature detection, segmentation or classification \cite{kuruoglu2004modeling,achim2006sar,moser2006sar,karakucs2018generalized}. Over the last three decades, despeckling approaches were mostly implemented in transform domains, such as the discrete wavelet transform (DWT)~\cite{achim2003sar,pizurica2001despeckling,datcu2007wavelet}. The idea behind these approaches is to apply a direct transform on the observed noisy SAR images, estimate the speckle-free transform coefficients, and finally apply the inverse transform on the despeckled coefficients \cite{achim2003sar}. In the literature, regularization based approaches have been used in conjunction with SAR despeckling methods with $TV$ \cite{huang2009new, bioucas2010multiplicative} and $L_1$ \cite{foucher2008sar, hao2012multiplicative} norms.

Let us consider an observed SAR image $Y$, affected by multiplicative speckle noise, $V$. We can write
\begin{align}\label{equ:multip}
    Y = XV,
\end{align}
where $X$ is the speckle-free SAR image. The multiplicative speckle image formation model given in (\ref{equ:multip}) is often transformed into an additive one by taking the logarithm of  both sides as
\begin{align}\label{equ:additive} 
\log(Y) &= \log(XV)\\
\log(Y) &= \log(X) + \log(V)\\
\label{equ:additive2} \Tilde{Y} &= \Tilde{X} + \Tilde{V},
\end{align}
where $\Tilde{Y}$, $\Tilde{X}$ and $\Tilde{V}$ refer to the logarithms of $Y$, $X$ and $V$, respectively. The DWT is a linear operation. Hence, when applied to (\ref{equ:additive2}) we get additive terms corresponding to noisy wavelet coefficients ($\Gamma_{(i)}$) at each resolution level and for all orientations that can be written as the sum of the transformations of the speckle-free signal ($\Psi_{(i)}$) and the noise components ($\nu_{(i)}$) as \cite{achim2003sar}
\begin{align}\label{equ:wavelet2}
W\Tilde{Y} &= W\Tilde{X} + W\Tilde{V},\\
    \label{equ:wavelet}\Gamma_{(i)} &= \Psi_{(i)} + \nu_{(i)}, \quad \text{where } \quad i = 1, 2, 3.
\end{align}

\begin{figure*}[htbp]
\centering
\includegraphics[width=.999\linewidth]{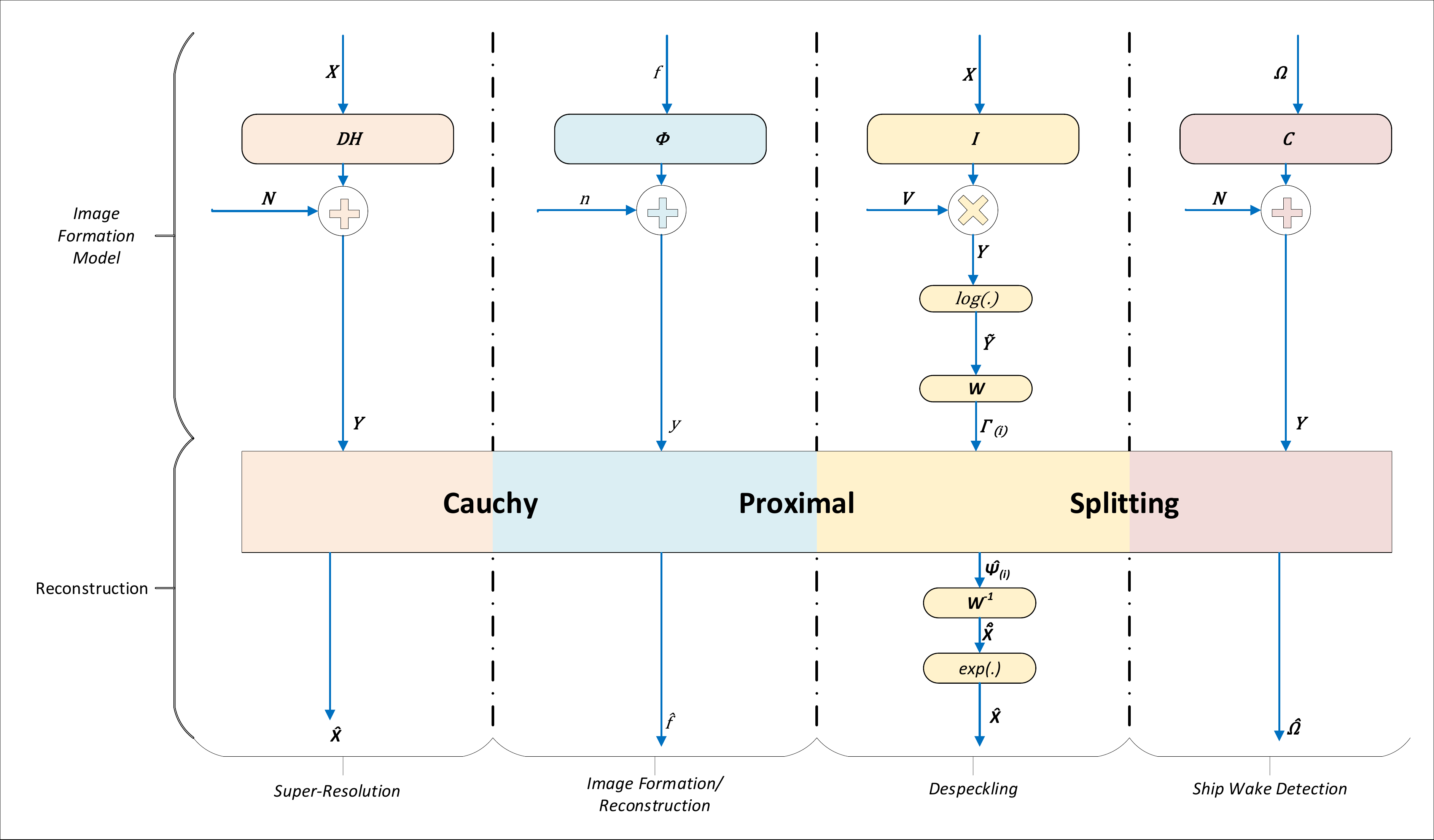}
\caption{Block diagram for the proposed solutions to SAR inverse problems.}
\label{fig:despeckling}
\end{figure*}

The despeckling model in this paper is depicted in Figure \ref{fig:despeckling}, where blocks $W$ and $W^{-1}$ represent the forward and inverse discrete wavelet transform operators.

\subsection{Ship Wake Detection}\label{sec:wake}
In SAR images, a moving ship in deep sea typically creates three different types of wakes: (i) \textit{turbulent wake}: the central dark streak, (ii) \textit{Narrow V-wake}: two bright arms lying either side of the turbulent wake, (iii) \textit{Kelvin wake}: two outer arms on each side of the turbulent wake, which limit the signatures of the moving ship.

Assuming that ship wakes can be modelled as linear structures, ship wake detection methods are generally based on the Radon transform, which creates bright peaks in the transform domain for bright lines, and troughs for dark lines. In the literature, the first Radon transform based ship wake detection method has been proposed by Murphy \cite{murphy1986linear}. Combining Wiener filtering with the Radon transform, Rey et al. \cite{rey1990} have proposed a method to increase the detectability of the peaks in Radon domain. Eldhuset \cite{eldhuset1996automatic} proposed a method for detection of ships and wakes automatically, whereby the detection performance is characterized by the number of lost and false wakes. A wavelet correlators based detection method has been proposed by Kuo and Chen \cite{kuo2003application} whilst Tunaley \cite{tunaley2003estimation} proposed a method based on a restricted search area in Radon domain. 
Zilman et al. \cite{zilman2004speed} have applied an enhancement operation to the Radon transform based on ship beam and speed estimation, and developed a SAR image simulator for moving vessels and their wakes in \cite{zilman2015detectability}. Graziano et. al. \cite{graziano1,graziano2} have proposed a wake detection methodology, based on noisy SAR images without performing any preliminary enhancement. Karakus et. al. \cite{karakucs2019ship, karakucs2019ship2} have proposed a method for ship wake detection, which involves solving an inverse problem based on the GMC penalty function. 
Furthermore in \cite{tianqi2020cauchySWD}, the Cauchy based penalty function was used for ship wake detection through a p-MCMC algorithm.
Here, we propose in fact an extension of the work in~\cite{tianqi2020cauchySWD}, whereby we use SAR images from several sources and observe the convergence condition detailed in \cite{karakucs2019cauchy1}.

Since we model ship wakes as linear features, the SAR image formation model can be expressed based on the Radon transform as \cite{karakucs2019ship2}
\begin{align}\label{equ:wake}
    Y = \mathcal{C}\Omega + N
\end{align}
where $Y$ is the $M \times M$ SAR image, $N$ is AWGN, $\mathcal{C} = \mathcal{R}^{-1}$ is the inverse Radon transform operator. $\Omega(r, \theta)$ refers to lines as a distance $r$ from the center of $Y$, and an orientation $\theta$ from the horizontal axis of $Y$. We use discrete operators $\mathcal{R}$ and $\mathcal{C}$ as described in \cite{kelley1993fast}.

Figure \ref{fig:despeckling} presents a block diagram representation of the proposed methodology for four example SAR imaging inverse problem discussed in this chapter.

\section{Cauchy-based Regularization}\label{sec:Cauchyprox}
In this section, we propose the use of the Cauchy distribution in the form of a non-convex penalty function for the purpose of solving imaging inverse problems. The Cauchy distribution is one of the special members of the $\alpha$-stable distribution family which is known to be heavy-tailed and promote sparsity in various applications. Contrary to the general $\alpha$-stable family, it has a closed-form probability density function, which is given by \cite{wan2011segmentation} 
\begin{align} \label{equ:Cauchy}
    p(X) \propto \frac{\gamma}{\gamma^2+X^2}
\end{align}	
where $\gamma$ is the scale (or the dispersion) parameter, which controls the spread of the distribution. By replacing $p(X)$ in (\ref{equ:mini1}) with the Cauchy prior given in (\ref{equ:Cauchy}), we have 
\begin{align}\label{equ:miniCauchy}
    \hat{X}_{\text{Cauchy}} = \arg\min_X \frac{\|Y - \mathcal{A}X\|_2^2}{2\sigma^2} - \sum_{i,j}\log\left(\frac{\gamma}{\gamma^2+X_{i,j}^2}\right),
\end{align}
which is the Cauchy regularized minimization in $X$, with the proposed non-convex Cauchy based penalty function
\begin{align}\label{equ:neglogCuachy}
\psi(x) = -\log\left(\dfrac{\gamma}{\gamma^2+x^2}\right).
\end{align} 


In order to solve the minimisation problem in (\ref{equ:miniCauchy}) by using proximal algorithms such as the forward-backward (FB) or the alternating direction method of multipliers (ADMM), \textit{the proximal operator} of the Cauchy based penalty function in (\ref{equ:neglogCuachy}) should be defined.
In a related recent publication \cite{karakucs2019cauchy1}, we showed that a closed-form expression for such a proximal operator does exist and can be derived starting from 

\begin{align}\label{equ:proxCauchy}
    prox_{Cauchy}^{\mu}(x) = \arg\min_u \left\{\frac{\|x - u \|_2^2}{2\mu} - \log\left(\frac{\gamma}{\gamma^2+u^2}\right) \right\}
\end{align}

The solution to this minimization problem can be obtained by taking the first derivative of (\ref{equ:proxCauchy}) in terms of $u$ and setting it to zero. Hence we have 
\begin{align}\label{equ:proxCauchy2}
    u^3-xu^2+(\gamma^2+2\mu)u-x\gamma^2 = 0.
\end{align}

The solution to the cubic function given in (\ref{equ:proxCauchy2}) can be obtained through Cardano’s method as \cite{wan2011segmentation}
\begin{align}
    p &\gets \gamma^2 + 2\mu - \frac{x^2}{3},\\
q &\gets x\gamma^2 + \frac{2x^3}{27} - \frac{x}{3}\left(\gamma^2 + 2\mu\right),\\
s &\gets \sqrt[3]{q/2 + \sqrt{p^3/27 + q^2/4}},\\
t &\gets \sqrt[3]{q/2 - \sqrt{p^3/27 + q^2/4}},\\
\label{equ:cardano}z &\gets \frac{x}{3} + s + t.
\end{align}
where $z$ is the solution to $prox_{Cauchy}^{\mu}(x)$.

\subsection{Cauchy proximal splitting (CPS) algorithm}
The use of a proximal operator corresponding to the proposed penalty function would enable the use of a proximal splitting algorithm to solve the optimization problem in (\ref{equ:miniCauchy}).
In particular, an optimization problem of the form
\begin{align}\label{equ:FB1}
    \arg\min_x (f_1 + f_2)(x)
\end{align}
can be solved via the FB algorithm. From the definition \cite{combettes2011proximal}, provided $f_2:\mathbb{R}^N \rightarrow \mathbb{R}$ is $L$-Lipschitz differentiable with Lipschitz constant $L$ and $f_1:\mathbb{R}^N \rightarrow \mathbb{R}$, then (\ref{equ:FB1}) can be solved iteratively as
\begin{align}\label{equ:FB2}
    x^{(n+1)} = prox_{f_1}^{\mu} \left( x^{(n)} - \mu\bigtriangledown f_2(x^{(n)})  \right)
\end{align}
where the step size $\mu$ is set within the interval $\left(0, \frac{2}{L} \right)$. In our case, the function $f_2$ is the data fidelity term and takes the form of $\frac{\|y - \mathcal{A}x\|_2^2}{2\sigma^2}$ from (\ref{equ:miniCauchy}) whilst the function $f_1$ corresponds to the Cauchy based penalty function $\psi$, which in~\cite{karakucs2019cauchy1} we have proved to be twice continuously differentiable. 

Observing (\ref{equ:miniCauchy}), it can be easily deduced that since the penalty function $\psi$ is non-convex, the overall cost function is also non-convex in general. Hence, in order to avoid local minimum point estimates, one should ensure convexity of the proximal splitting algorithm employed. To this effect, we have formulated the following theorem in \cite{karakucs2019cauchy1}, which we recall here for completeness.

\begin{thm}[Theorem 2. in \cite{karakucs2019cauchy1}]
 \label{thm:theorem2}
Let the twice continuously differentiable and non-convex regularization function $\psi$ be the function $f_1$ and the $L$-Lipschitz differentiable data fidelity term $\frac{\|y - \mathcal{A}x\|_2^2}{2\sigma^2}$ be the function $f_2$. The iterative FB sub-solution to the optimization problem in (\ref{equ:miniCauchy}) is
\begin{align}\label{equ:thm2}
    x^{(n+1)} = prox_{Cauchy}^{\mu} \left( x^{(n)} - \frac{\mu\mathcal{A}^T(\mathcal{A}x^{(n)} - y)}{\sigma^2}  \right)
\end{align}
where $\bigtriangledown f_2(x^{(n)}) = \frac{\mathcal{A}^T(\mathcal{A}x^{(n)} - y)}{\sigma^2}$. If the condition
\begin{align}\label{equ:Proofthm2}
    \gamma \geq \frac{\sqrt{\mu}}{2}
\end{align}
holds, then the sub-solution of the FB algorithm is strictly convex, and the FB iteration in (\ref{equ:thm2}) converges to the global minimum.
\end{thm}

For the proof of the theorem, we refer the reader to \cite{karakucs2019cauchy1}.
In order to comply with the condition imposed by the theorem, two approaches are possible: 
\begin{enumerate}
    \item The step size $\mu$ can be set following estimation of $\gamma$ directly from the observations,
    \item The scale parameter $\gamma$ can be set, for cases when the Lipschitz constant $L$ is computed or if estimating $\gamma$ requires computationally expensive calculations.
\end{enumerate}
In this paper, we adopt the second option, i.e. (calculate $L$) $\rightarrow$ (set $\mu$) $\rightarrow$ (set $\gamma$). 

Based on Theorem \ref{thm:theorem2}, in Algorithm \ref{alg:FB} we provide our proposed FB-based proximal splitting method for solving (\ref{equ:miniCauchy}).

\begin{algorithm}[ht!]
\caption{Algorithmic representation of Cauchy Proximal Splitting}\label{alg:FB}
\setstretch{1.2}
\begin{algorithmic}[1]
\State \textbf{Input:} $\text{SAR data, }Y \text{ and } MaxIter$
\State \textbf{Input:} $\mu\in\left(0, \frac{2}{L}\right) \text{ and } \gamma\geq\frac{\sqrt{\mu}}{2}$
\State \textbf{Set:} $i\gets0 \text{ and } X^{(0)}$
  \Do
    \State $u^{(i)} \gets X^{(i)} - \mu \mathcal{A}^T(\mathcal{A}X^{(i)} - Y)$
    \State $X^{(i+1)} \gets prox_{Cauchy}^{\mu}(u^{(i)}) \text{ via (\ref{equ:cardano})}$
    \State $i++$
  \doWhile{$\dfrac{\|X^{(i)} - X^{(i-1)}\|}{\|X^{(i-1)}\|} > \varepsilon \text{ or } i<MaxIter$}
\end{algorithmic}
\end{algorithm} 

The notations in  Algorithm \ref{alg:FB} are based on the generic inverse problem given in (\ref{equ:IP}) for the Cauchy based penalty function. For each application discussed in Sections \ref{sec:SR}, \ref{sec:Formation}, \ref{sec:despeckling} and \ref{sec:wake}, the generic variables $Y$, $X$ and the forward operator $\mathcal{A}$ in Algorithm \ref{alg:FB} should be substituted with the corresponding variables for the inverse problems given in (\ref{equ:SRformation}), (\ref{equ:SARRecons}), (\ref{equ:wavelet}) and (\ref{equ:wake}) respectively, according to Table \ref{tab:relation}. In addition, for each corresponding image formation model, the cost functions to be minimized are also presented in Table \ref{tab:costFuncs}.

\begin{table}[htbp]
\caption{Cost functions for SAR imaging inverse problems.}
\label{tab:costFuncs}
    \centering
    \begin{tabularx}{0.76\linewidth}{@{} CCR @{}}
    \toprule
    Equation & & Cost Function\\
    \toprule
    (\ref{equ:SRformation})     &$\rightarrow$& $\frac{\|Y - DHX\|_2^2}{2\sigma^2} - \log p(X)$ \\
      (\ref{equ:SARRecons})   &$\rightarrow$& $\frac{\|y - \Phi f\|_2^2}{2\sigma^2} - \log p(f)$  \\
       (\ref{equ:wavelet})  &$\rightarrow$& $\frac{\|\Gamma_{(i)} - \Psi_{(i)}\|_2^2}{2\sigma^2} - \log p(\Psi_{(i)})$ \\
         (\ref{equ:wake}) &$\rightarrow$& $\frac{\|Y - \mathcal{C}\Omega\|_2^2}{2\sigma^2} - \log p(\Omega)$ \\
         \bottomrule
    \end{tabularx}
\end{table}

Finally, please note that as long as the the data fidelity term is convex and $L$-Lipschitz differentiable, for the Cauchy-based penalty function in (\ref{equ:neglogCuachy}), the FB algorithm proposed here can be replaced by other proximal splitting algorithms such as (ADMM) or Douglas-Rashford (DR) and convergence is still going to be guaranteed according to Theorem 1.

\section{Results and Discussions}\label{sec:results}
In this section, we show results obtained when employing our proposed Cauchy proximal operator and corresponding splitting algorithm to the four inverse problems introduced in Section \ref{sec:SARIP}.
In the sequel, we describe  separately the simulation experiments and data sets utilized, and discuss the results for each example.

\subsection{Super-Resolution}
In the first set of simulations, we investigated the single image super-resolution problem. The data set for these simulations comprises of five X-band, HH polarized, Stripmap SAR products from TerraSAR-X \cite{terraSARX}, all of which of 700$\times$700 pixels. In order to obtain the LR images (of size 350$\times$350), we used the degradation model given in (\ref{equ:SRformation}) with a point spread function which was modelled as a symmetric, 5$\times$5 Gaussian low-pass filter with standard deviation of 2. Down-sampling by a factor of 2 was applied, while the AWGN corresponds to a blurred-signal-to-noise-ratio (BSNR) of 30 dB.

\begin{figure}[ht!]
\centering
\subfigure[]{\includegraphics[width=.6\linewidth]{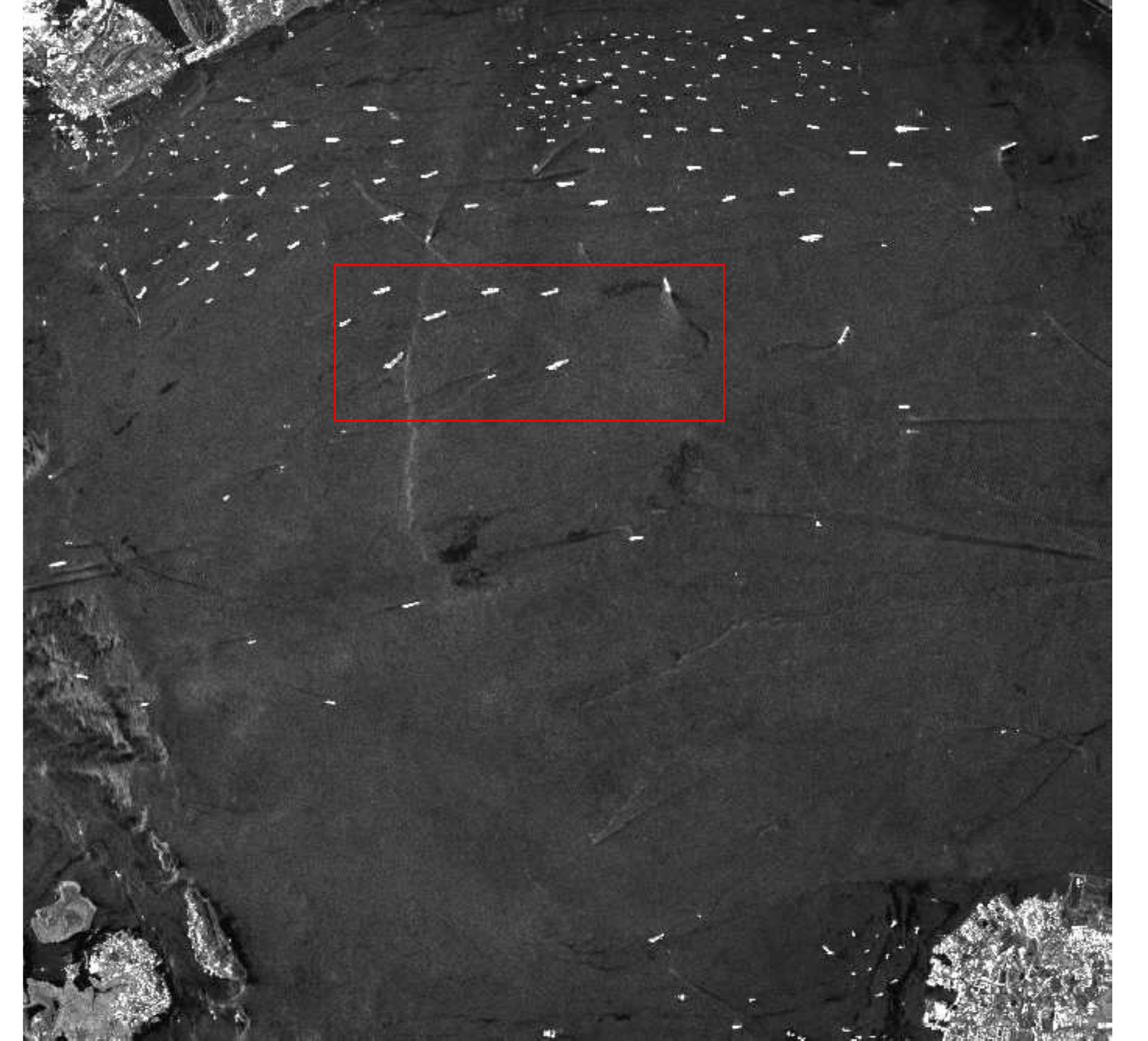}}\hfill\\
\subfigure[]{\includegraphics[width=.32\linewidth]{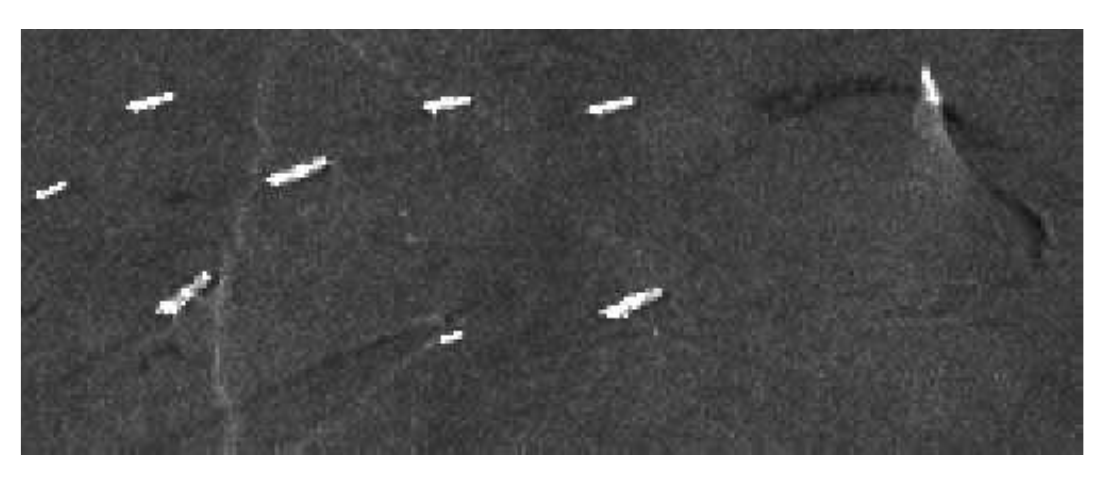}}
\subfigure[]{\includegraphics[width=.32\linewidth]{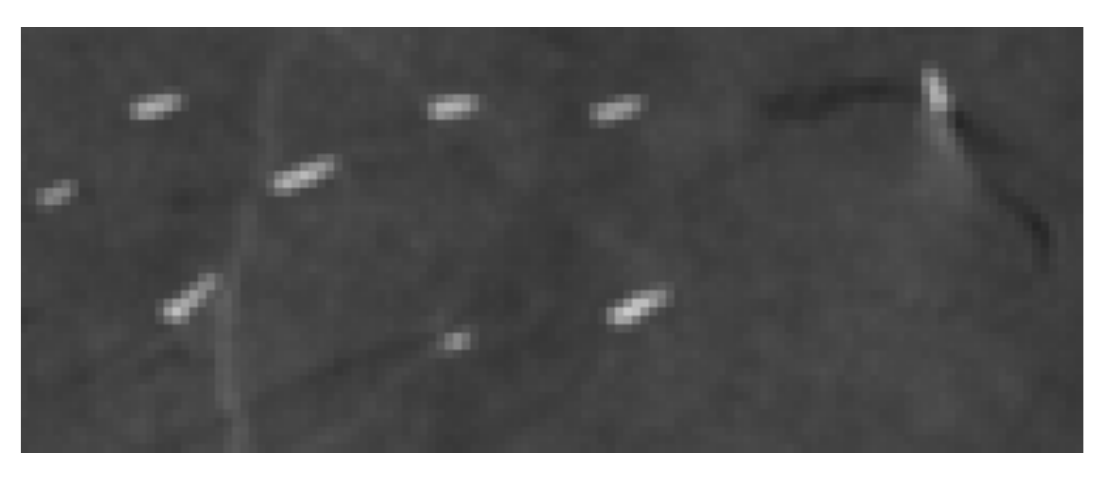}}
\subfigure[]{\includegraphics[width=.32\linewidth]{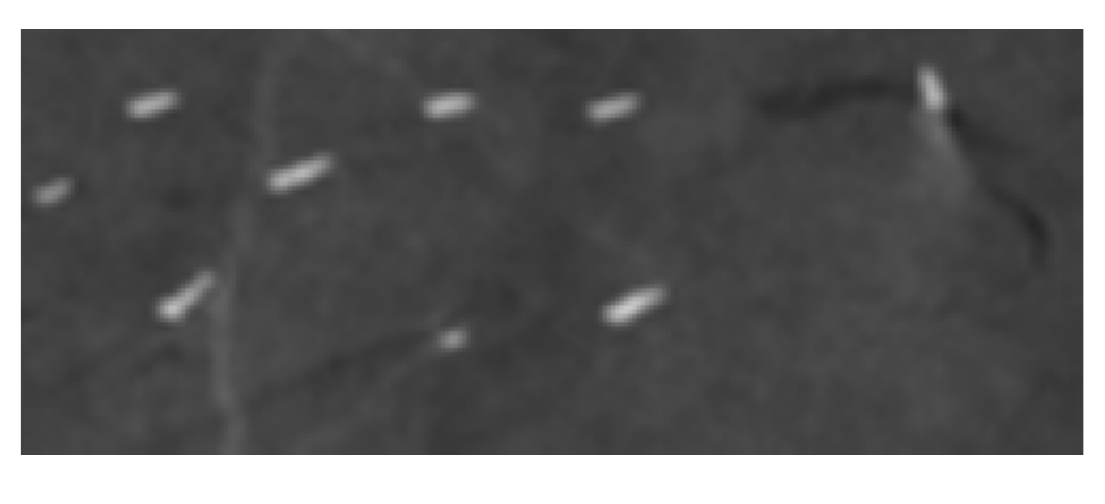}}\hfill
\subfigure[]{\includegraphics[width=.32\linewidth]{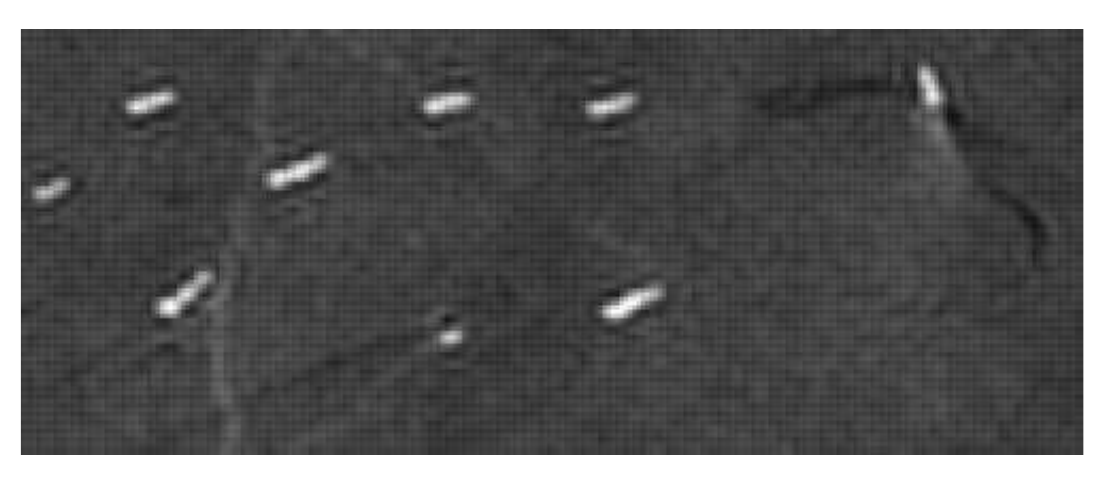}}
\subfigure[]{\includegraphics[width=.32\linewidth]{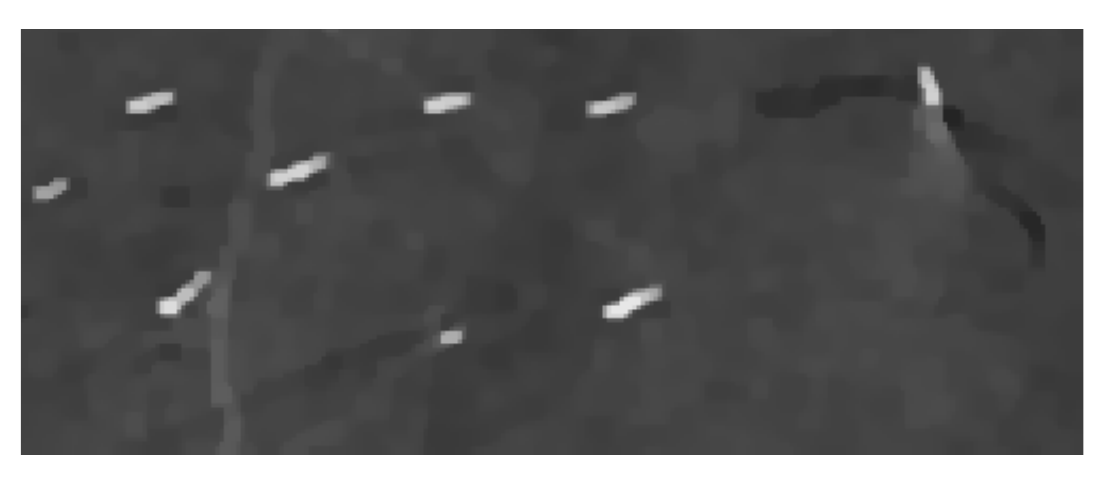}}
\subfigure[]{\includegraphics[width=.32\linewidth]{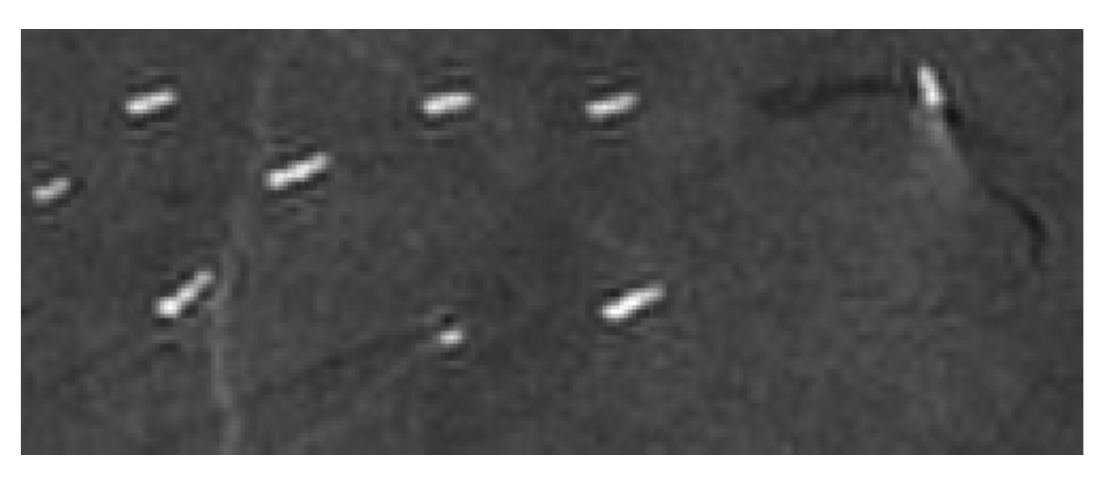}}\hfill
\caption{Super-resolution results for Image-1. (a) Original image, (b) Cropped original image, (c) Low resolution image, super-resolved images by (d) Bicubic, (e) L1, (f) TV and (g) Cauchy.}
\label{fig:sr1}\vspace{0.7cm}
\end{figure}

\begin{figure}[ht!]
\centering
\subfigure[]{\includegraphics[width=.20\linewidth]{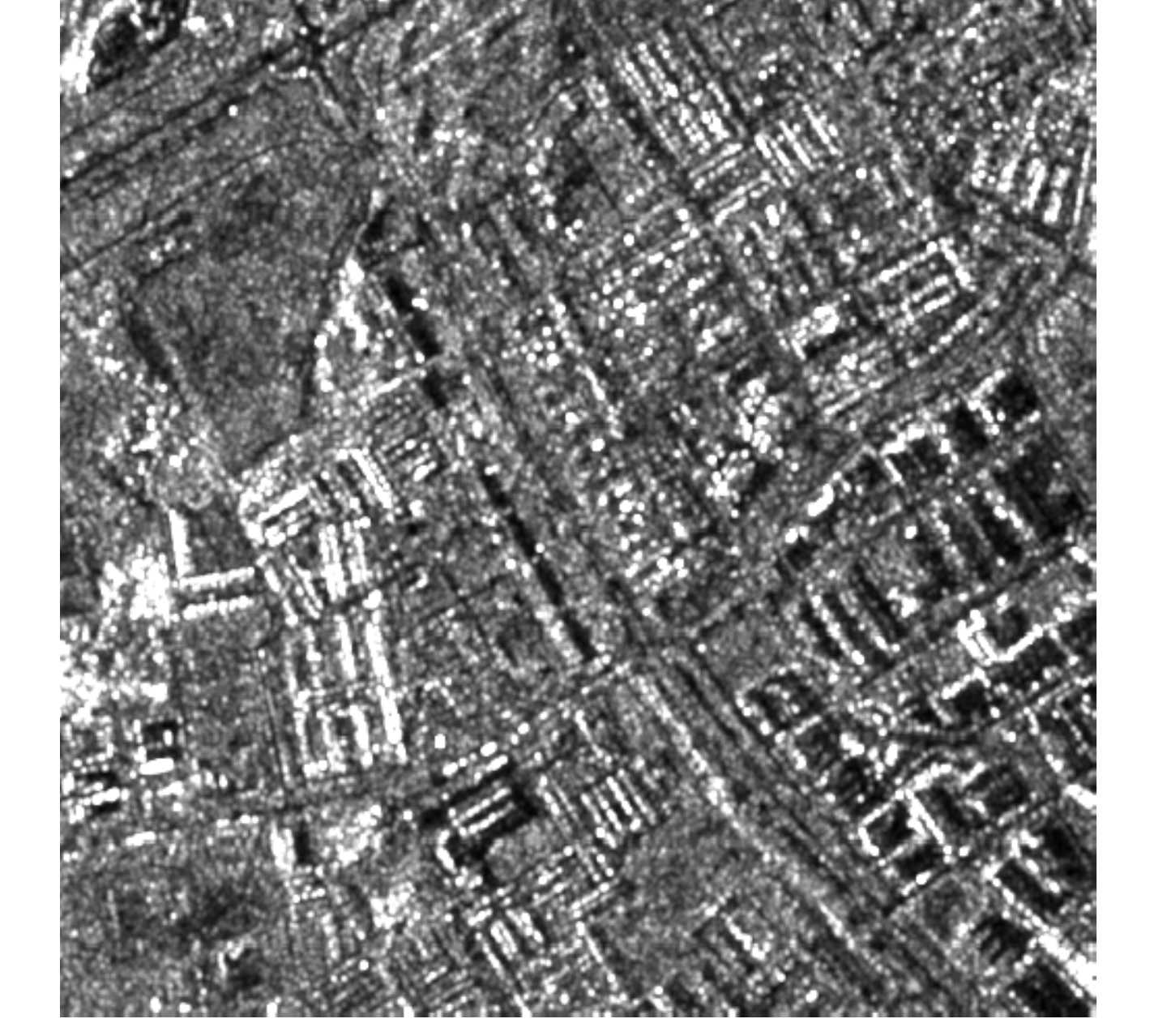}}
\subfigure[]{\includegraphics[width=.38\linewidth]{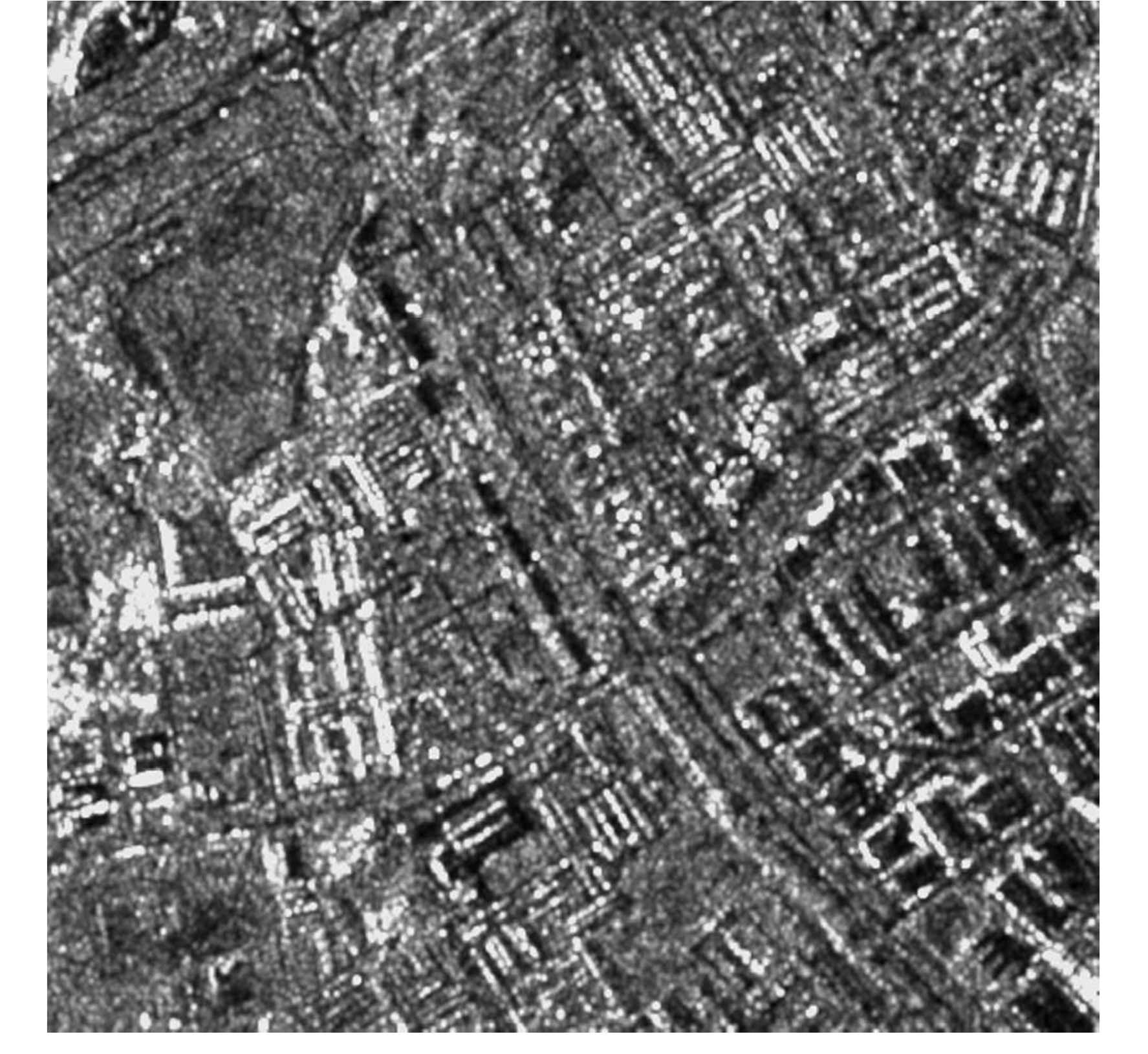}}
\subfigure[]{\includegraphics[width=.38\linewidth]{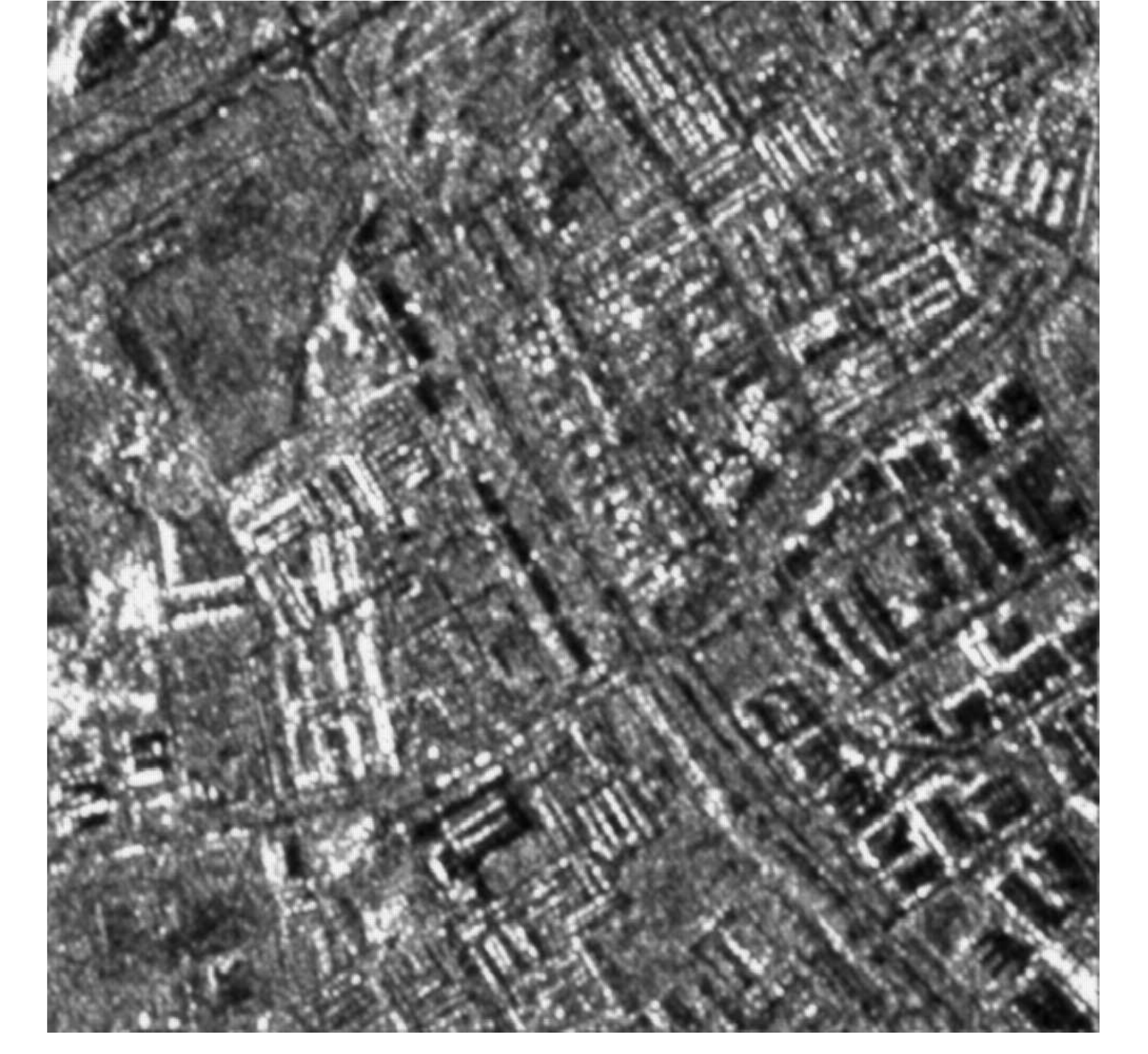}}
\subfigure[]{\includegraphics[width=.38\linewidth]{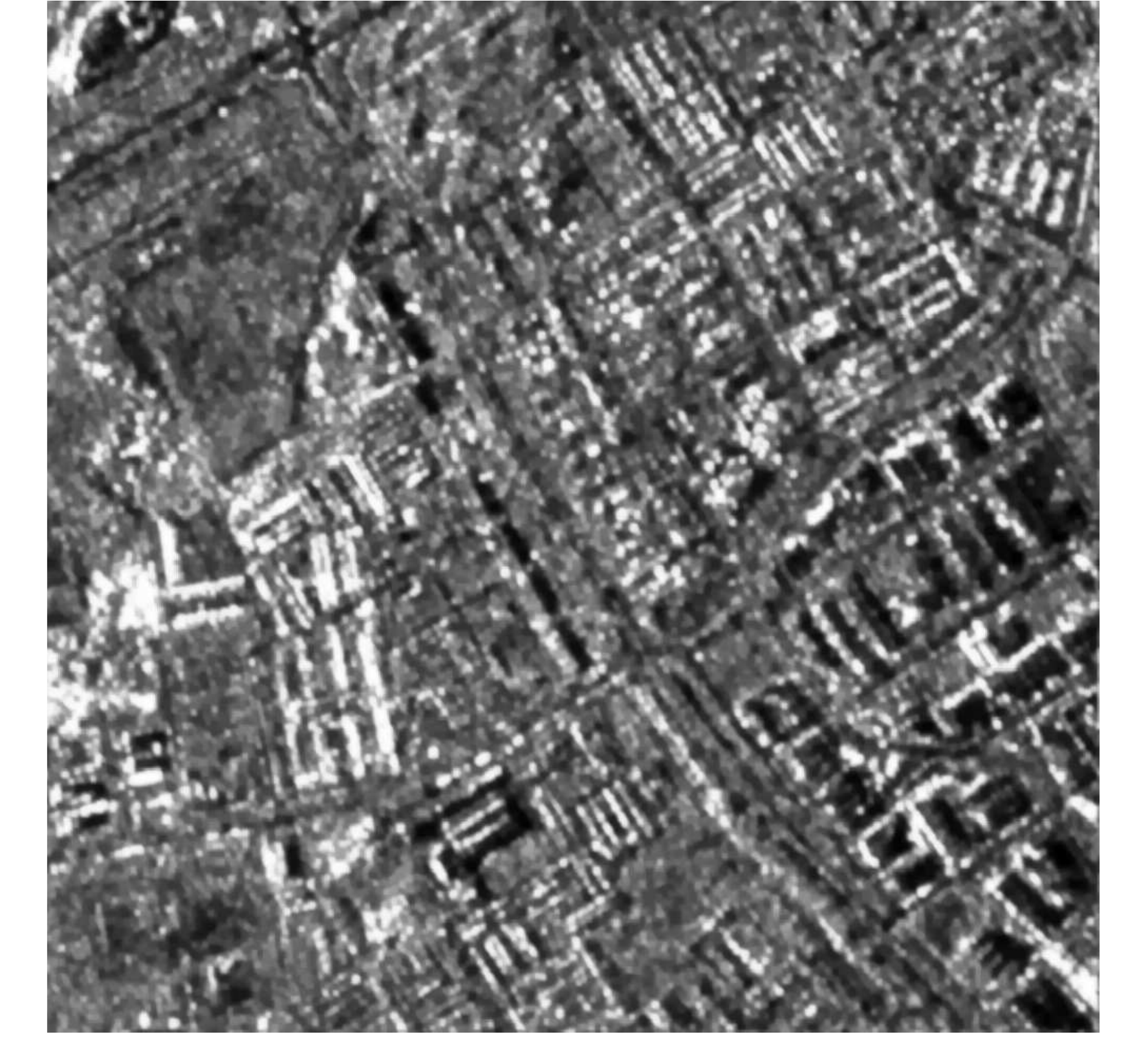}}
\subfigure[]{\includegraphics[width=.38\linewidth]{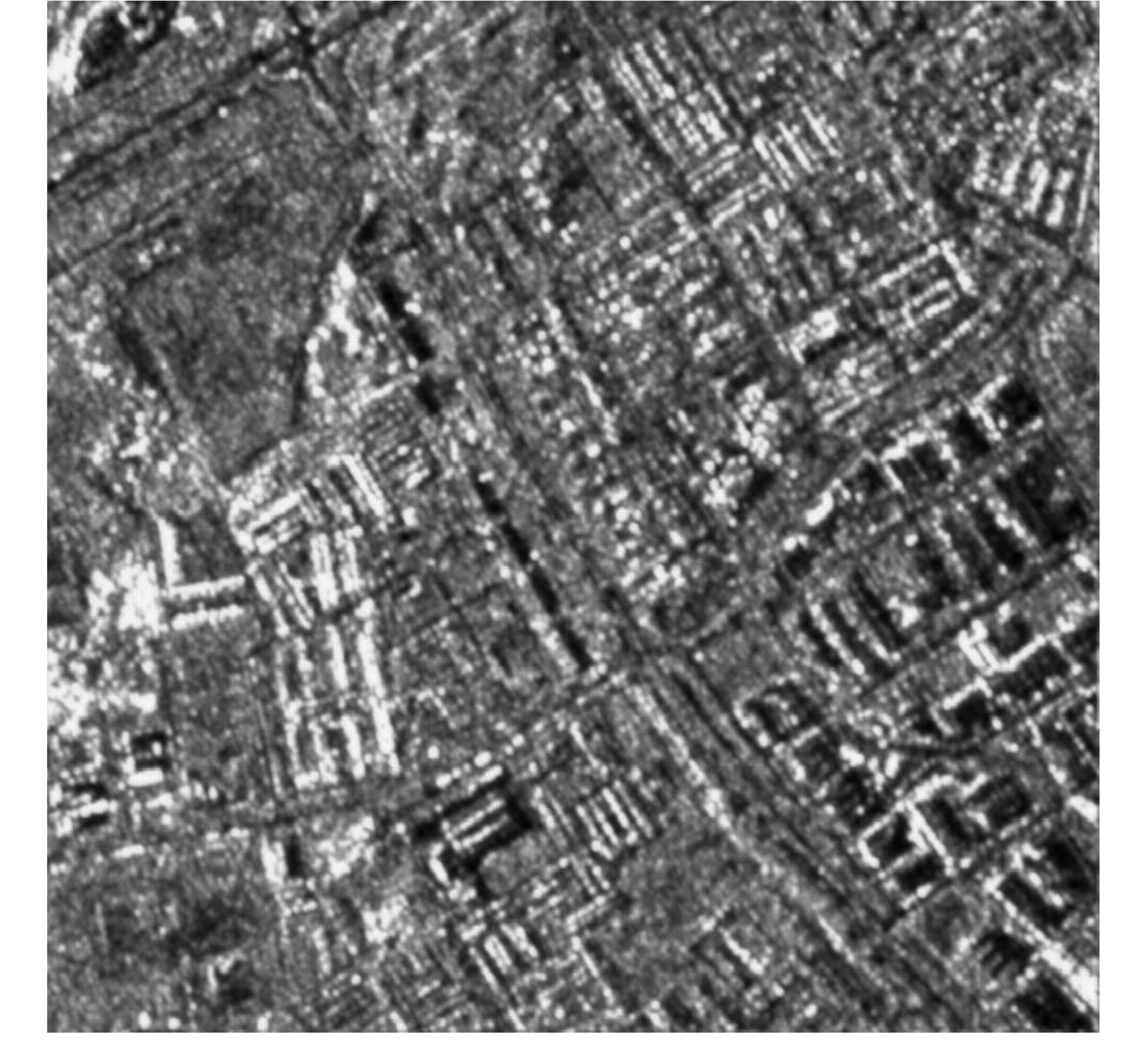}}
\caption{Super-resolution results for a real 500$\times$500 SAR Image in (a). 1000$\times$1000 super-resolved images by (b) Bicubic, (c) L1 (d) TV, (e) Cauchy.}
\label{fig:sr2}
\end{figure}

The performance of the proposed super-resolution algorithm was compared to methods based on the $L_1$ and $TV$ regularization functions, as well as the standard method of SR by bicubic interpolation. The performance of all methods was evaluated in terms of the peak signal to noise ratio (PSNR), structural similarity index (SSIM) and root mean square error (RMSE). In addition to simulations, 
for subjective evaluation, a real X-band TerraSAR-X image of size 500$\times$500, representing an urban scene was also utilized, and HR images of size 1000$\times$1000 were obtained for each method. Results are presented in Table \ref{tab:SR}, and Figures \ref{fig:sr1} and \ref{fig:sr2}.
\begin{table}[htbp]
\setstretch{0.7}\small
  \centering
  \caption{SAR Super-resolution performance quantification in terms of PSNR, SSIM and RMSE.}
    \begin{tabularx}{0.96\linewidth}{@{} ClCCC @{}}
    \toprule
          &       & PSNR  & SSIM  & RMSE \\
          \toprule
    \multirow{4}[0]{*}{Image-1} & Bicubic & 24.149 & 0.617 & 0.062 \\
          & L1    & 25.218 & 0.599 & 0.055 \\
          & TV    & 25.413 & 0.640 & 0.054 \\
          & Cauchy & \textbf{25.727} & \textbf{0.686} & \textbf{0.052} \\
          \midrule
    \multirow{4}[0]{*}{Image-2} & Bicubic & 28.042 & 0.458 & 0.040 \\
          & L1    & 28.203 & 0.507 & 0.039 \\
          & TV    & 28.163 & 0.459 & 0.039 \\
          & Cauchy & \textbf{28.562} & \textbf{0.526} & \textbf{0.037} \\
          \midrule
    \multirow{4}[0]{*}{Image-3} & Bicubic & 23.851 & 0.549 & 0.064 \\
          & L1    & 24.641 & 0.596 & 0.059 \\
          & TV    & 24.626 & 0.594 & 0.059 \\
          & Cauchy & \textbf{25.048} & \textbf{0.650} & \textbf{0.056} \\
          \midrule
    \multirow{4}[0]{*}{Image-4} & Bicubic & 22.293 & 0.492 & 0.077 \\
          & L1    & 23.281 & 0.568 & 0.069 \\
          & TV    & 23.180 & 0.542 & 0.069 \\
          & Cauchy & \textbf{23.654} & \textbf{0.616} & \textbf{0.066} \\
          \midrule
    \multirow{4}[0]{*}{Image-5} & Bicubic & 22.433 & 0.584 & 0.076 \\
          & L1    & 23.538 & 0.599 & 0.067 \\
          & TV    & 23.479 & 0.638 & 0.067 \\
          & Cauchy & \textbf{23.898} & \textbf{0.667} & \textbf{0.064} \\
          \bottomrule
    \end{tabularx}%
  \label{tab:SR}%
\end{table}%

From Table \ref{tab:SR}, we can clearly see that our proposed Cauchy based penalty function leads to the highest PSNR/SSIM and lowest RMSE values for all the images. Even though the results are close for all the methods, the proposed method achieves a PSNR gain of around 0.5 dBs over the second best method, which is generally either $TV$ or the $L_1$ regularization based method. Furthermore, $TV$ and $L_1$ results are similar whilst bicubic results fall short of all other methods, as expected.

When examining reconstruction results in the enlarged area in Figure \ref{fig:sr1}, it is obvious that the bicubic reconstruction result is very blurry compared to the others. Furthermore, even though it looks smoother and less noisy than all other methods, the $TV$-based SR reconstruction approach discards lots of background details, e.g. sea surface waves. These structures are clearer in the Cauchy case in Figure \ref{fig:sr1}-(g), which further highlights the reconstruction performance of the proposed method. Reconstruction results in Figure \ref{fig:sr2}, which correspond to direct SR without down-sampling are visually consistent with those achieved on simulated LR SAR images.

\subsection{Image Formation/Reconstruction}
In the second set of simulations, we tested the proposed penalty function in SAR image reconstruction for three different SAR data sets: (i) GOTCHA volumetric SAR data set \cite{casteel2007challenge}, (ii) Backhoe Data Dome \cite{naidu2004data}, and (iii) Civilian Vehicle Data Dome \cite{dungan2010civilian}.

The Air Force Research Laboratory (AFRL) released GOTCHA data set \cite{casteel2007challenge} as a challenge problem for 2D/3D imaging of targets from a volumetric data set in an urban environment in 2007. The scene includes several targets and civilian vehicles. The data set consists of fully polarimetric data from 8 passes and covers full 360 degrees of azimuth. We utilized the data for the pass 1 with HH polarisation and for full azimuth of 360 degrees with 15 degrees separations. The scene has a size of 100 m$\times$100 m with a pixel spacing of 20 cm, which results in a 501$\times$501 pixel SAR image.

The Backhoe data dome data set \cite{naidu2004data} was also released by the AFRL in 2004 for a synthetically generated data dome of a backhoe target. The data set consists of 110 degrees azimuth cut between 350 to 100 degrees at 0 and 30 degrees elevations at 6GHz bandwidth. To form the image, we used all 110 degrees azimuth and 0 degrees elevation. The scene has a size of 10 m$\times$10 m with a pixel spacing of 2 cm, which results in a 501$\times$501 pixel SAR image.

Civilian Vehicle Data Dome data set \cite{dungan2010civilian} was released by the Ohio State University in 2010 and includes a set of synthetically generated data domes of various civilian vehicles. The data set consists of full 360 degrees azimuth at 30 and 60 degrees elevations with 5.35GHz bandwidth. We used 30 degrees elevation data for full 360 degrees azimuth for the civilian vehicles Tacoma, Jeep93 and Camry. The scene has a size of 10 m$\times$10 m with a pixel spacing of 2 cm, which results in a 501$\times$501 pixel SAR image.

\begin{figure*}[!ht]
\centering
\subfigure[]{\includegraphics[width=.24\linewidth]{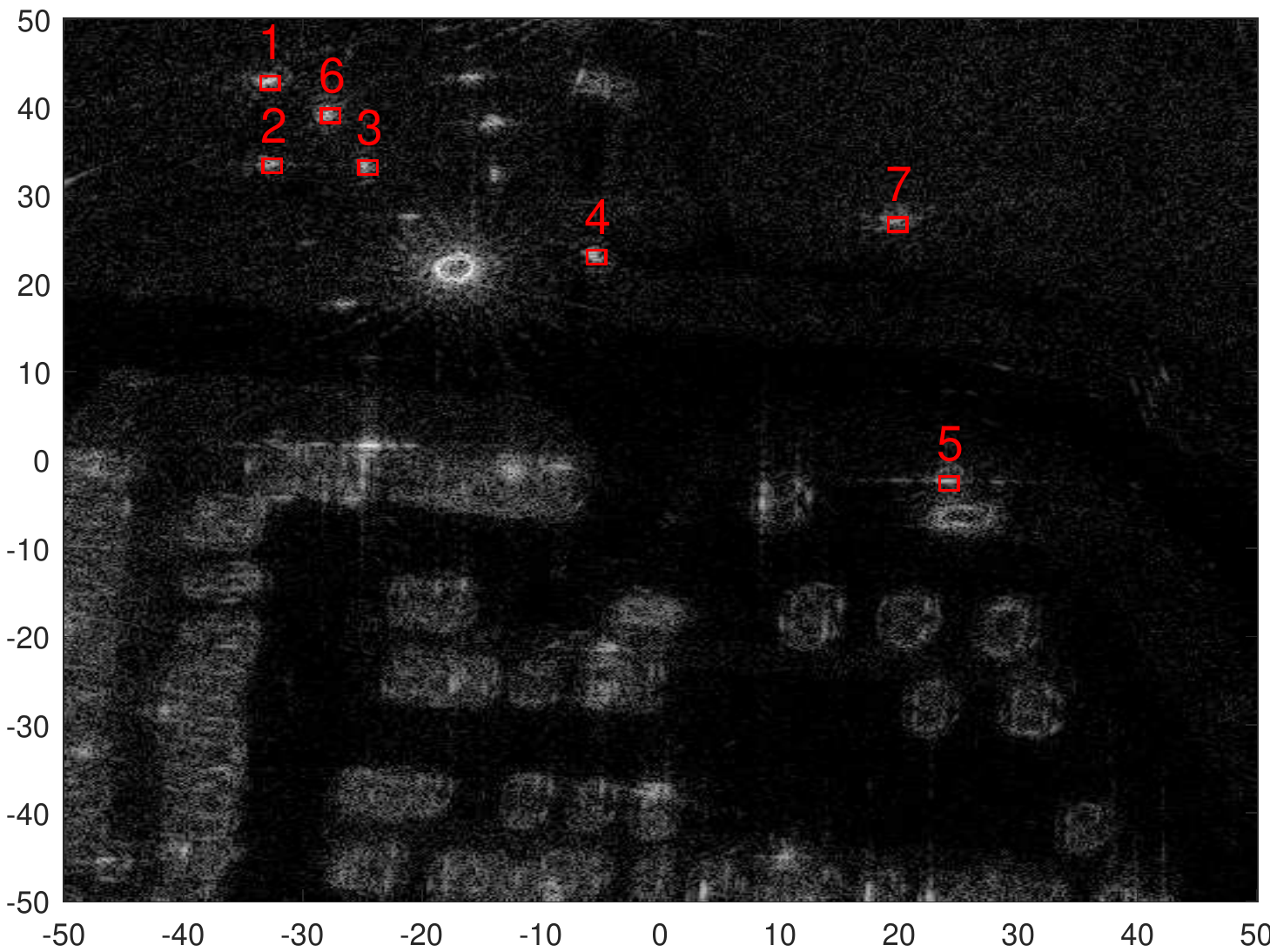}}
\subfigure[]{\includegraphics[width=.24\linewidth]{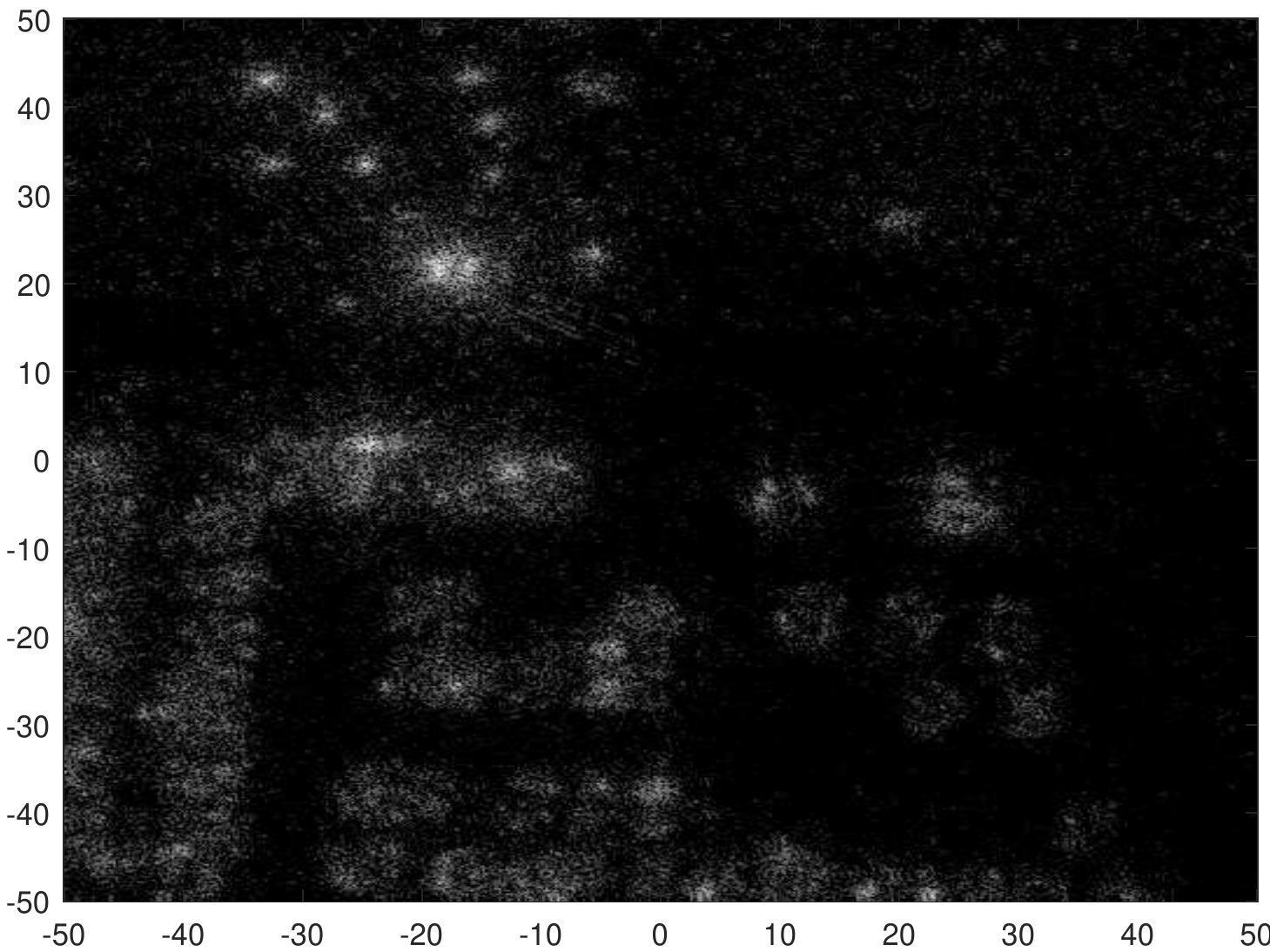}}
\subfigure[]{\includegraphics[width=.24\linewidth]{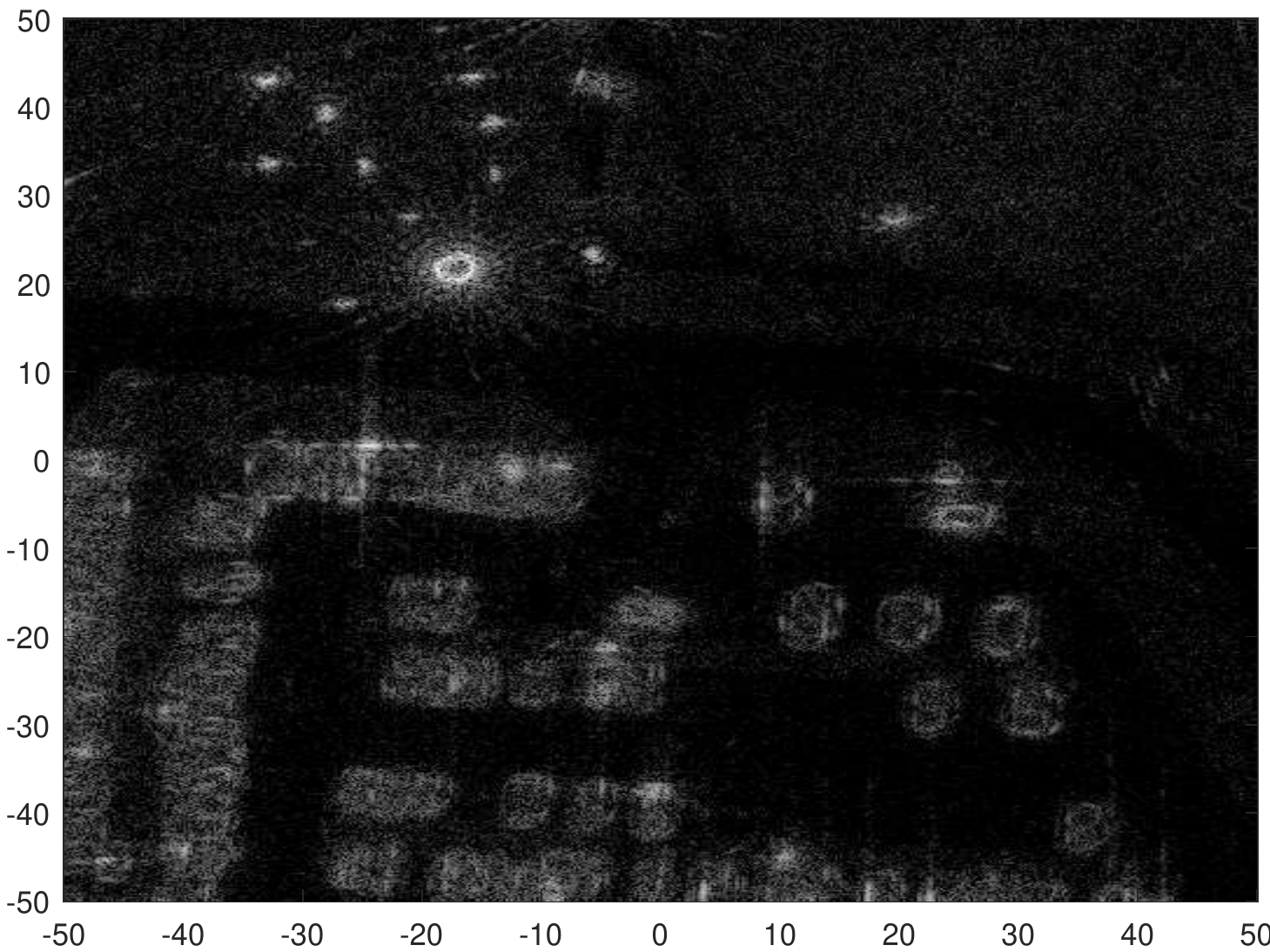}}
\subfigure[]{\includegraphics[width=.24\linewidth]{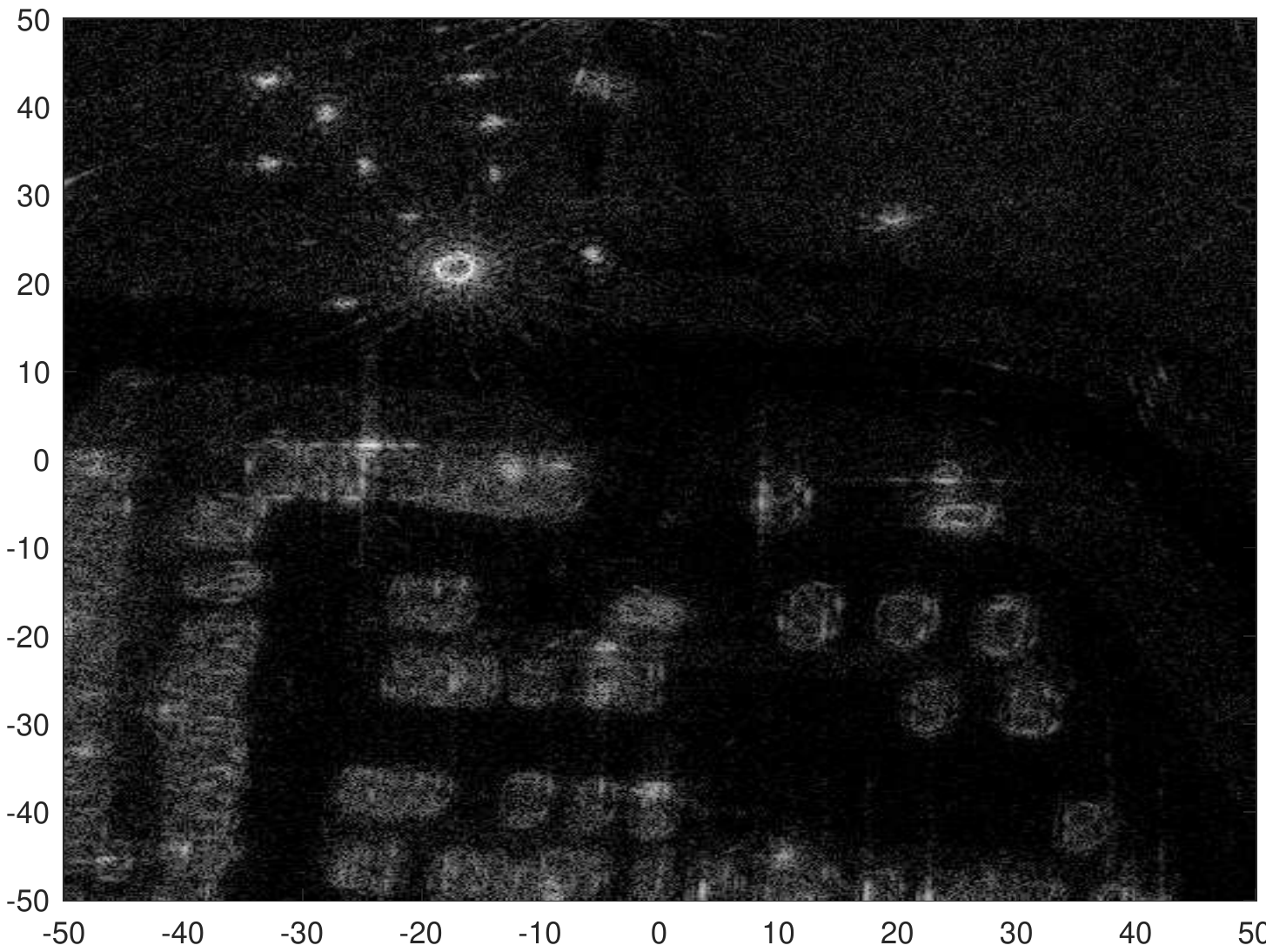}}\hfill

\centering
\subfigure[]{\includegraphics[width=.24\linewidth]{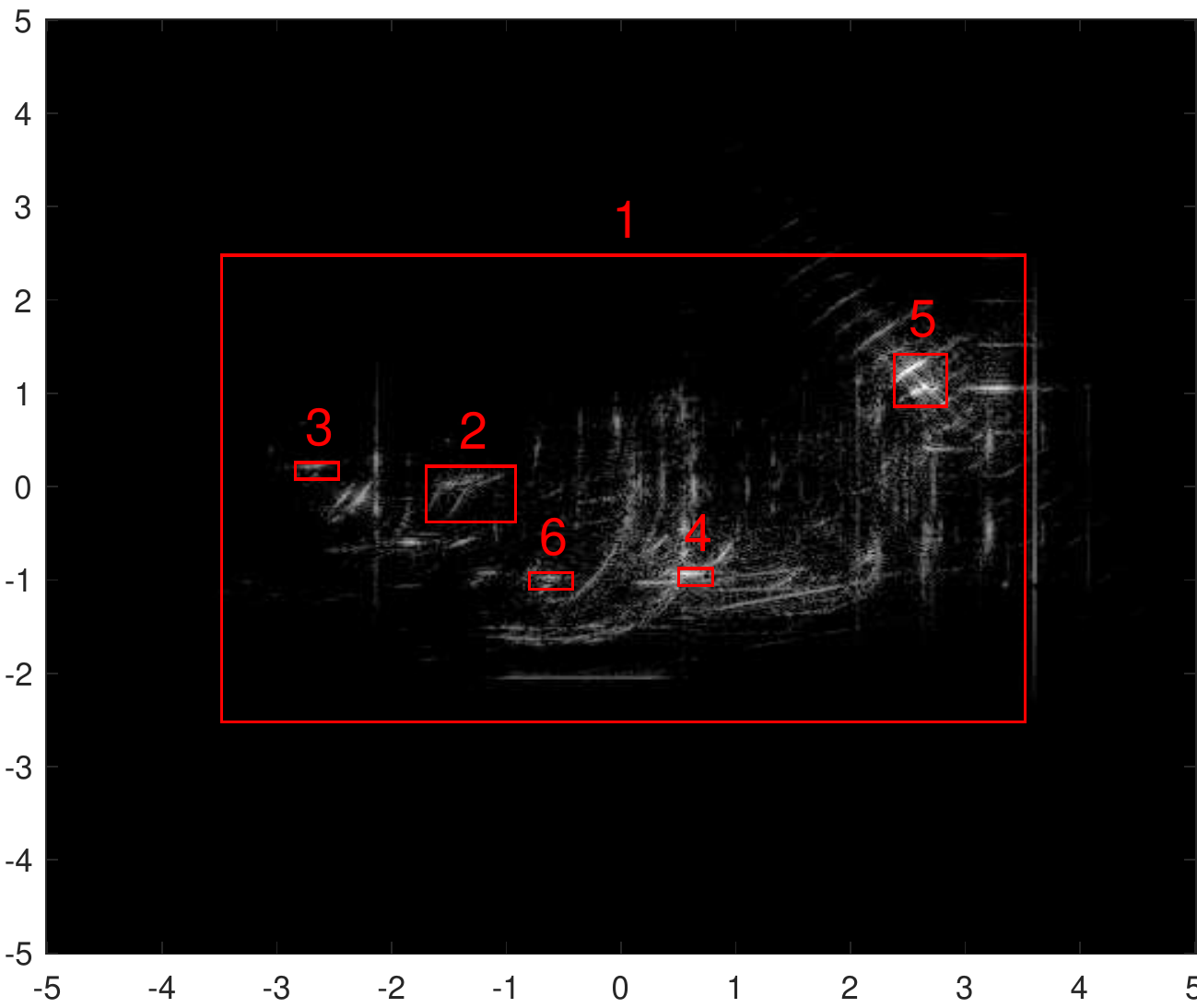}}
\subfigure[]{\includegraphics[width=.24\linewidth]{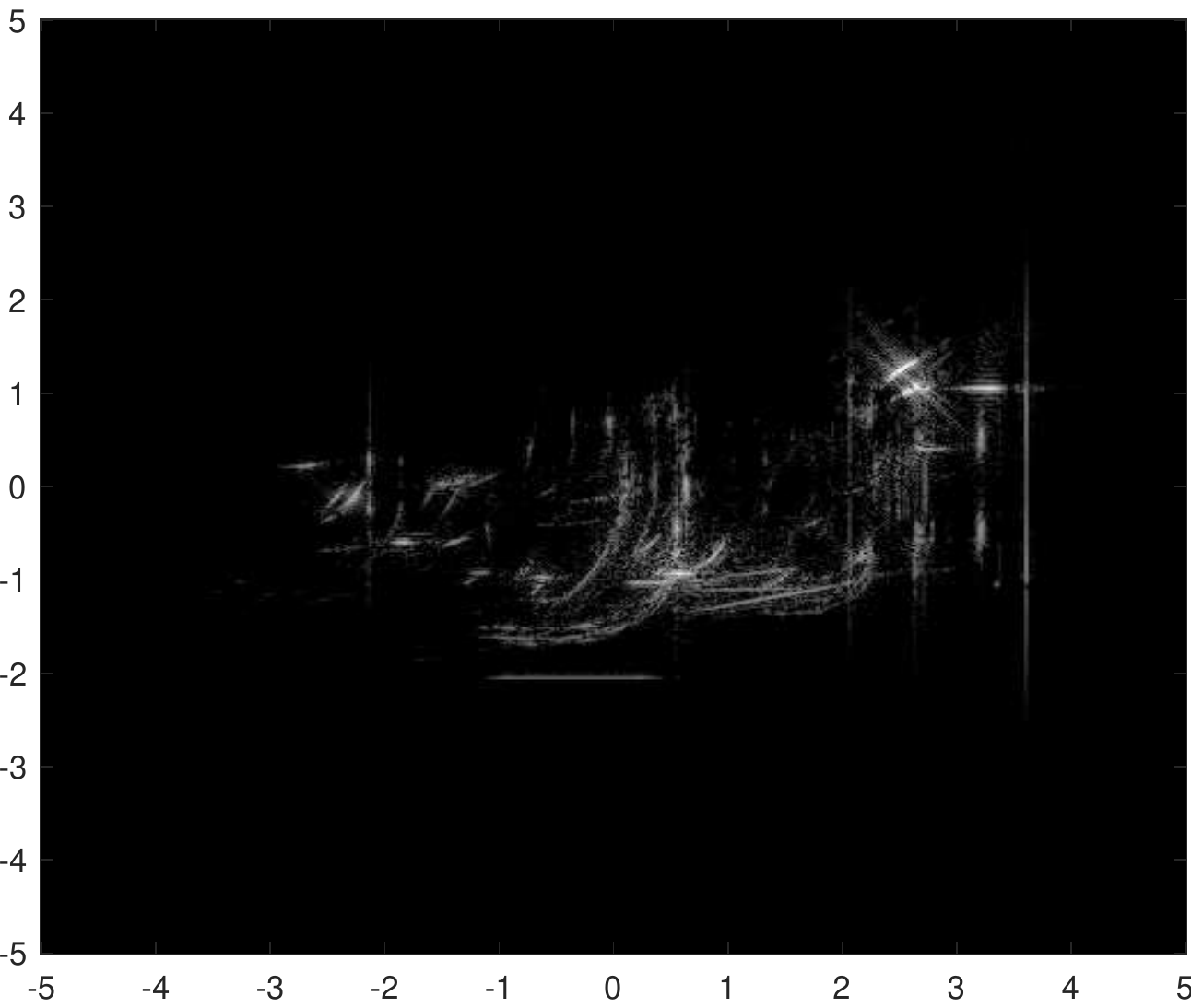}}
\subfigure[]{\includegraphics[width=.24\linewidth]{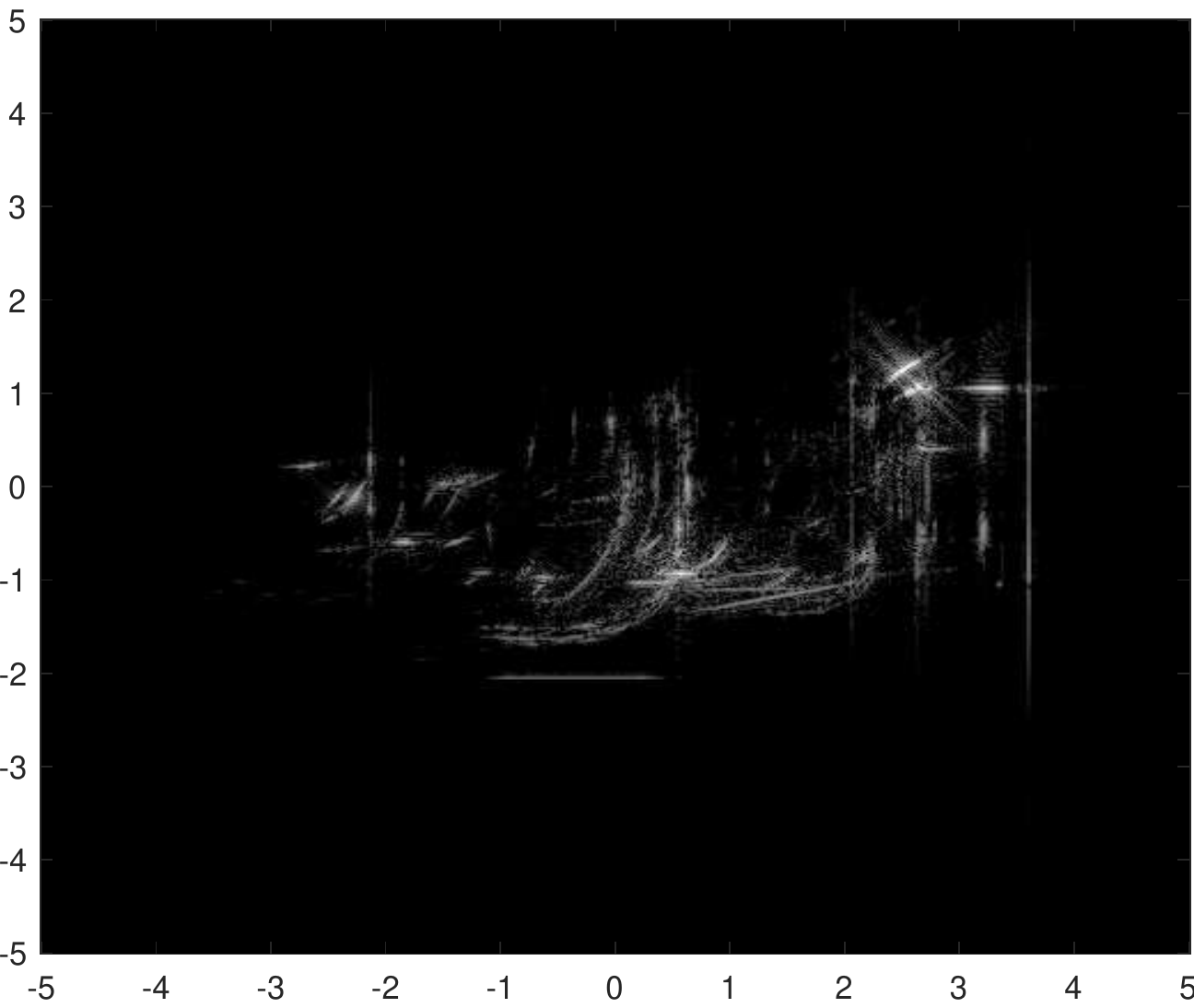}}
\subfigure[]{\includegraphics[width=.24\linewidth]{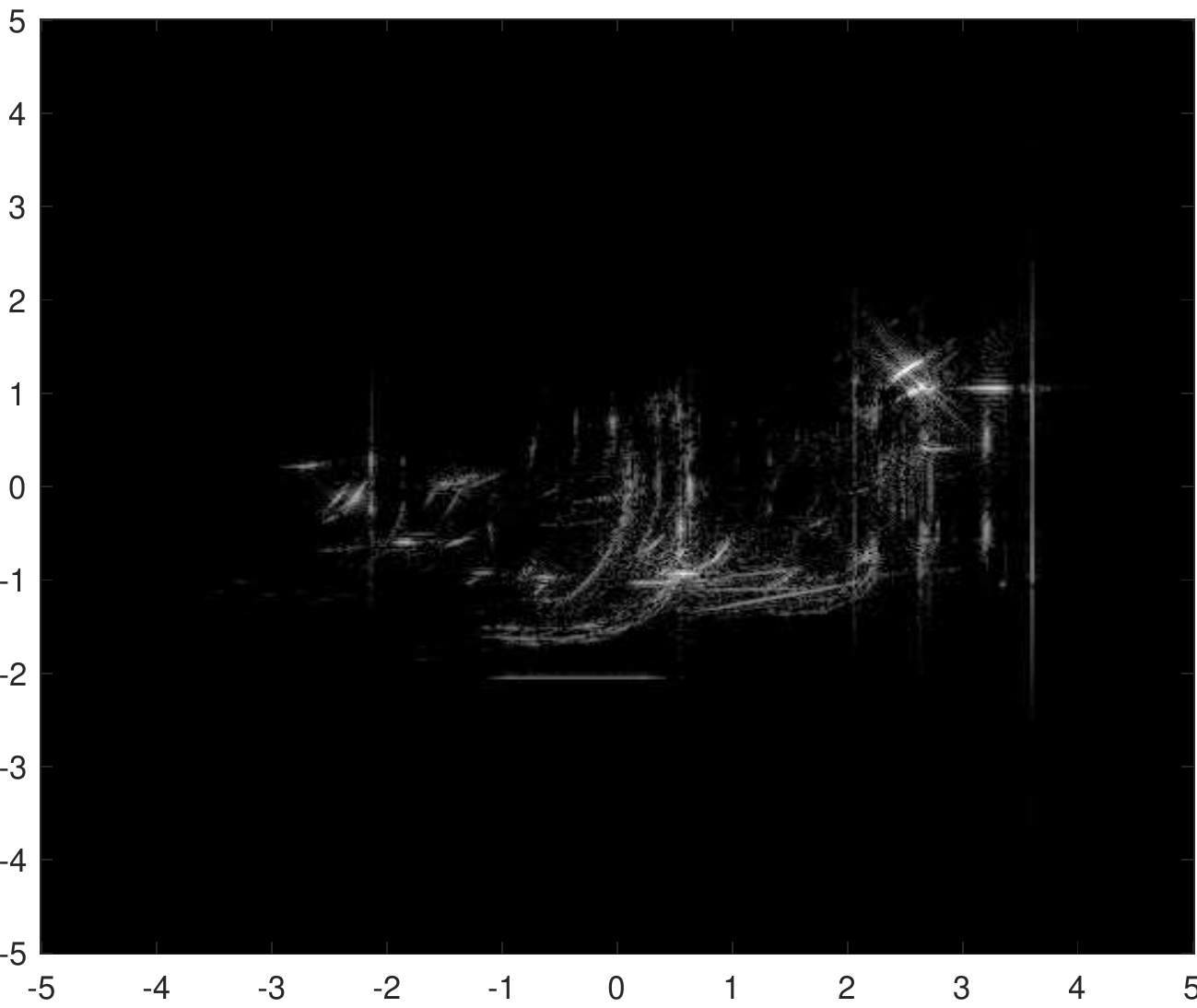}}\hfill

\centering
\subfigure[]{\includegraphics[width=.24\linewidth]{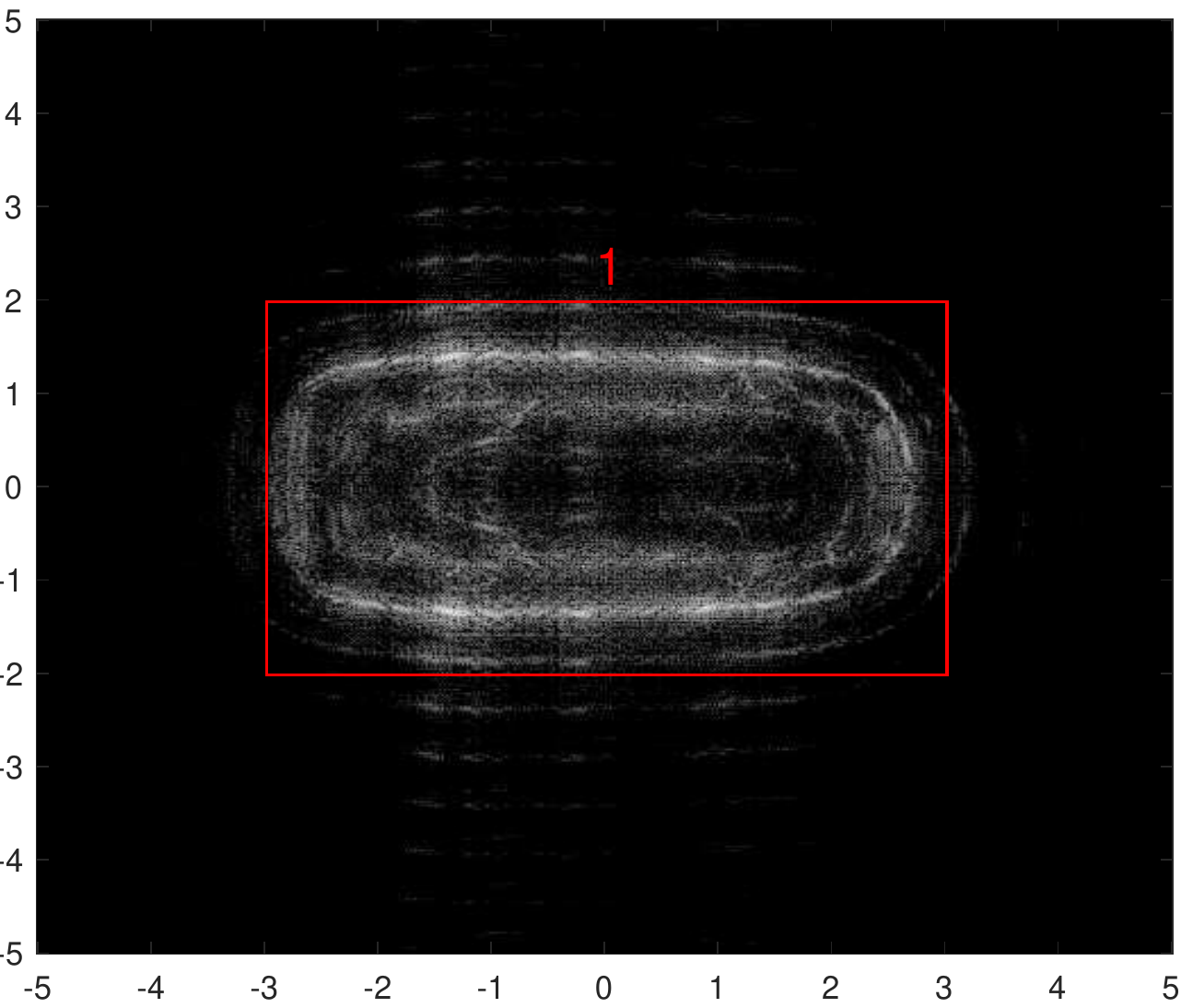}}
\subfigure[]{\includegraphics[width=.24\linewidth]{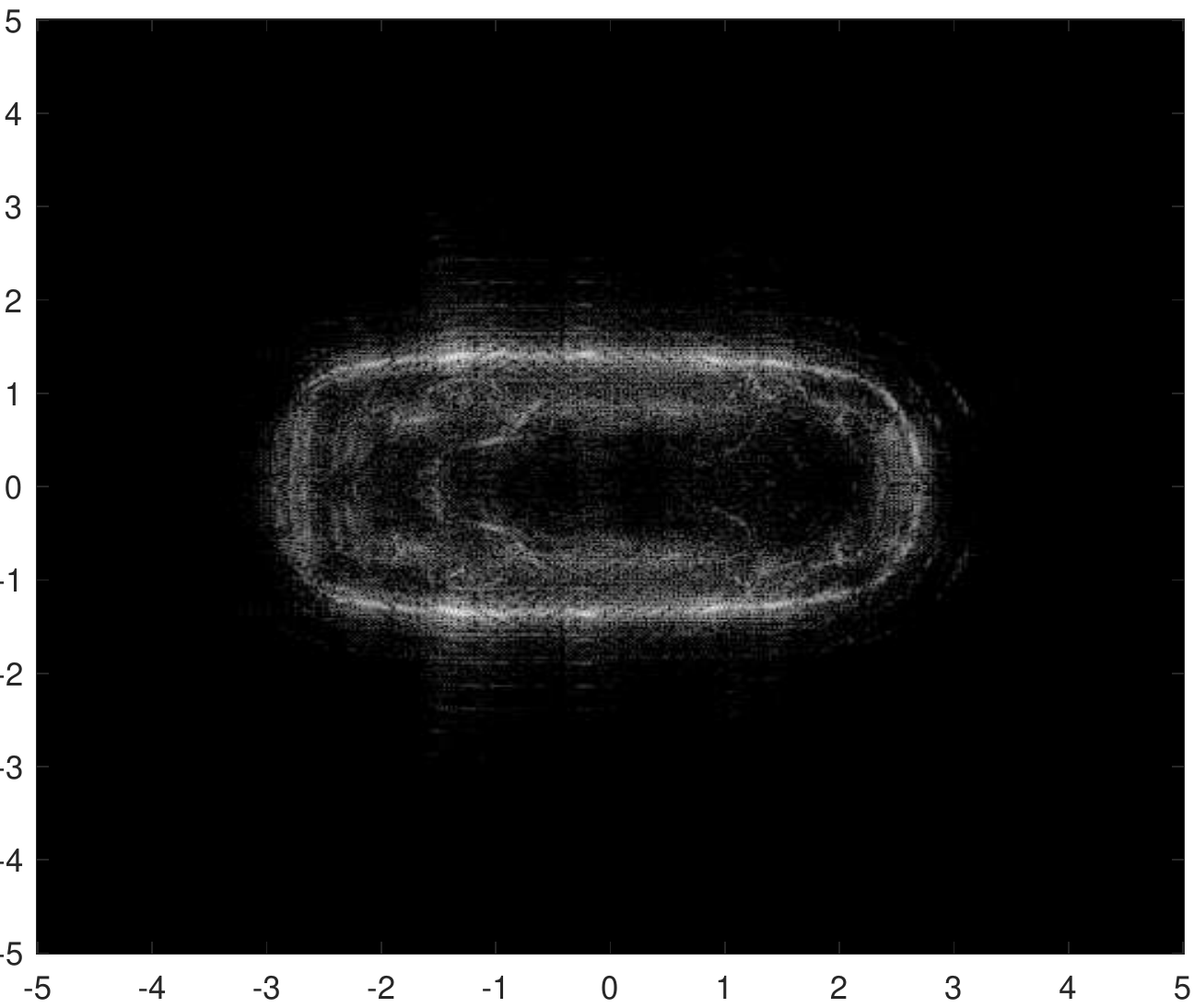}}
\subfigure[]{\includegraphics[width=.24\linewidth]{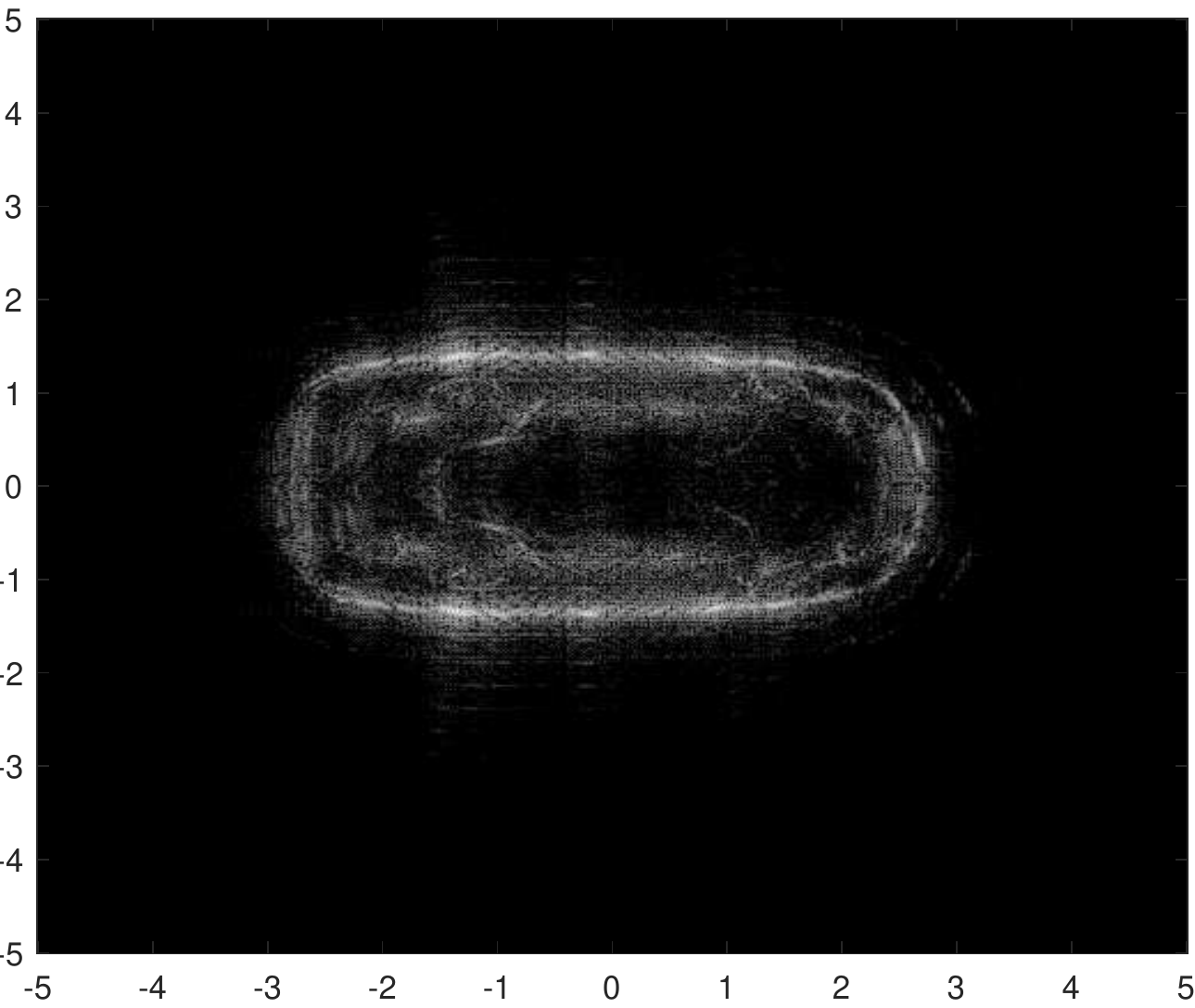}}
\subfigure[]{\includegraphics[width=.24\linewidth]{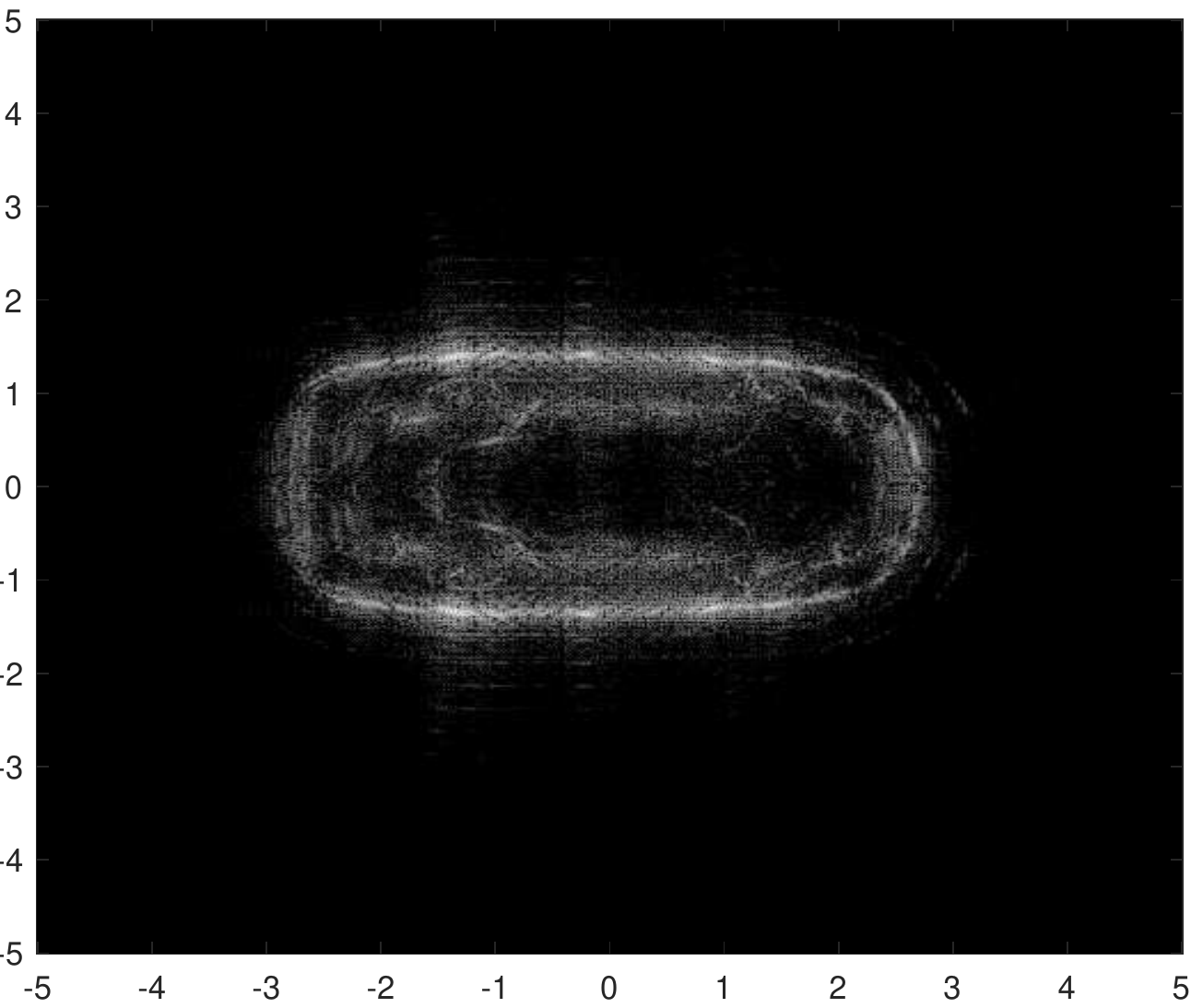}}

\caption{Reconstruction results for Gotcha, Backhoe and Civilian Vehicles (Camry) data sets at $1^{st}$, $2^{nd}$ and $3^{rd}$ rows, respectively. (a), (e) and (i) Matched filter. (b), (f) and (j) L1. (c), (g) and (k) GMC. (d), (h) and (l) Cauchy. Numbered rectangles in the first column refer to targets analysed in terms of RE values in Table \ref{tab:RE}.}
\label{fig:recons1}
\end{figure*}

The performance of the proposed method was compared to $L_1$ and GMC regularization-based methods, as well as the reconstructed image by the traditional back-projection (BP) method \cite{gorham2010sar}. We used the relative error (RE) \cite{wei2019improved}, which measures the bias between the reconstructed and matched-filtered results (i.e. the BP results) and is calculated as \cite{wei2019improved}
\begin{align}
    RE = \left| 10\log_{10} \left(\dfrac{\|\hat{X} \|^2}{\|X_{MF} \|^2} \right) \right|
\end{align}
where $\hat{X}$ refers to the reconstructed image 
and $X_{MF}$ is the back-projected SAR image. The lower the RE value 
the better the corresponding image reconstruction performance.

\begin{table}[htbp]
\setstretch{0.7}\small
  \centering
  \caption{Performance comparison in terms of RE values for various targets.}
    \begin{tabularx}{0.96\linewidth}{@{} CCCCCC @{}}
    \toprule
    \multirow{2}[0]{*}{Data set} && Target & L1    & GMC   & Cauchy \\
    &&&&&\\
    \toprule
    \multirow{7}[0]{*}{Gotcha} && T1    & 0.671 & 0.140 & \textbf{0.129} \\
          && T2    & 4.068 & 0.726 & \textbf{0.701} \\
          && T3    & 1.416 & 0.346 & \textbf{0.308} \\
          && T4    & 4.254 & 0.539 & \textbf{0.514} \\
          && T5    & 6.279 & 0.950 & \textbf{0.933} \\
          && T6    & 4.179 & \textbf{0.065} & 0.084 \\
          && T7    & 5.414 & \textbf{0.780} & 0.781 \\
          \midrule
    \multirow{6}[0]{*}{Backhoe} && T1    & 0.396 & 0.358 & \textbf{0.351} \\
          && T2    & 0.096 & 0.051 & \textbf{0.044} \\
          && T3    & 0.092 & 0.068 & \textbf{0.061} \\
          && T4    & 0.850 & 0.846 & \textbf{0.836} \\
          && T5    & 0.102 & 0.097 & \textbf{0.089} \\
          && T6    & 0.038 & 0.019 & \textbf{0.012} \\
          \midrule
          \multirow{3}[0]{*}{Civilian Vehicle} && Tacoma & 0.154 & 0.134 & \textbf{0.120} \\
          && Jeep93 & 0.843 & \textbf{0.781} & 0.782 \\
          && Camry & 0.408 & 0.374 & \textbf{0.370} \\

          \bottomrule
    \end{tabularx}%
  \label{tab:RE}%
\end{table}%

From GOTCHA and Backhoe data sets several targets were selected whilst for civilian vehicles the whole vehicle was selected as a single target for the comparisons performed in this section. These correspond to the red-numbered rectangles on the images in the first column of Figure \ref{fig:recons1}. All the reconstruction results 
are given in Figure \ref{fig:recons1}. All images were normalized between -50 and 0 dBs for a better visual analysis. Table \ref{tab:RE} presents RE values for all the methods and given targets for each data sets. The measurement matrix $\Phi$ given in (\ref{equ:SARRecons}) is the re-projection operation which was proposed in \cite{kelly2011iterative}, whilst $\Phi^T$ is the back-projection operation \cite{gorham2010sar}.

\begin{table*}[ht]
\setstretch{0.7}\small
  \centering
  \caption{SAR Despeckling performance comparison for various images and different simulated speckle noise.}
    \begin{tabularx}{0.96\textwidth}{@{} Cl | CCCC | CCCC @{}}
    \toprule
          &       & \multicolumn{4}{c|}{Gamma Speckle} & \multicolumn{4}{c}{Lognormal Speckle} \\ \toprule
          &       & \multicolumn{2}{c}{L = 5} & \multicolumn{2}{c|}{L = 15} & \multicolumn{2}{c}{L = 5} & \multicolumn{2}{c}{L = 15} \\ \toprule
          &       & \multicolumn{1}{c}{PSNR} & \multicolumn{1}{c}{S/MSE} & \multicolumn{1}{c}{PSNR} & \multicolumn{1}{c|}{S/MSE} & \multicolumn{1}{c}{PSNR} & \multicolumn{1}{c}{S/MSE} & \multicolumn{1}{c}{PSNR} & \multicolumn{1}{c}{S/MSE} \\
    \toprule
    \multirow{4}[2]{*}{Image-1} & Noisy & 18.767 & 6.989 & 23.372 & 11.758 & 18.891 & 6.997 & 23.419 & 11.766 \\
          & L1    & 17.417 & 6.270 & 18.088 & 6.942 & 17.523 & 6.376 & 18.105 & 6.959 \\
          & TV    & 22.364 & \textbf{11.218} & 23.317 & 12.172 & 22.561 & 11.416 & 23.337 & 12.192 \\
          & Cauchy & \textbf{22.370} & 11.051 & \textbf{25.087} & \textbf{13.804} & \textbf{22.783} & \textbf{11.454} & \textbf{25.178} & \textbf{13.895} \\
    \midrule
    \multirow{4}[2]{*}{Image-2} & Noisy & 15.829 & 6.990 & 20.080 & 11.764 & 16.179 & 6.994 & 20.210 & 11.762 \\
          & L1    & 13.239 & 6.172 & 13.878 & 6.813 & 13.336 & 6.269 & 13.891 & 6.826 \\
          & TV    & 16.746 & 9.680 & 17.576 & 10.511 & 16.895 & 9.830 & 17.592 & 10.527 \\
          & Cauchy & \textbf{18.284} & \textbf{10.746} & \textbf{20.654} & \textbf{13.241} & \textbf{18.669} & \textbf{11.112} & \textbf{20.732} & \textbf{13.318} \\
    \midrule
    \multirow{4}[2]{*}{Image-3} & Noisy & 15.595 & 6.989 & 19.997 & 11.761 & 15.850 & 6.987 & 20.038 & 11.759 \\
          & L1    & 14.197 & 6.088 & 14.818 & 6.709 & 14.290 & 6.180 & 14.833 & 6.724 \\
          & TV    & 18.610 & 10.501 & 19.375 & 11.266 & 18.761 & 10.651 & 19.389 & 11.280 \\
          & Cauchy & \textbf{18.810} & \textbf{10.667} & \textbf{21.211} & \textbf{13.096} & \textbf{19.172} & \textbf{11.025} & \textbf{21.282} & \textbf{13.166} \\
    \midrule
    \multirow{4}[2]{*}{Image-4} & Noisy & 16.683 & 6.988 & 21.009 & 11.761 & 17.003 & 6.987 & 21.146 & 11.759 \\
          & L1    & 13.563 & 5.789 & 14.098 & 6.325 & 13.645 & 5.870 & 14.108 & 6.335 \\
          & TV    & 16.010 & 8.238 & 16.547 & 8.775 & 16.108 & 8.336 & 16.556 & 8.784 \\
          & Cauchy & \textbf{17.993} & \textbf{9.831} & \textbf{21.413} & \textbf{12.400} & \textbf{18.289} & \textbf{10.116} & \textbf{21.585} & \textbf{12.512} \\
    \midrule
    \multirow{4}[2]{*}{Image-5} & Noisy & 16.180 & 6.995 & 20.317 & 11.759 & 16.533 & 6.991 & 20.433 & 11.762 \\
          & L2    & 14.119 & 6.417 & 14.836 & 7.134 & 14.228 & 6.525 & 14.849 & 7.147 \\
          & TV    & \textbf{19.939} & \textbf{12.238} & 21.185 & 13.484 & \textbf{20.183} & \textbf{12.482} & 21.206 & 13.505 \\
          & Cauchy & 19.547 & 11.512 & \textbf{22.628} & \textbf{14.734} & 20.013 & 11.961 & \textbf{22.741} & \textbf{14.840} \\
    \bottomrule
    \end{tabularx}%
  \label{tab:despeckling}%
\end{table*}%

On examining Figure \ref{fig:recons1}, for all data sets it is clear that all methods achieve improvements over the traditional BP technique. Specifically, GMC and Cauchy reconstruction results are very similar, whereas $L_1$ reconstruction exhibits a slightly worse performance, especially for the GOTCHA (cf. Figure \ref{fig:recons1}-(d)). 

For the chosen targets from each data set, Table \ref{tab:RE} shows a better performance for the proposed method over GMC and $L_1$. It is obvious from the Table \ref{tab:RE} that the GMC and Cauchy based methods achieve very close RE results, but Cauchy is the one achieving the smallest RE values for most targets. For all the data sets, $L_1$ falls short in terms of RE values. These results show that the GMC and Cauchy based penalty function can lead to better estimation than the $L_1$-based approach. 

In addition to the smaller RE values and good visual results, the Cauchy based method is also much less computational expensive than its $L_1$ and GMC counterparts. Specifically, GMC is twice slower than $L_1$ and Cauchy in a single iteration, however it converges in less number of iterations than $L_1$, which makes the reconstruction time for GMC and $L_1$ very similar. Despite having similar duration for a single iteration, the Cauchy based penalty necessitates around 10-15 iterations to reach the results with corresponding $\varepsilon$ value of $10^{-3}$, whilst GMC and $L_1$ need around 250-300 and 450-500 iterations, respectively. This is a very significant gain for a single iteration of the FB algorithm. As an example, for the GOTCHA data set, due to the high number of samples, a single iteration takes around 55-60 seconds. Overall, The proposed Cauchy based method converges in around 10-12 minutes whilst it takes around 8 hours for $L_1$ and GMC.

\subsection{Despeckling}
\begin{figure*}[ht]
\centering
\subfigure[]{\includegraphics[width=.19\linewidth]{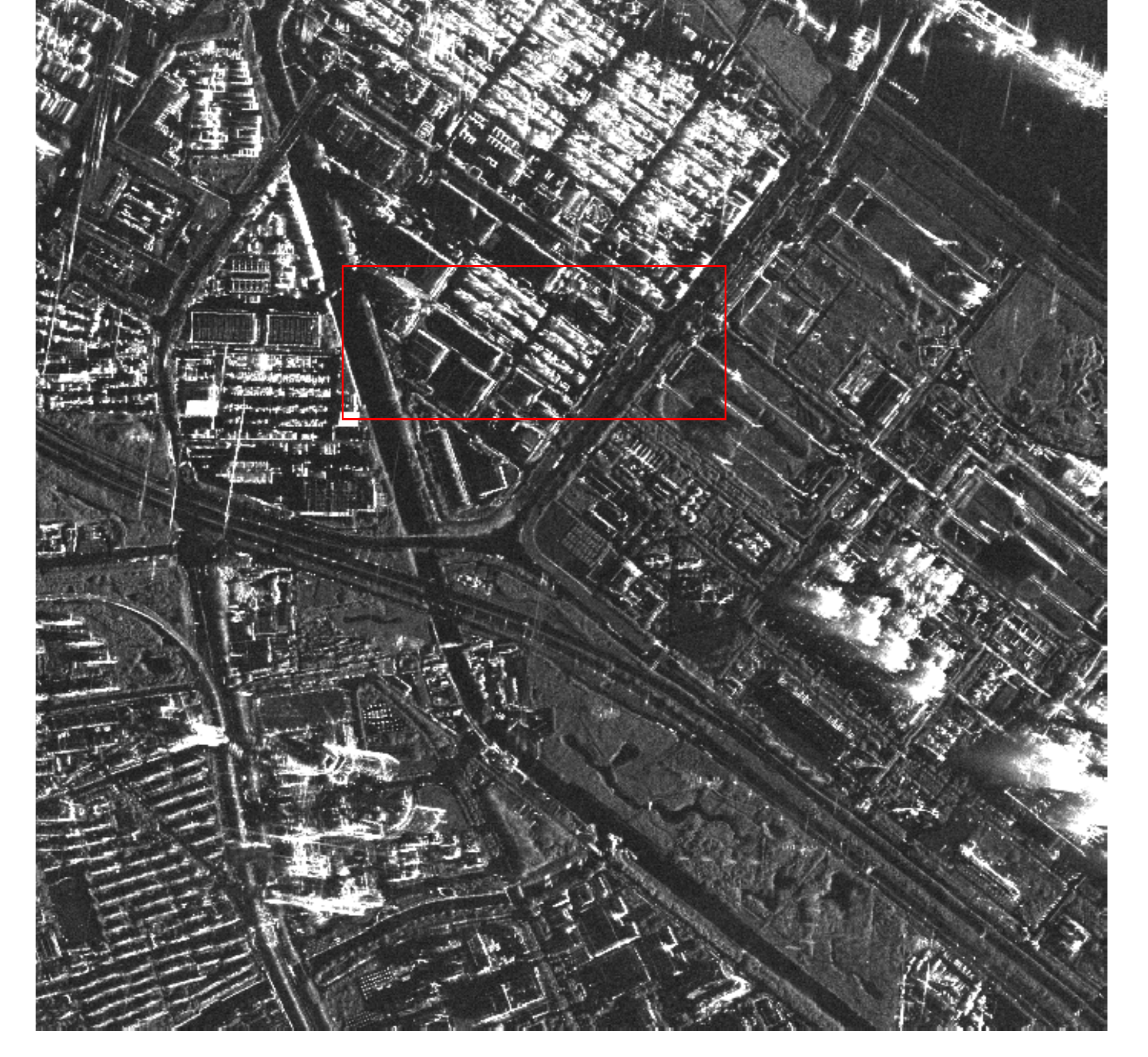}}
\subfigure[]{\includegraphics[width=.19\linewidth]{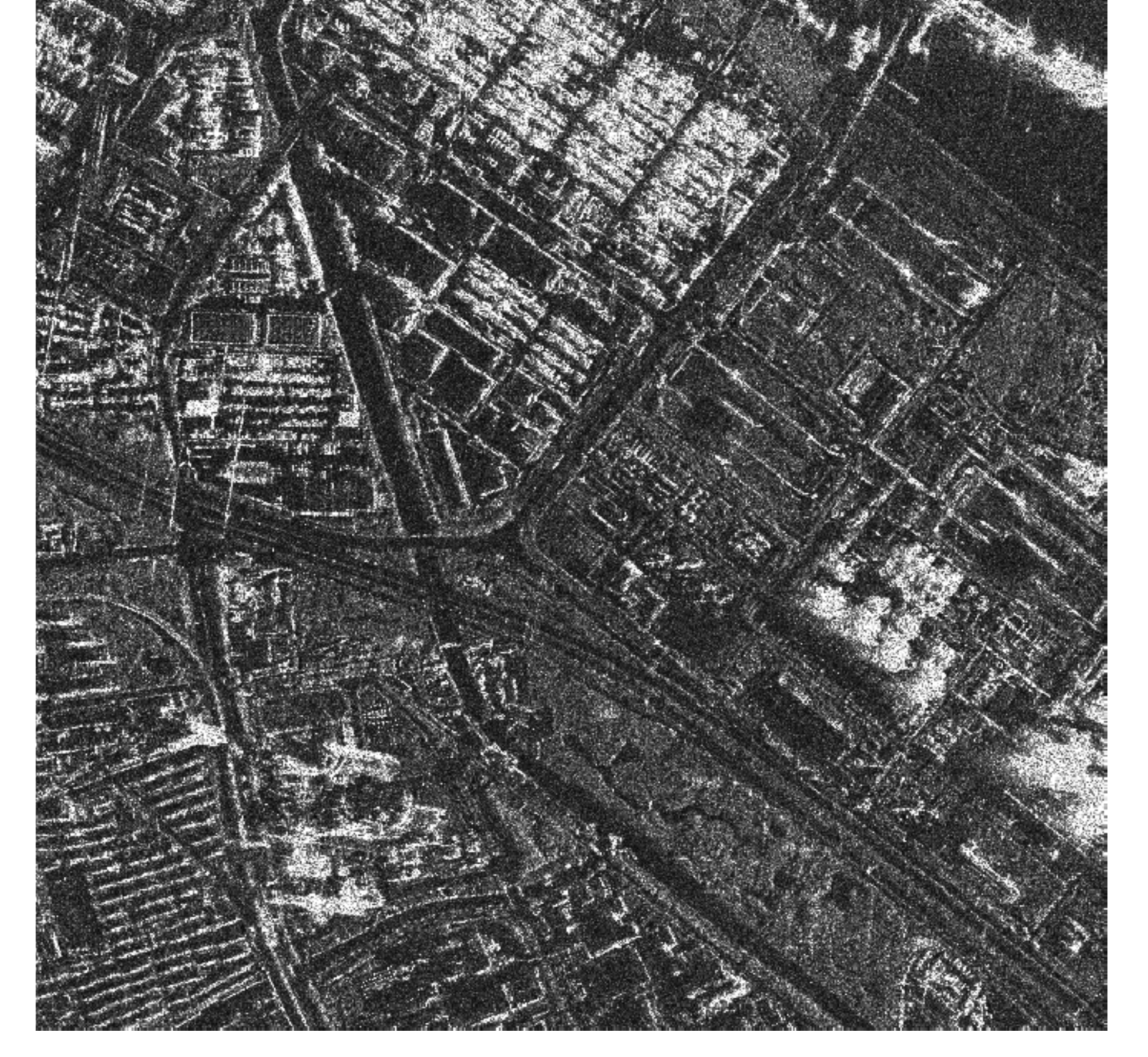}}
\subfigure[]{\includegraphics[width=.19\linewidth]{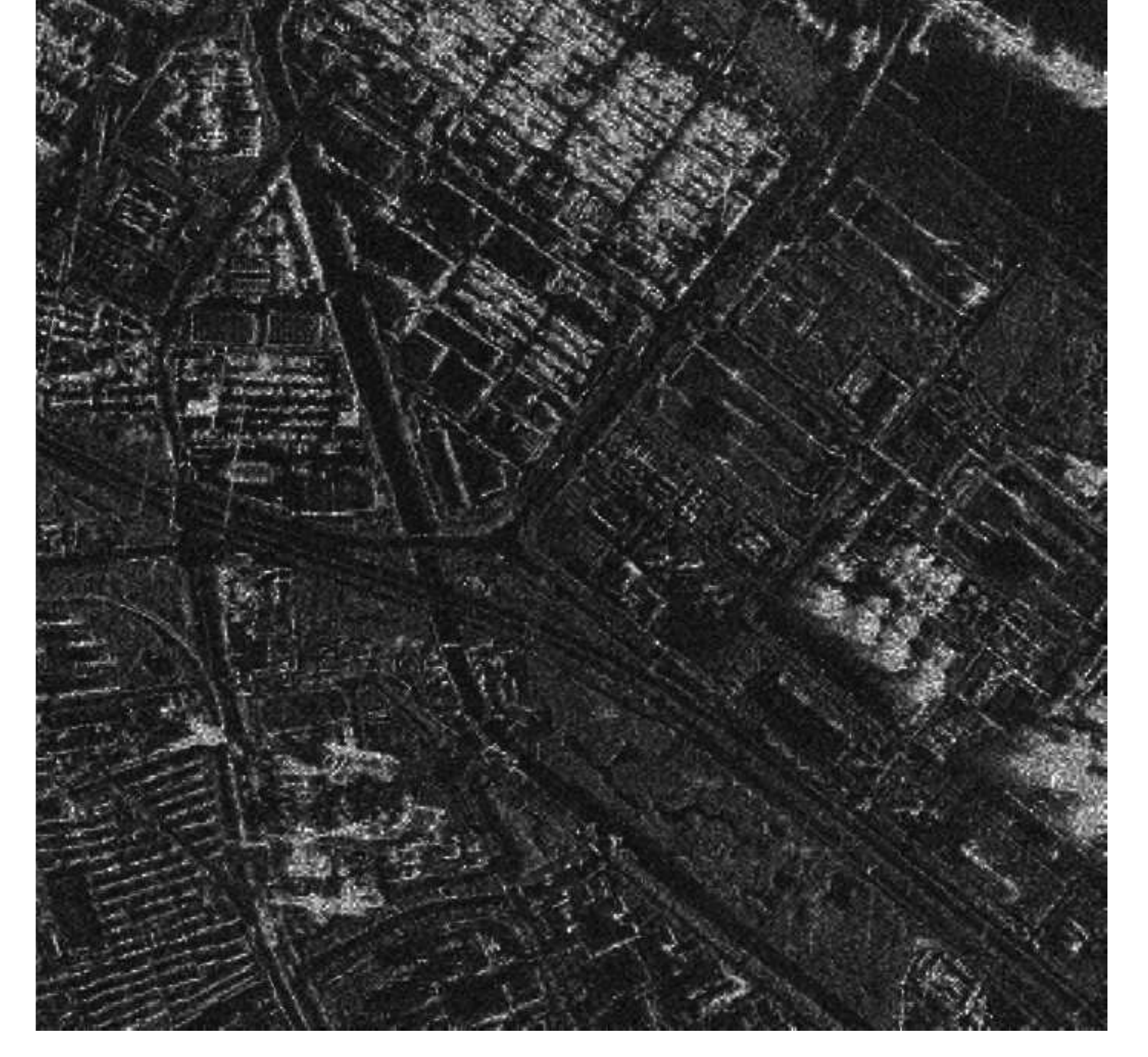}}
\subfigure[]{\includegraphics[width=.19\linewidth]{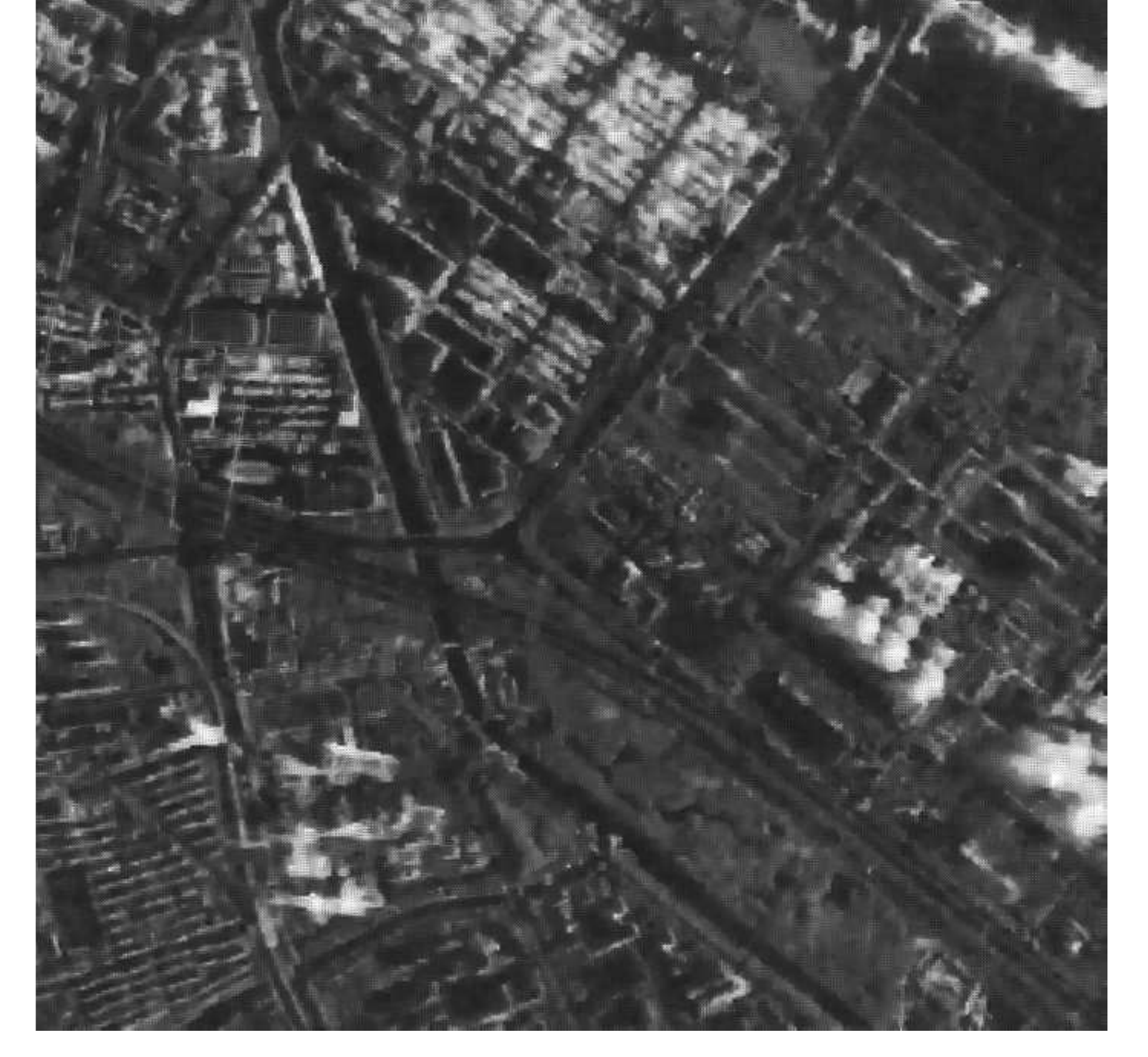}}
\subfigure[]{\includegraphics[width=.19\linewidth]{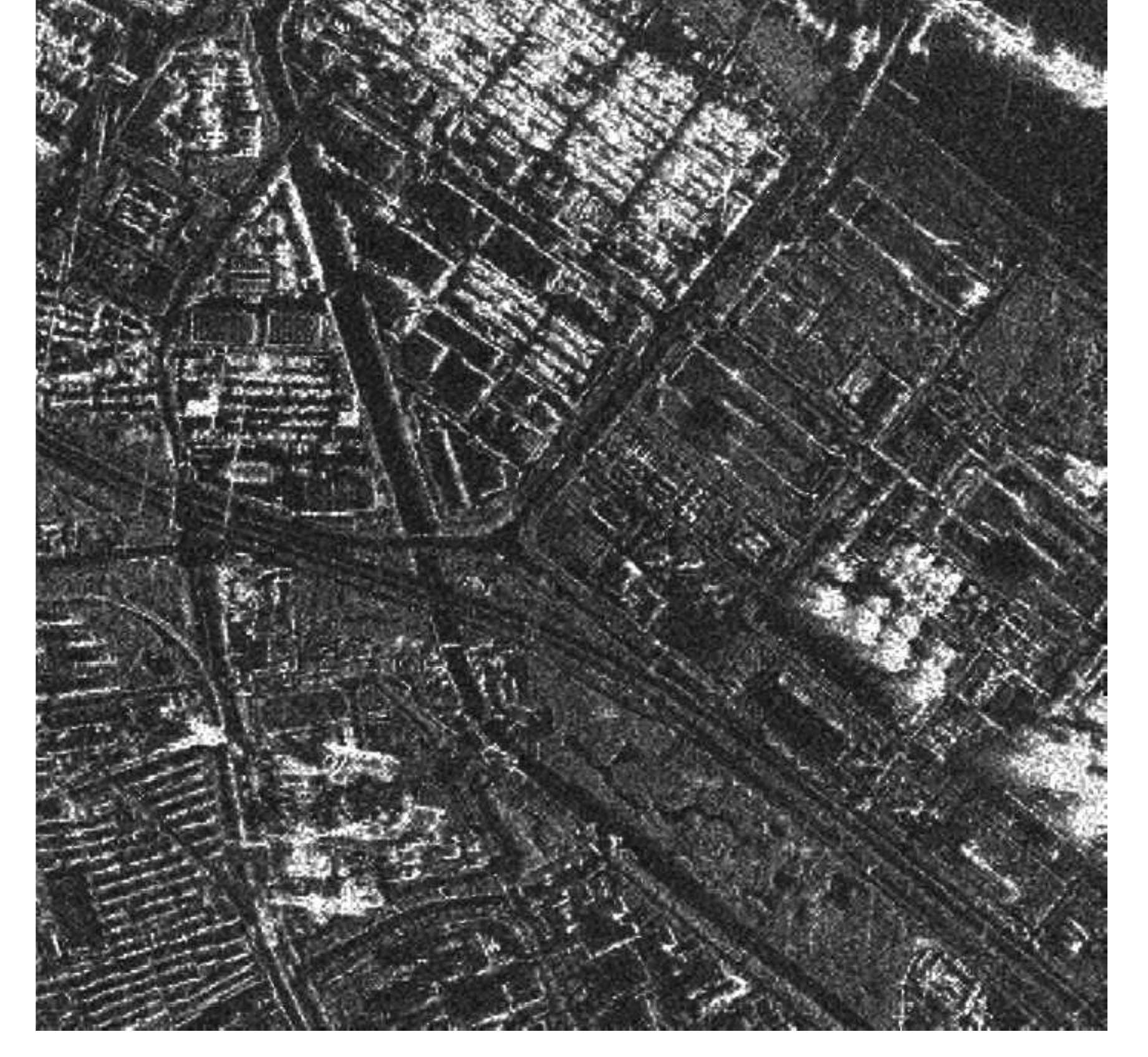}}\hfill\\

\subfigure[]{\includegraphics[width=.19\linewidth]{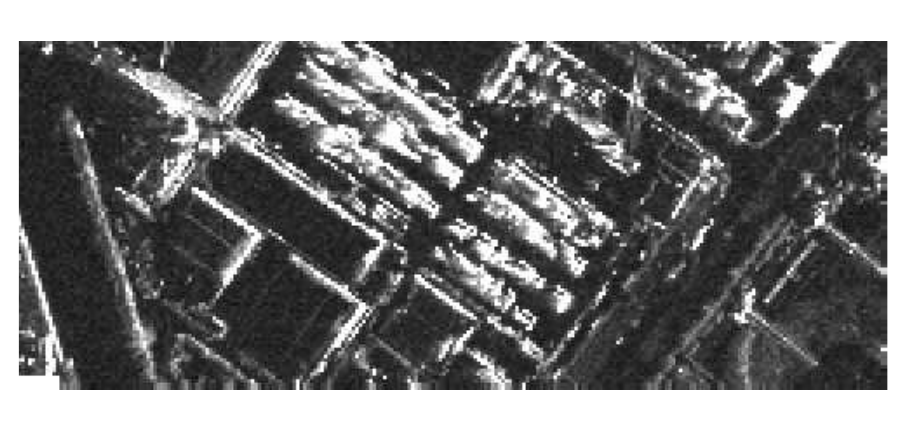}}
\subfigure[]{\includegraphics[width=.19\linewidth]{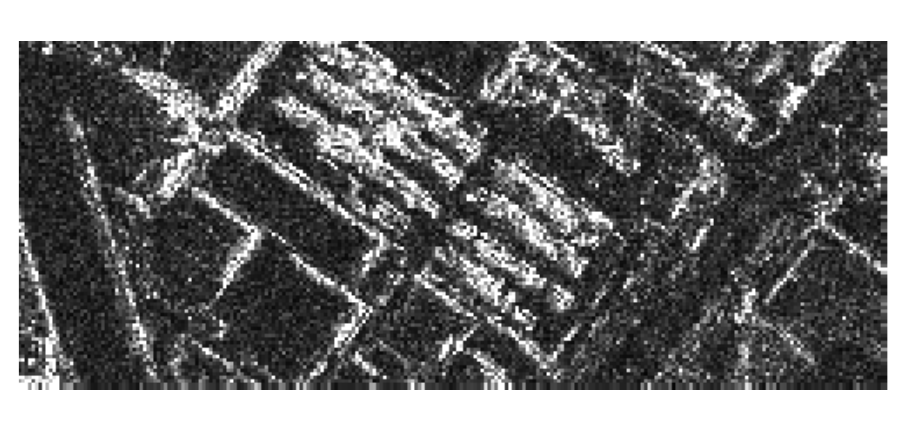}}
\subfigure[]{\includegraphics[width=.19\linewidth]{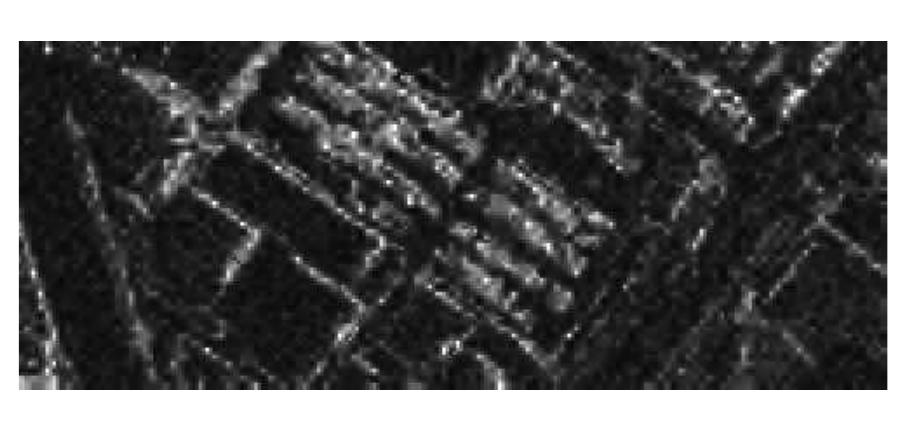}}
\subfigure[]{\includegraphics[width=.19\linewidth]{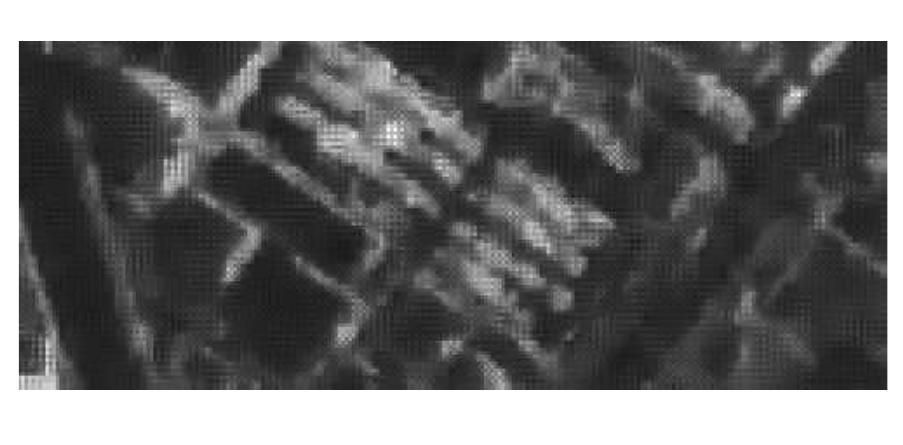}}
\subfigure[]{\includegraphics[width=.19\linewidth]{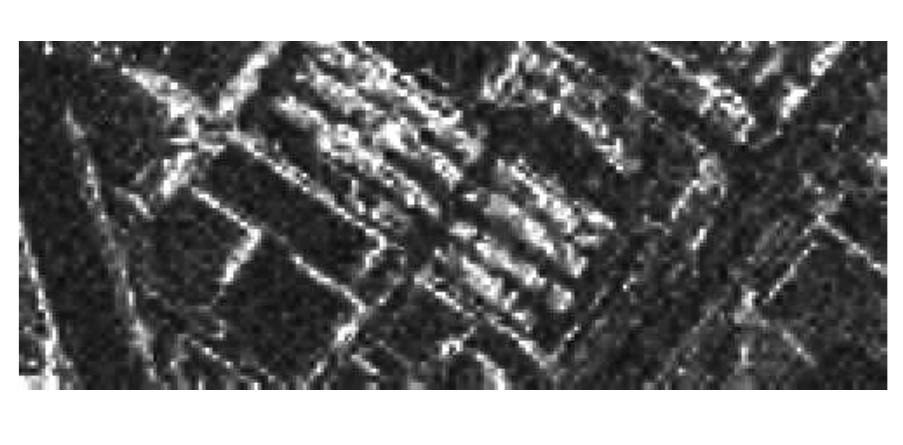}}\hfill\\

\subfigure[]{\includegraphics[width=.19\linewidth]{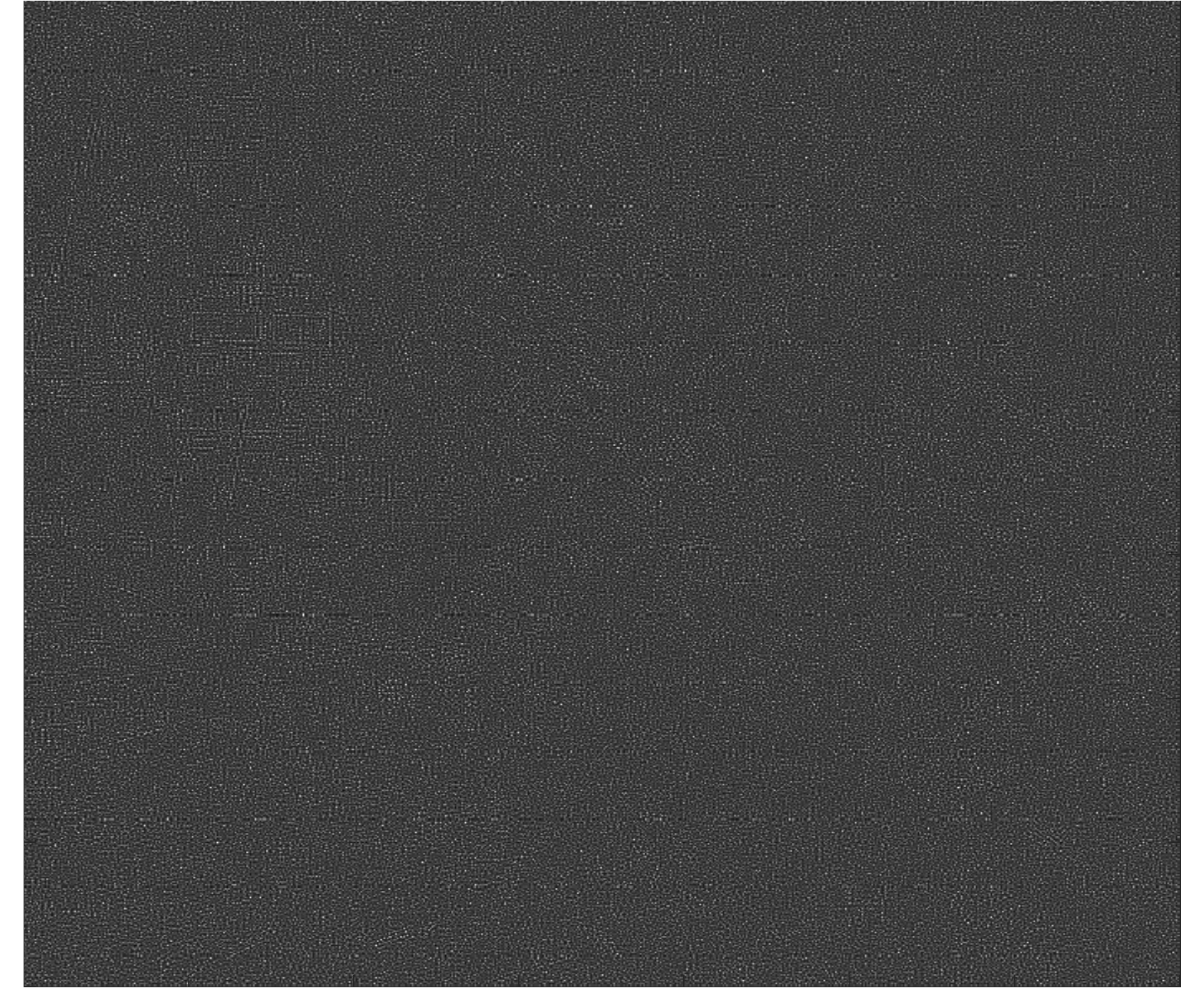}}
\subfigure[]{\includegraphics[width=.19\linewidth]{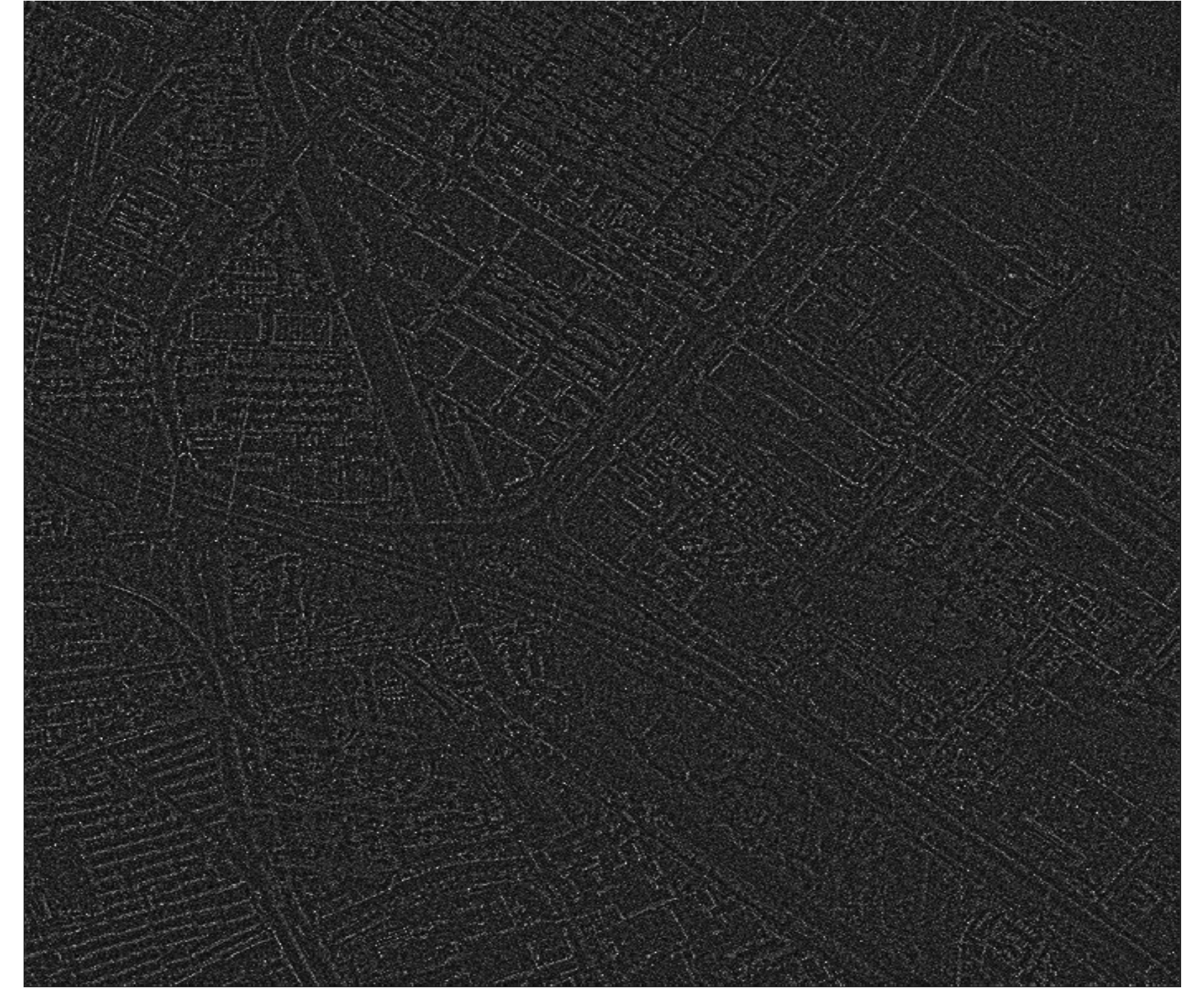}}
\subfigure[]{\includegraphics[width=.19\linewidth]{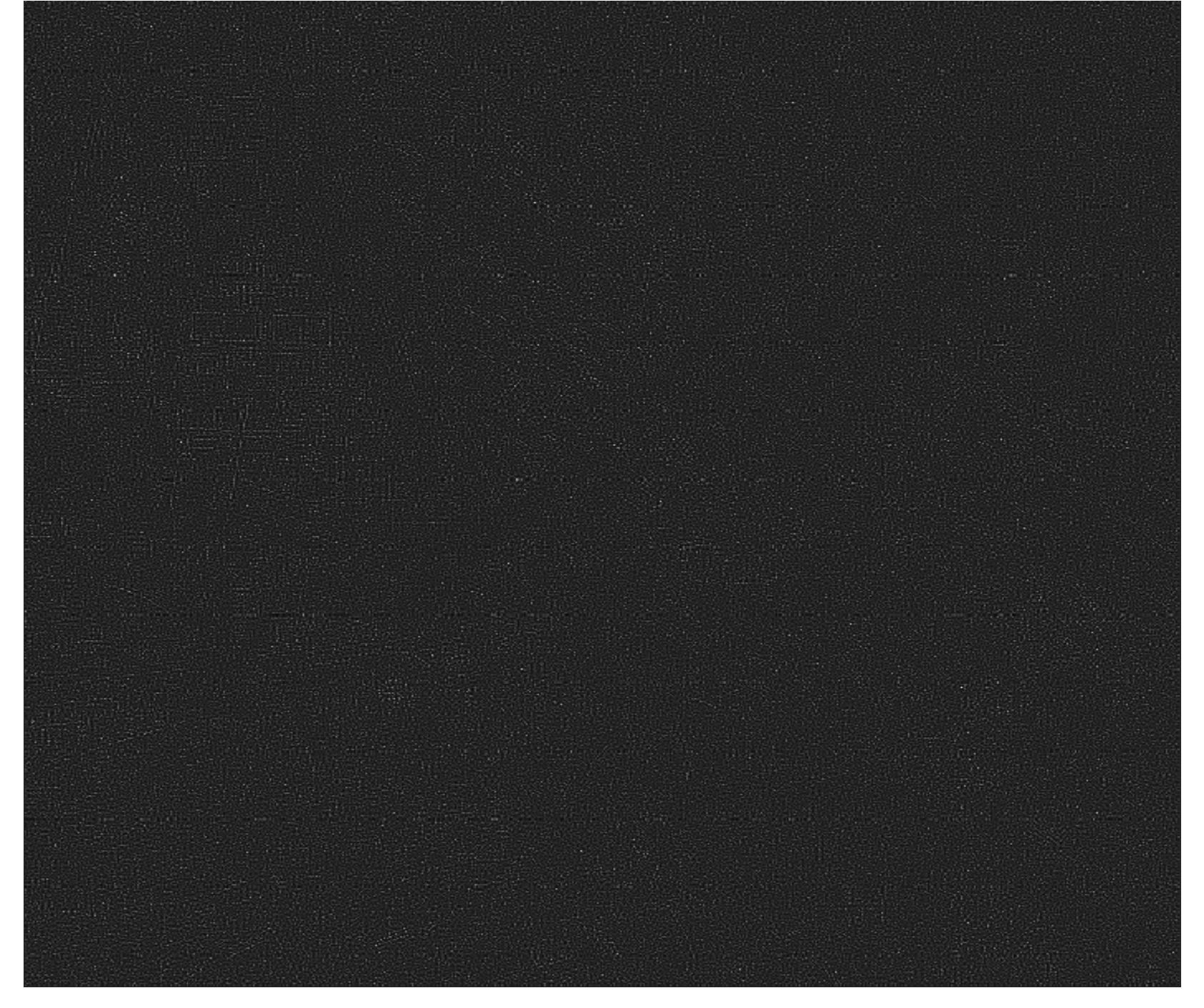}}\hfill
\caption{Despeckling results for Image-4 under Gamma speckle with L = 5. (a) and (f) Original image, (b) and (g) Speckle image, (c) and (h) L1, (d) and (i) TV, (e) and (j) Cauchy. (k), (l) and (m) are ratio images for L1, TV and Cauchy, respectively. Images on the second row represent zoomed-in images of the rectangle given in (a).}
\label{fig:ds1}
\end{figure*}

\begin{figure*}[ht!]
\centering
\subfigure[]{\includegraphics[width=.24\linewidth]{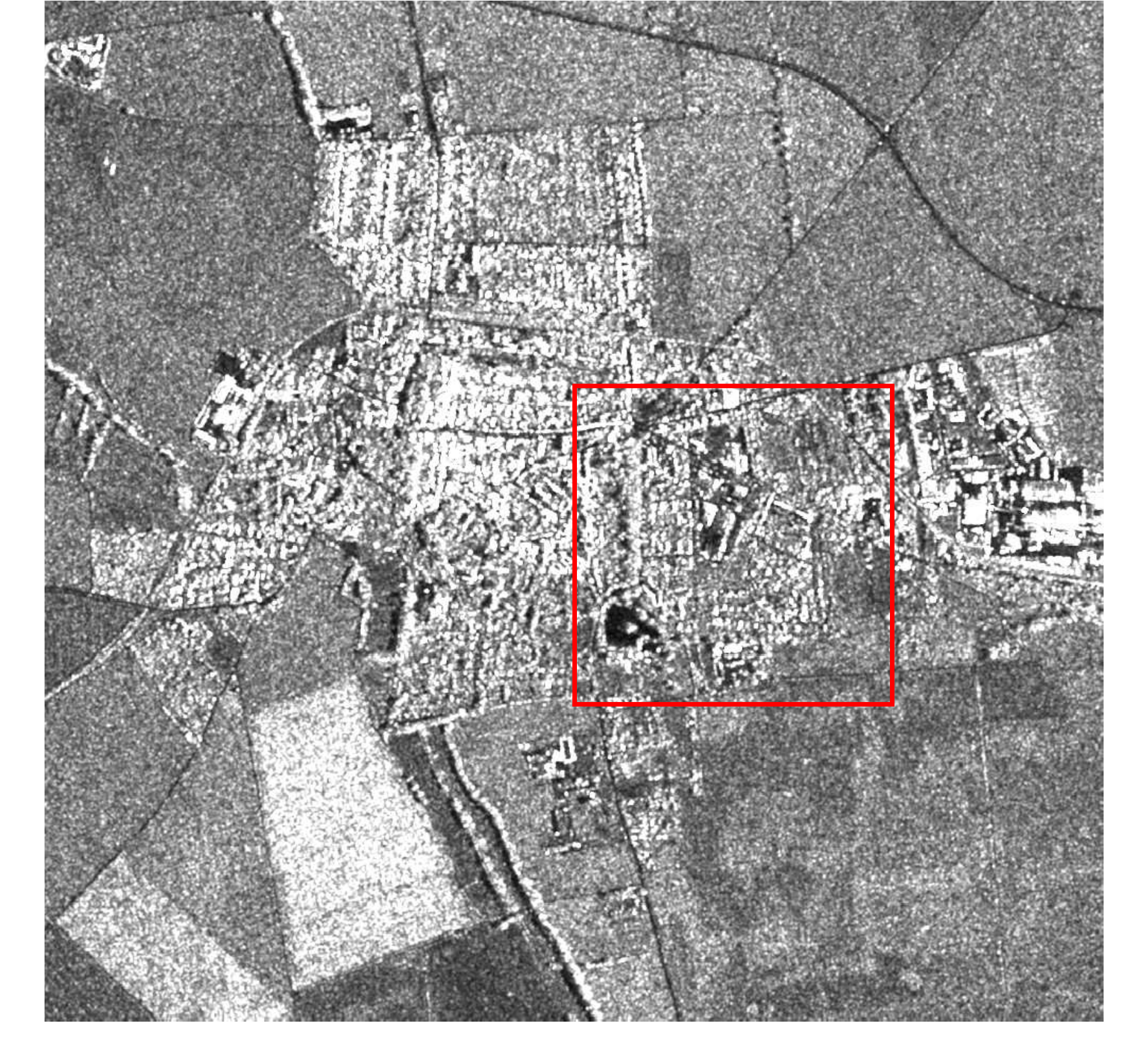}}
\subfigure[]{\includegraphics[width=.24\linewidth]{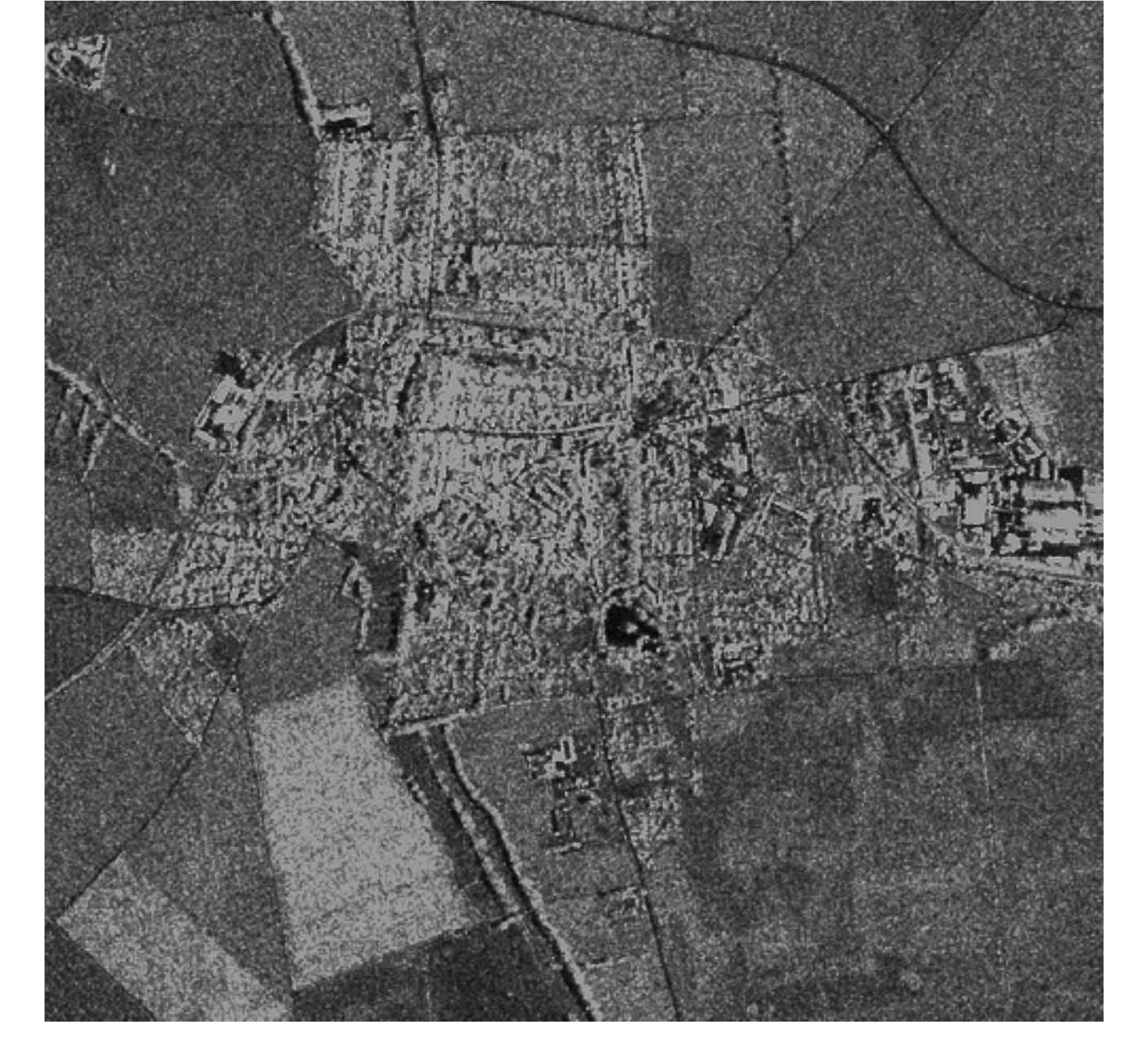}}
\subfigure[]{\includegraphics[width=.24\linewidth]{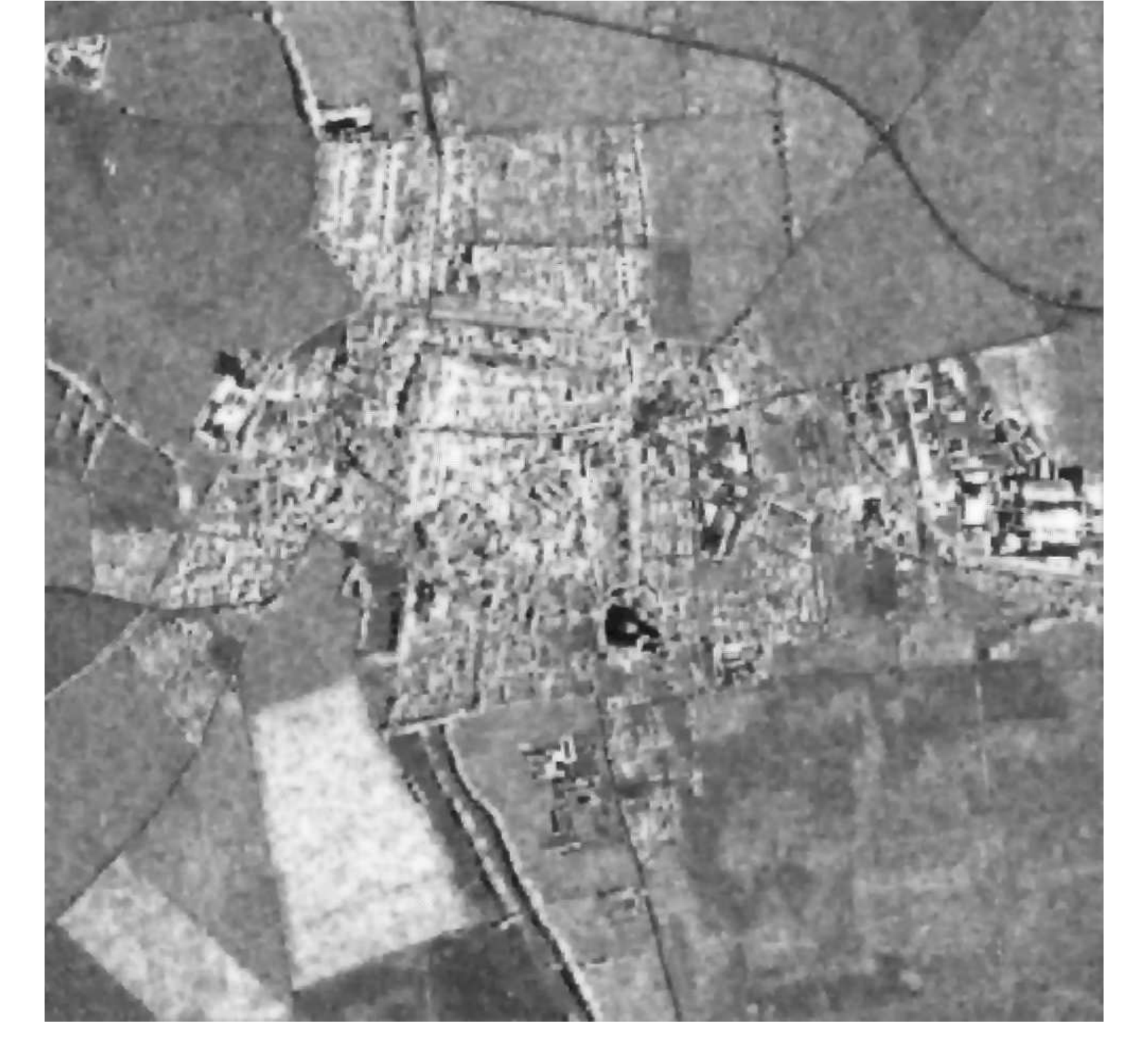}}
\subfigure[]{\includegraphics[width=.24\linewidth]{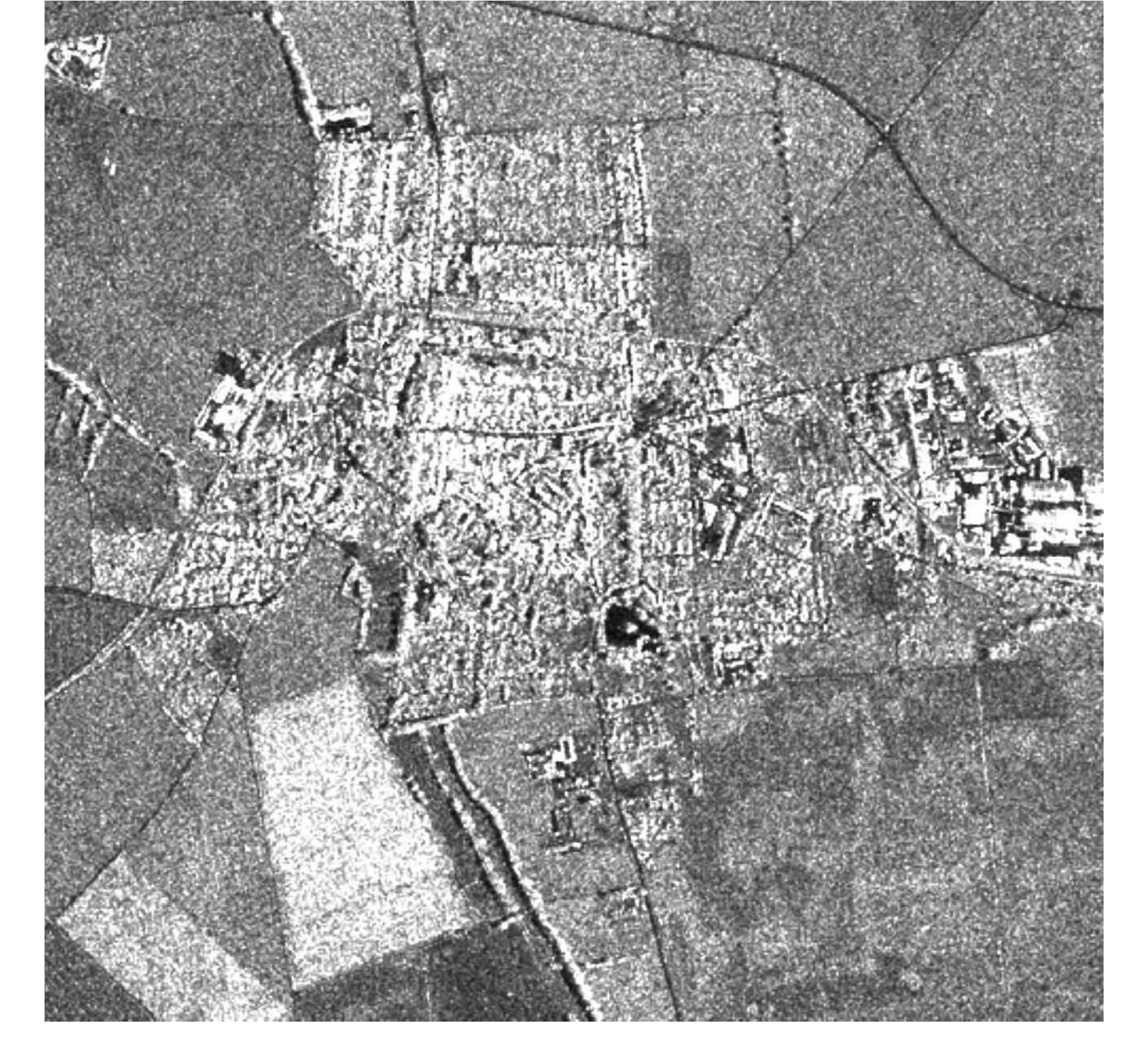}}\hfill

\subfigure[]{\includegraphics[width=.18\linewidth]{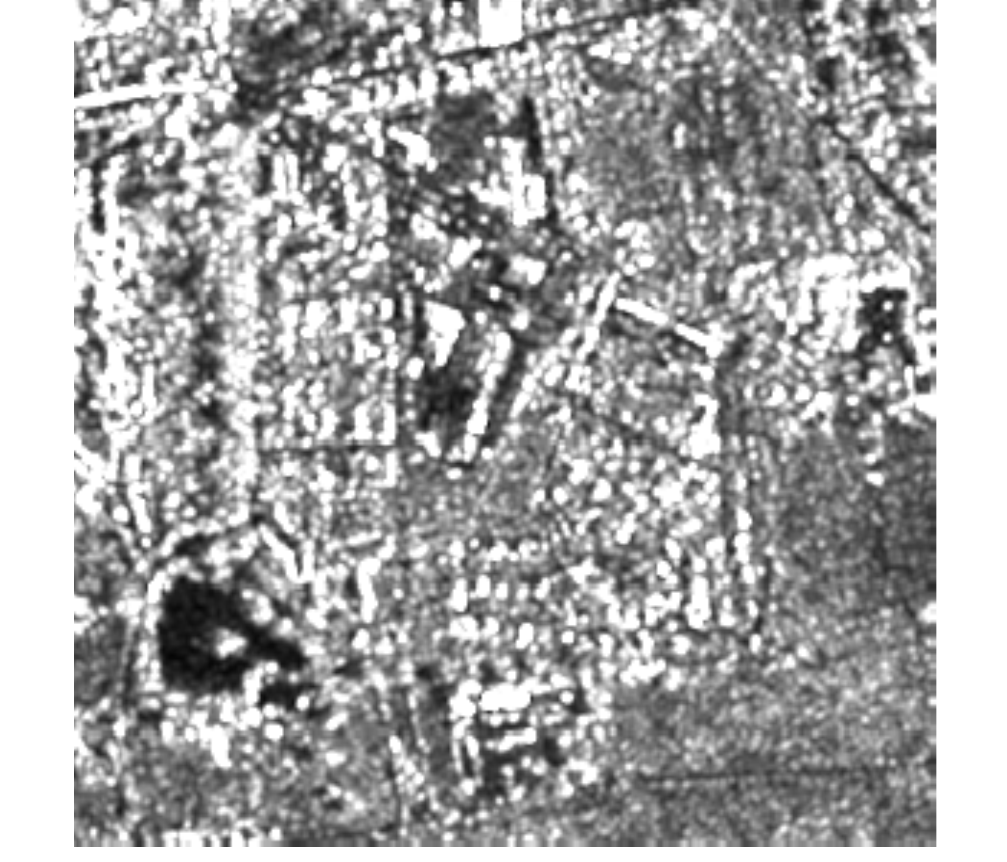}}
\subfigure[]{\includegraphics[width=.18\linewidth]{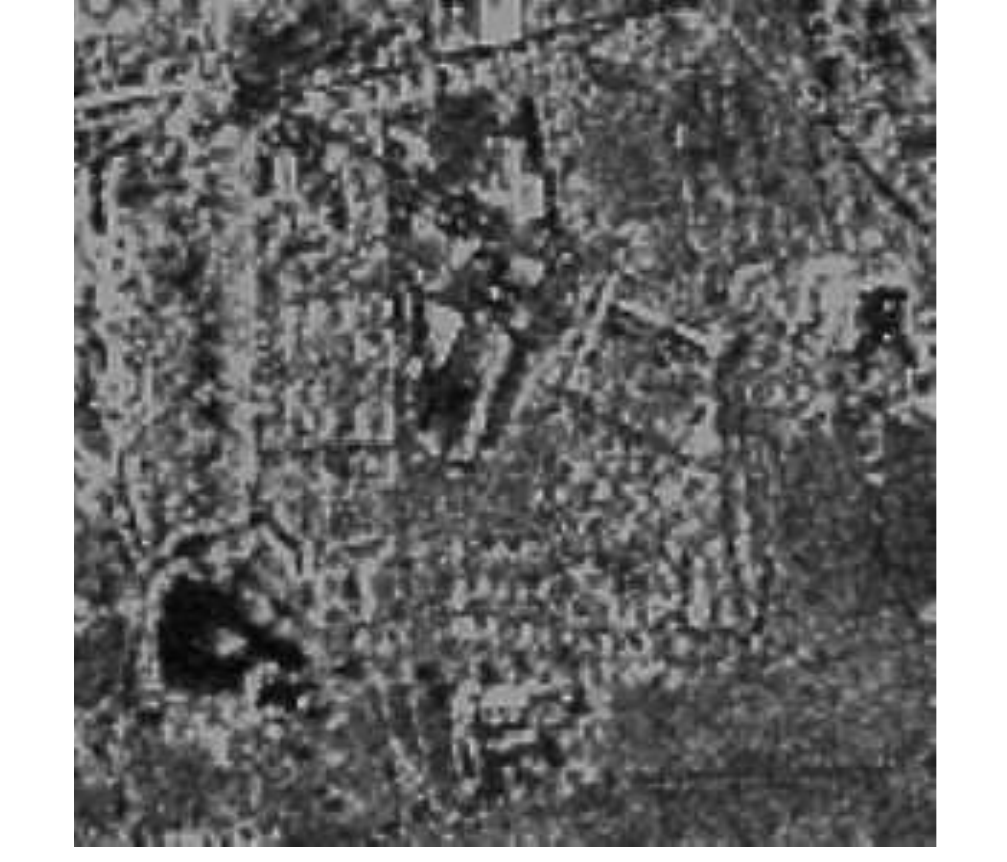}}
\subfigure[]{\includegraphics[width=.18\linewidth]{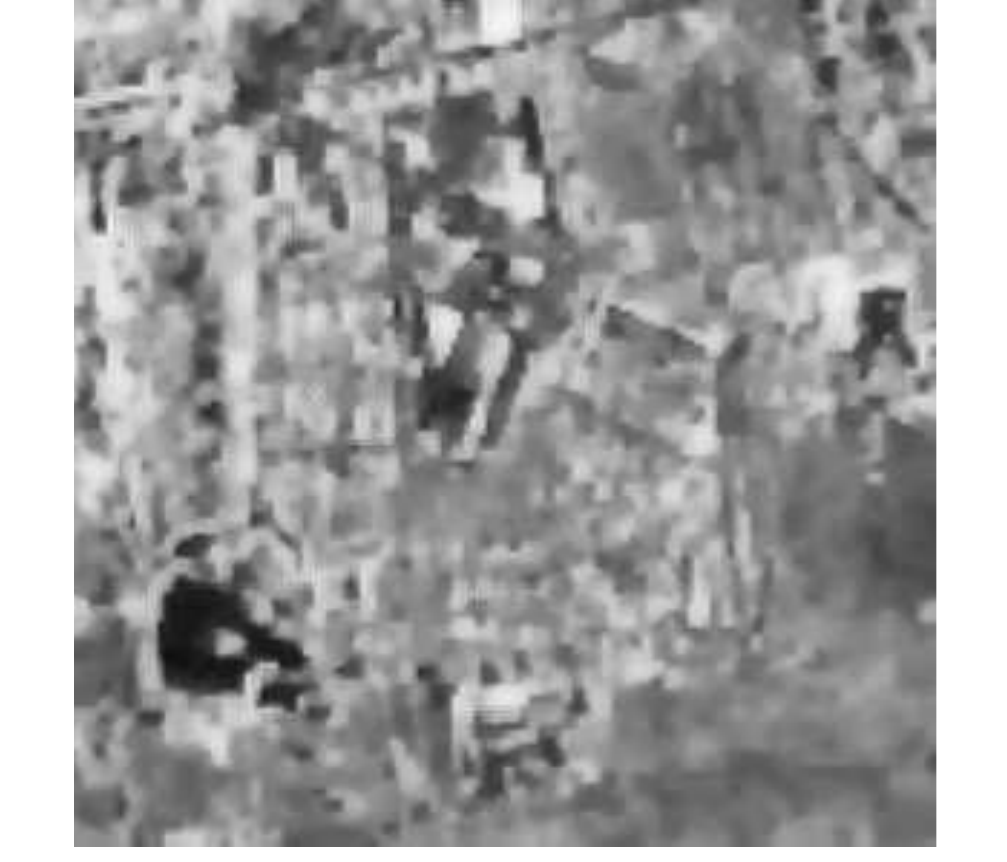}}
\subfigure[]{\includegraphics[width=.18\linewidth]{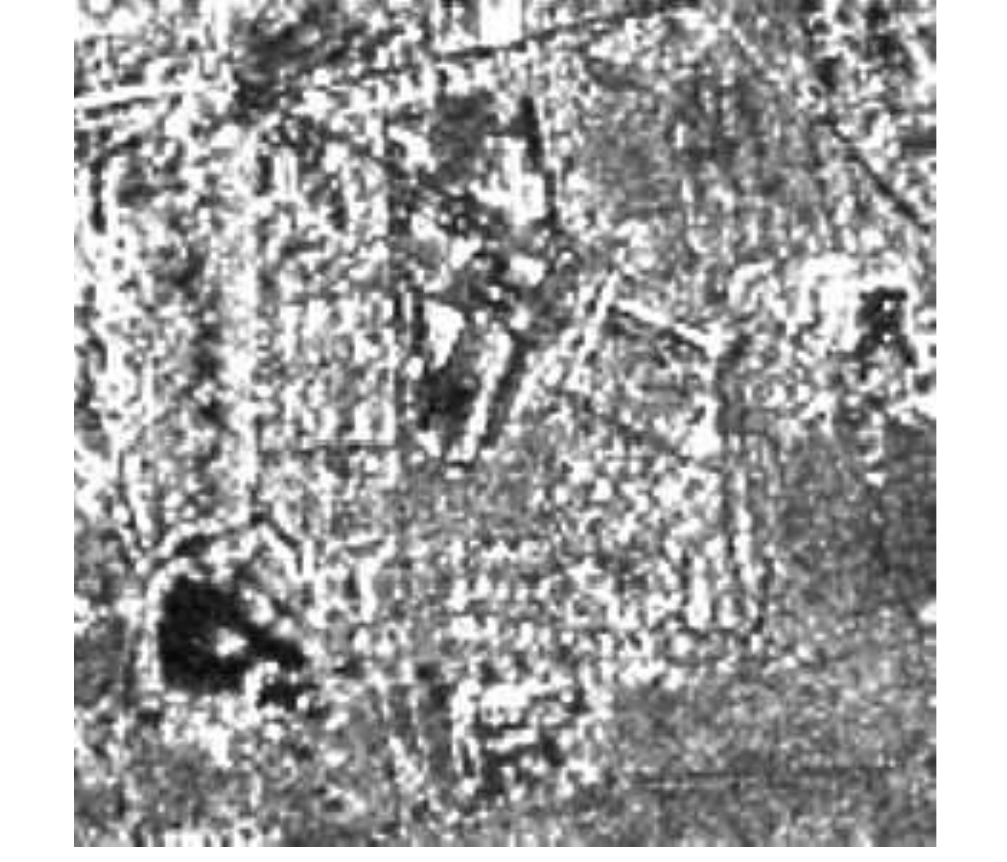}}\hfill\\
\subfigure[]{\includegraphics[width=.24\linewidth]{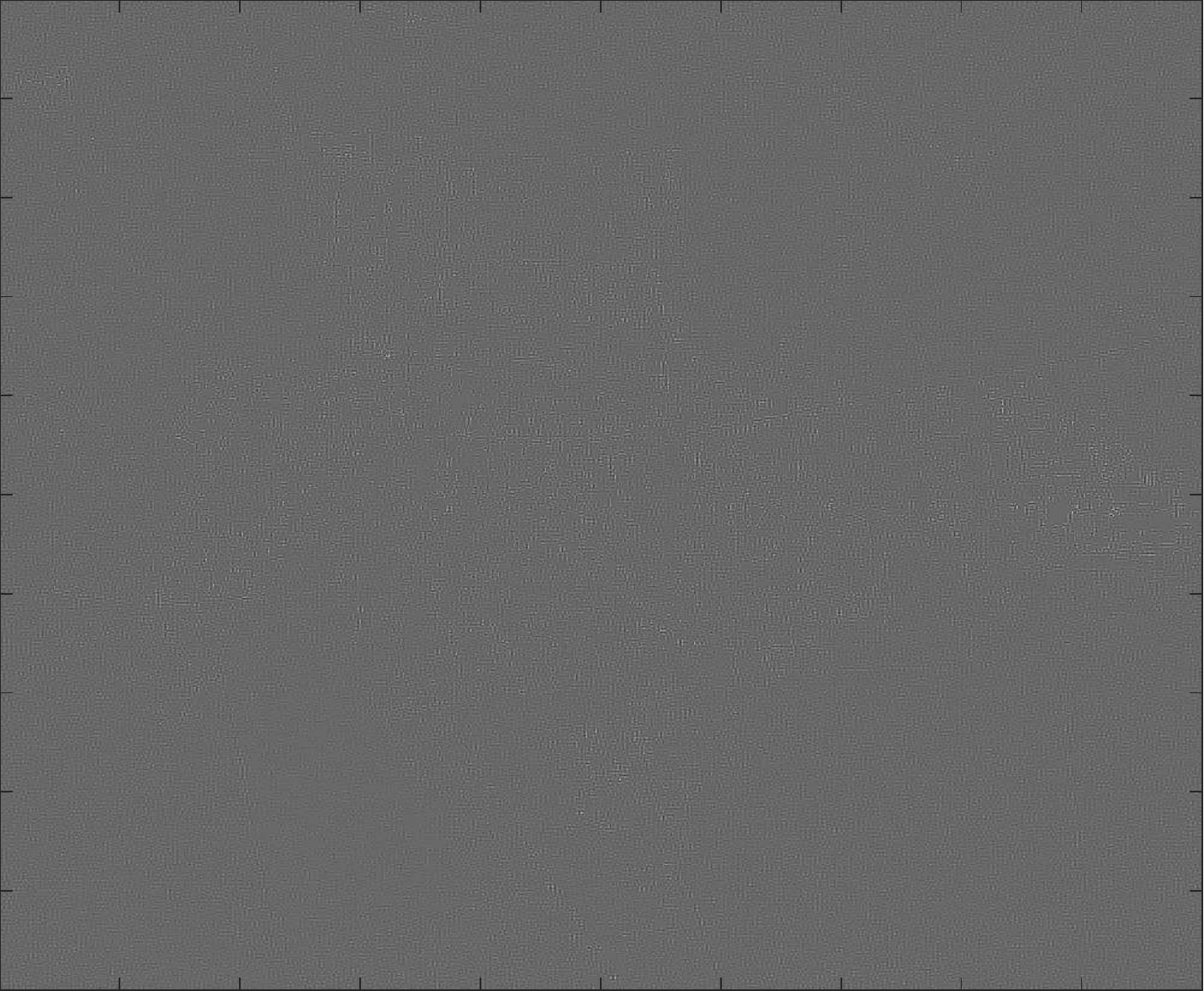}}
\subfigure[]{\includegraphics[width=.24\linewidth]{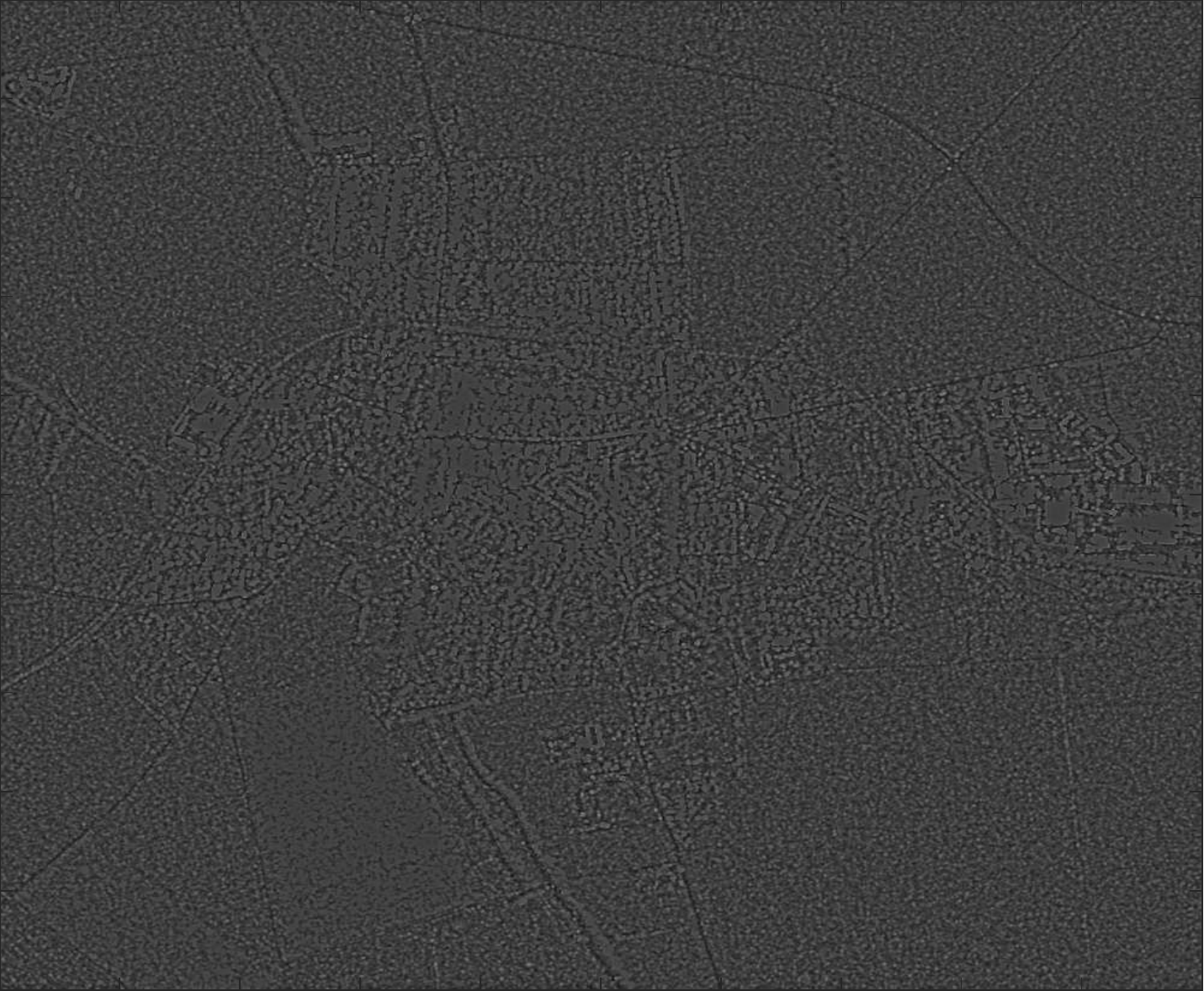}}
\subfigure[]{\includegraphics[width=.24\linewidth]{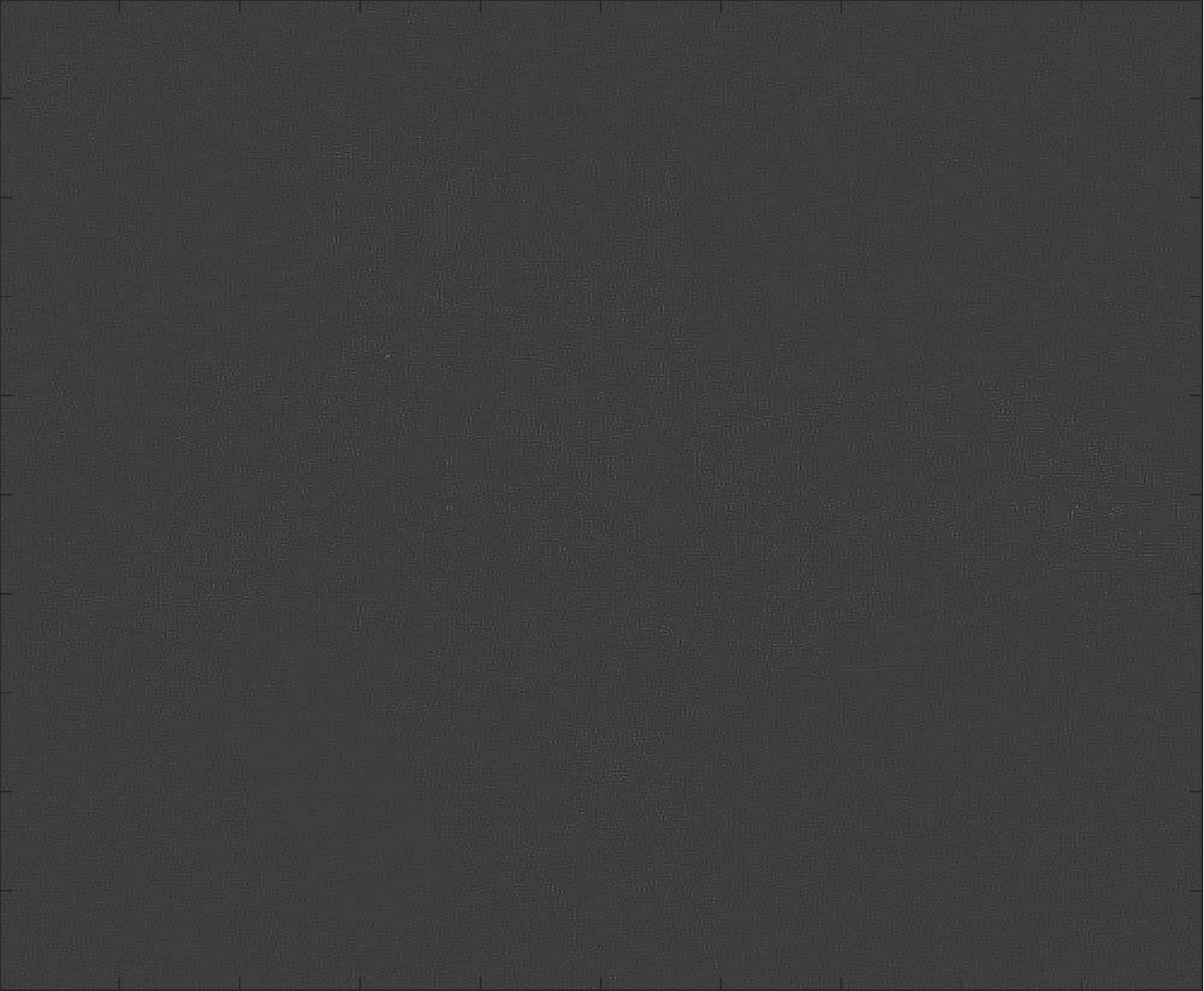}}\hfill
\caption{Visual despeckling results for a real SAR image. (a) and (e) Speckle image, (b) and (f) L1, (c) and (g) TV, (d) and (h) Cauchy. (i), (j) and (k) are ratio images for L1, TV and Cauchy, respectively. Images on the second row represent zoomed-in image of the rectangle given in (a).}
\label{fig:ds2}
\end{figure*}

In the third set of simulations, the performance of the Cauchy based penalty function in SAR image despeckling application was tested. Five different speckle-free X-band, HH polarized, Stripmap SAR products from TerraSAR-X \cite{terraSARX} were used, with sizes varying between 700$\times$700 and 900$\times$900 pixels. All five speckle-free SAR images were then multiplied with log-normal \cite{gagnon1997speckle} and gamma noise sequences with number of looks, $L$, chosen to be 5 and 15, which correspond to a high and a low speckle noise levels, respectively. Speckle images for all noise cases were processed by using the despeckling method based on $L_1$, $TV$ and Cauchy penalty functions. The performance of the methods were then compared in terms of PSNR and signal-to-mean squared error (S/MSE) \cite{gagnon1997speckle} values, which are given in Table \ref{tab:despeckling}. In Figure \ref{fig:ds1}, despeckling results for Image-4 are depicted for Gamma distributed speckle with $L=5$. Figure \ref{fig:ds2} presents subjective results for a real SAR image with speckle noise.

Table \ref{tab:despeckling} shows that the proposed penalty function achieved the best despeckling results for both $L$ values and both speckle noise cases, for all but Image-5. $TV$ achieved better results for Image-5 for $L = 5$. When examining the visual results in Figure \ref{fig:ds1}, we can clearly see that the proposed method both reconstructs various structures in the image, and shows similar characteristics to the original speckle-free SAR image shown in Figure \ref{fig:ds1}-(a) when compared to $TV$. Even though $TV$ preserves building structures, as can be seen from the results in Figure \ref{fig:ds1}-(d) and \ref{fig:ds1}-(i), it discards a multitude of background details and leads to a very blurry final result.

In Figure \ref{fig:ds1}-(k), (l) and (m), we show ratio images corresponding to element-wise division of the original SAR image by each despeckled image result. On examining these figures, we can clearly see that the result obtained using the  $TV$ penalty includes a high number of image structures, which is indicative of poor performance in terms of structure preservation. $L_1$- and Cauchy-based results include less structures, with the former again coming across as the best.

Finally, when examining Figure \ref{fig:ds2}, which corresponds to a real SAR image, similar to the simulated speckle situation, it is obvious that the Cauchy-based method outperforms both reference methods.

\subsection{Ship Wake Detection}
For the final set of simulations, we studied the suitability of the Cauchy-based penalty function as a building block of a method for ship wake detection in maritime applications. Eleven different SAR images of the sea surface containing ship wakes, from four different satellite platforms, namely TerraSAR-X (3), COSMO-SkyMed (4), Sentinel-1 (2) and ALOS2 (2) were used. We then used these 11 wake images in the inverse problem formulation discussed in Section \ref{sec:wake}, followed by the ship wake detection procedure described in \cite{karakucs2019ship2}.

\begin{figure*}[ht!]
\centering
\subfigure[]{\includegraphics[width=.24\linewidth]{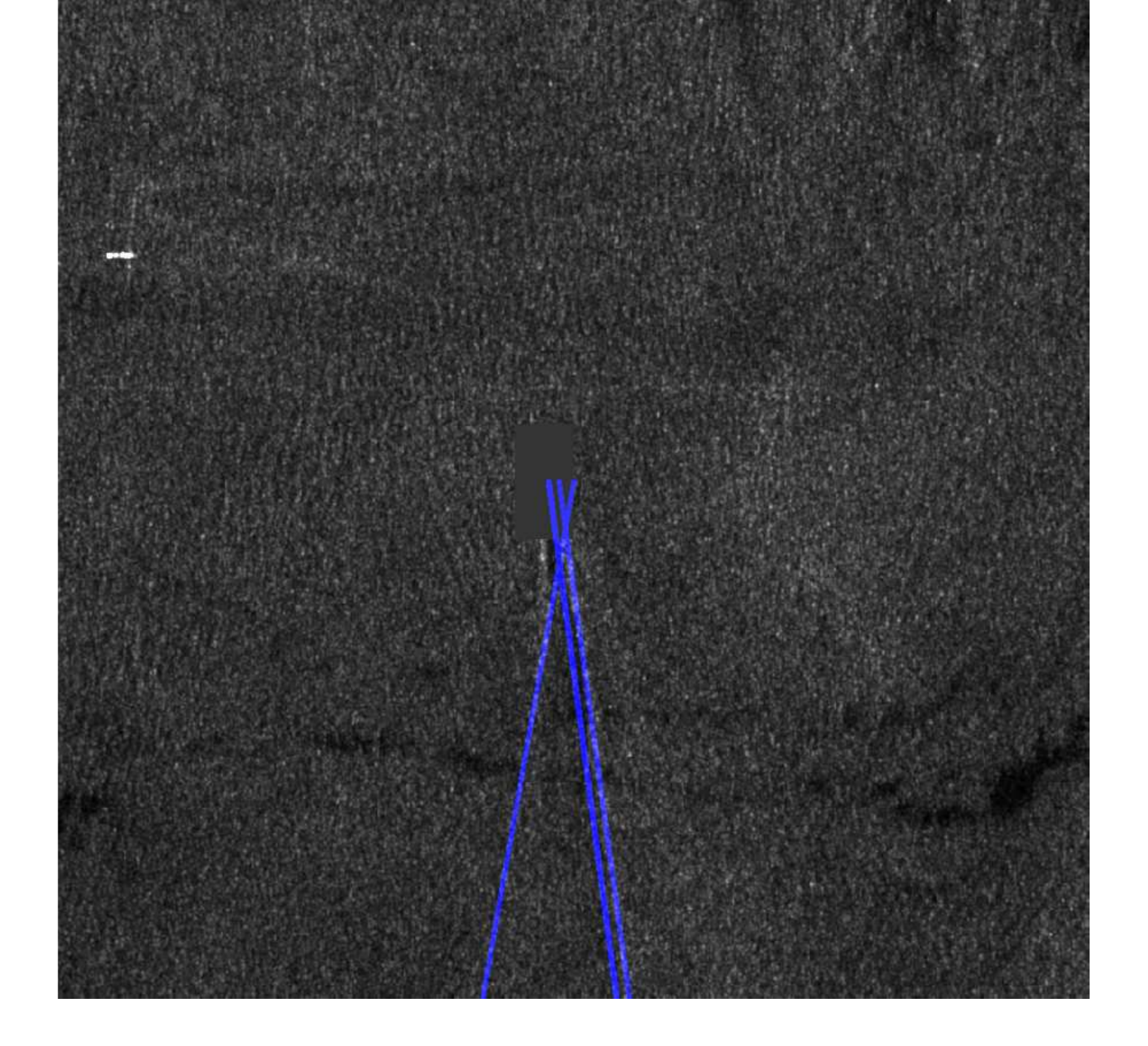}}
\subfigure[]{\includegraphics[width=.24\linewidth]{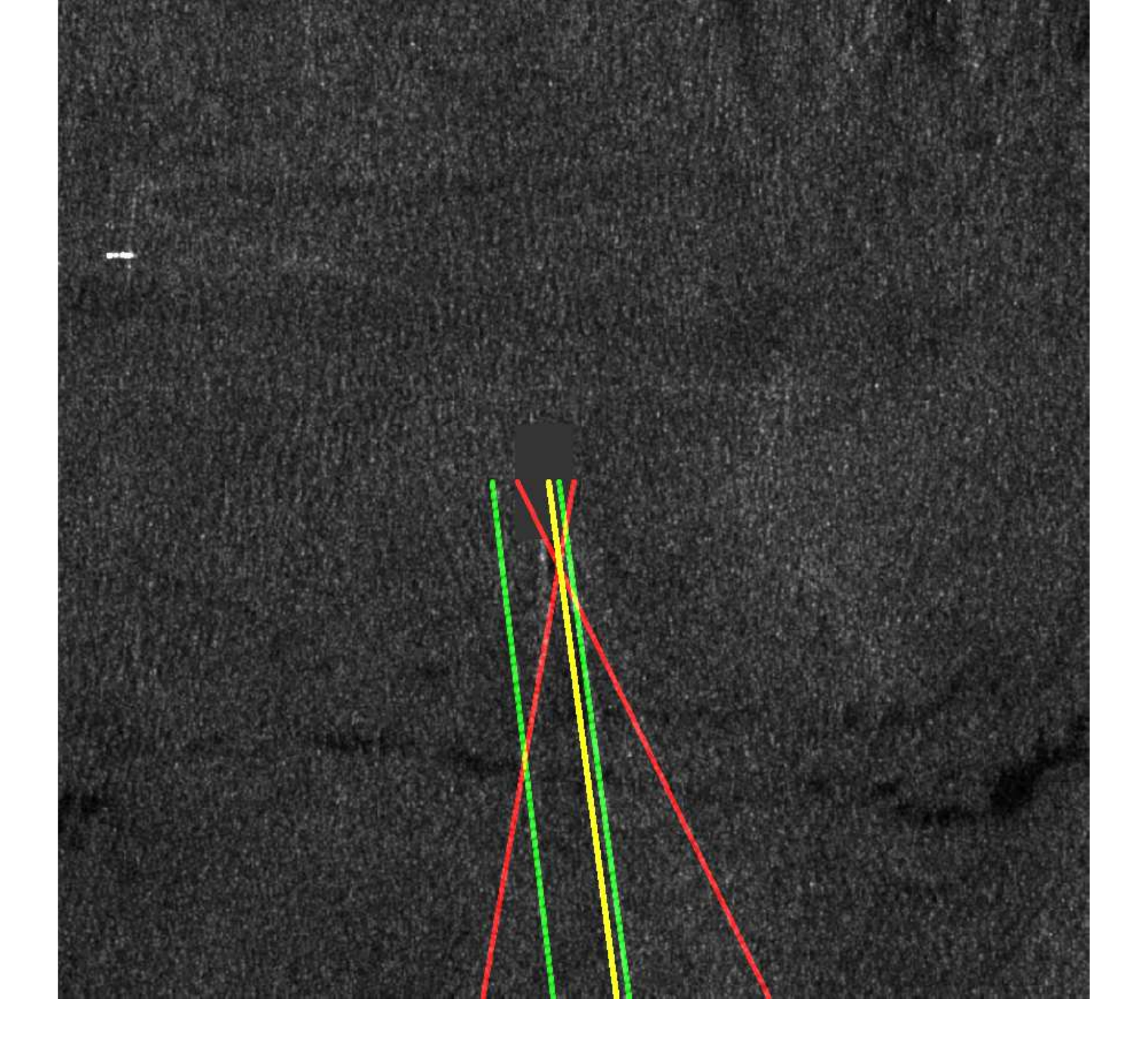}}
\subfigure[]{\includegraphics[width=.24\linewidth]{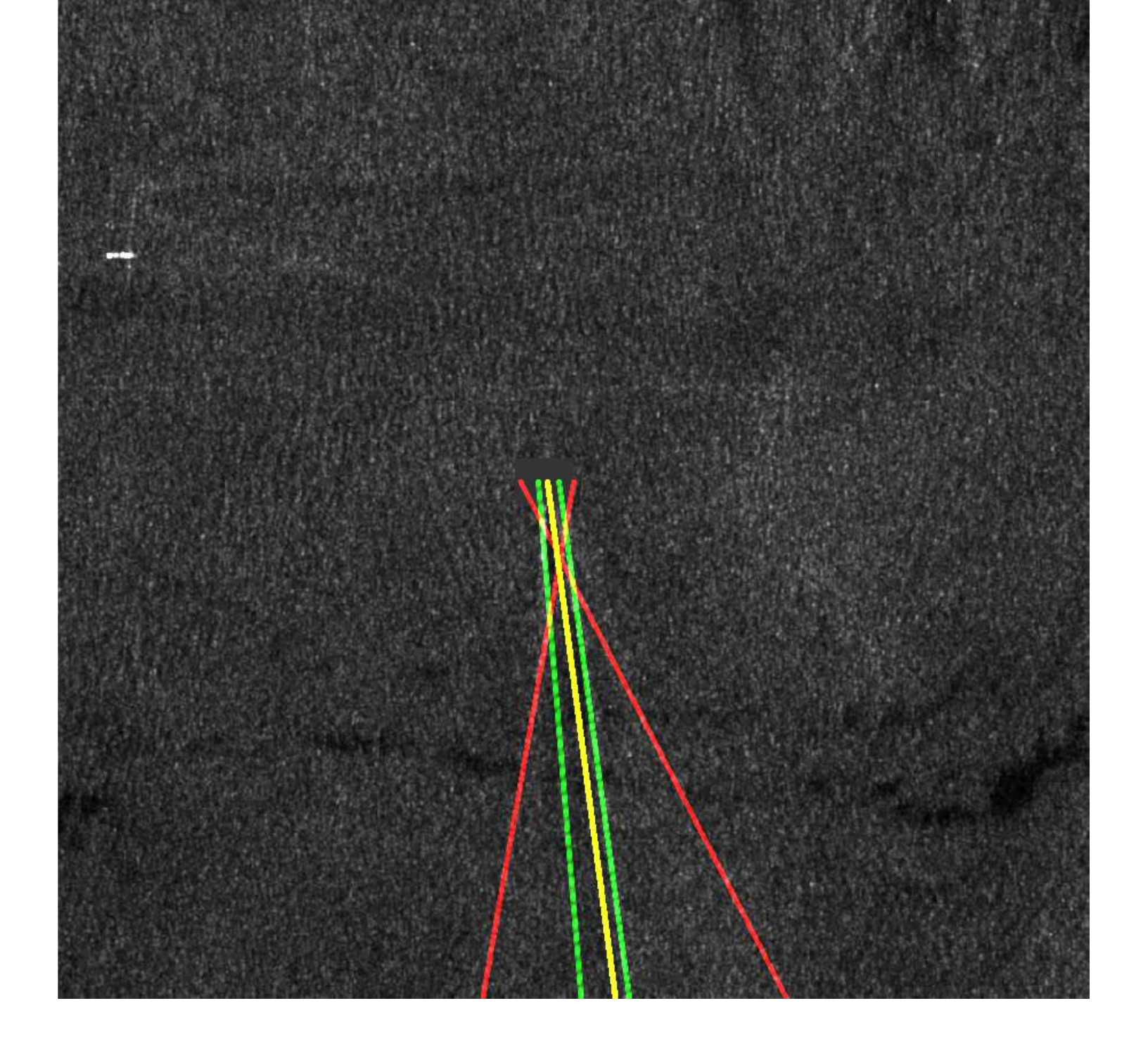}}
\subfigure[]{\includegraphics[width=.24\linewidth]{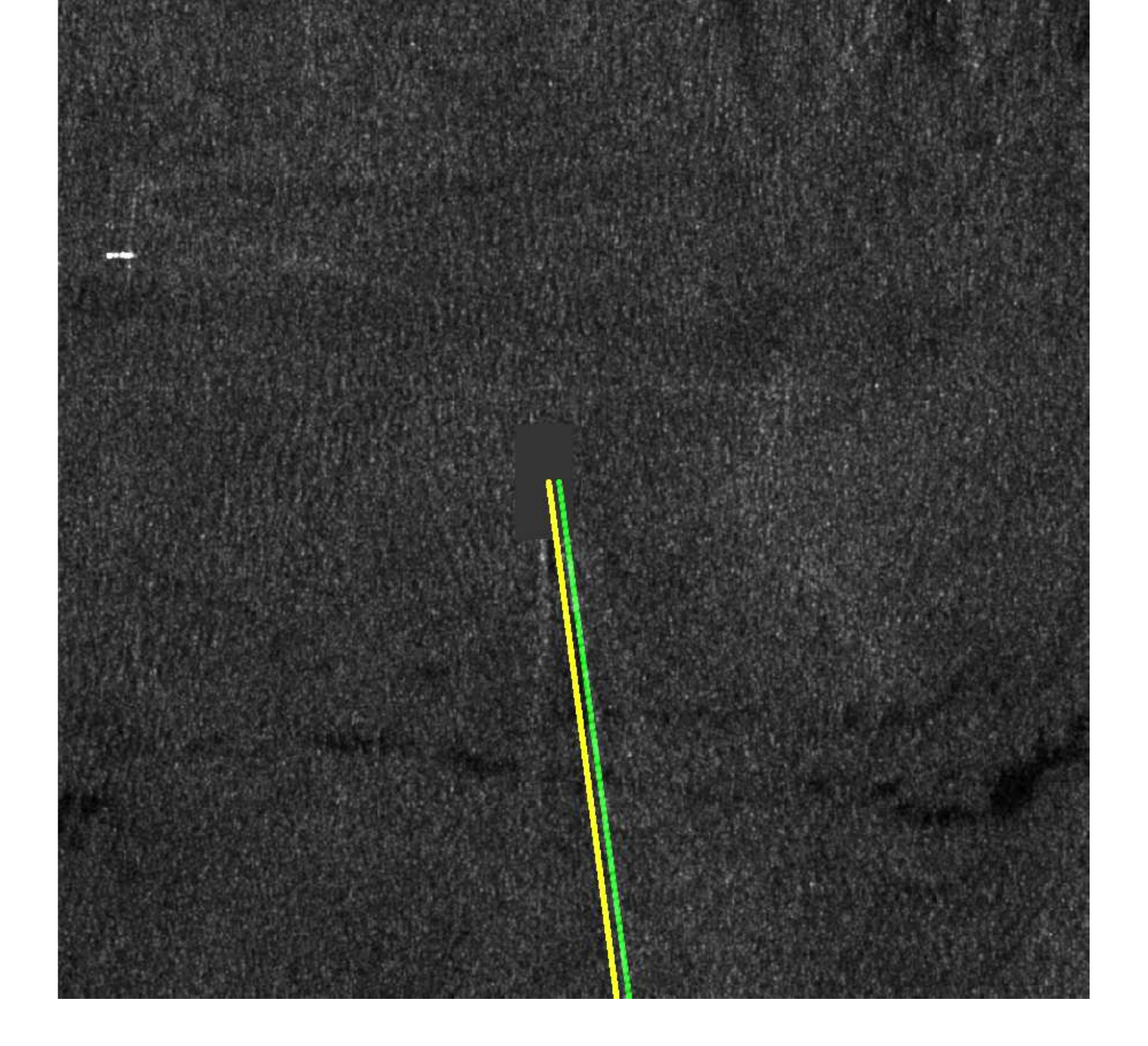}}
\caption{Ship wake detection results. (a) Ship centred original image with correct wake locations. Wake detection results for (b) GMC \cite{karakucs2019ship} (3 out of 5), (c) TV (3 out of 5) and (d) Cauchy (4 out of 5).}
\label{fig:swd}
\end{figure*}

For comparison, we chose the two best performing regularization functions in \cite{karakucs2019ship2}, which include the GMC and $TV$, and compared their wake detection performances with the proposed Cauchy based penalty function. For objective evaluation of the detection results, we used the receiver operation characteristics (ROC) of true positive (TP), true negative (TN), false positive (FP) and false negative (FN) as well as other common classification metrics such as accuracy, the $F_1$ score, positive likelihood ratio (LR+) and Youden's $J$ index \cite{karakucs2019ship2}. In Table \ref{tab:wake}, we present the results over all data sets in terms of the metrics defined above, whilst the visual evaluation of wake detection for a single image is illustrated in Figure \ref{fig:swd}.

The proposed method outperforms the reference methods by at least 3\% in terms of accuracy and to various degree in terms of the other performance metrics presented. Specifically, the GMC TP (correct detection) value is higher than that of Cauchy, however the Cauchy based penalty function leads to higher TN (correct discard) and less FP (false detection) values than the others, which makes it the most suitable method in ship wake detection overall.

In Figure \ref{fig:swd}, we provide visual results to assess ship wake detection performance. In Figure \ref{fig:swd}-(a) the original image (re-centered on the ship) is shown. There are three detected wakes, which are the turbulent wake, one arm of the narrow V-wake and one Kelvin arm as shown in Figure \ref{fig:swd}-(b). The GMC- and TV-based approaches detected all five hypothetical  wakes, resulting in a 60\% detection accuracy (two false detections). The proposed Cauchy-based method detects correctly 2 visible wakes (TP) and discards two invisible wakes (TN), with an erroneous discarding (FN) of the visible Kelvin arm, which corresponds to an 80\% detection accuracy.

\section{Conclusions}\label{sec:conc}
In this paper, we proposed the use of a proximal splitting algorithm which remains convex when used in conjunction with a non-convex Cauchy based penalty function for solving several SAR imaging inverse problems.
We followed the conditions defined in our previous work \cite{karakucs2019cauchy1} to guarantee convergence by establishing a relationship between the Cauchy scale parameter $\gamma$ and the proximal splitting algorithm step size parameter $\mu$.

We illustrated the proposed proximal splitting method in four different SAR imaging inverse problems, including super-resolution, image formation, despeckling and ship wake detection. In addition to presenting an easy-to-implement methodology for the use of the Cauchy based penalty function, we also demonstrated its superiority for all the examples considered. The Cauchy based penalty function achieved better image reconstruction performance compared to all alternative penalty functions investigated, including $L_1$, $TV$ and the GMC function. 


\begin{table*}[ht!]
  \centering
  \caption{Ship wake detection performance for various regularisation functions.}
    \resizebox{0.7\linewidth}{!}{\begin{tabular}{cccccccccccccccccccccccccccccccc}
    \toprule
          &       &       & \multicolumn{4}{c}{TP}        & \multicolumn{4}{c}{TN}        & \multicolumn{4}{c}{FP}        & \multicolumn{4}{c|}{FN}        & \multicolumn{4}{c}{Accuracy}  & \multicolumn{3}{c}{F1} & \multicolumn{3}{c}{LR+} & \multicolumn{3}{c}{Youden's J} \\
          \toprule
    \multicolumn{3}{c}{GMC \cite{karakucs2019ship2}} & \multicolumn{4}{c}{46.00\%}   & \multicolumn{4}{c}{30.00\%}   & \multicolumn{4}{c}{22.00\%}   & \multicolumn{4}{c|}{2.00\%}    & \multicolumn{4}{c}{78.18\%}   & \multicolumn{3}{c}{0.79} & \multicolumn{3}{c}{2.27} & \multicolumn{3}{c}{0.54} \\
    \multicolumn{3}{c}{TV} & \multicolumn{4}{c}{34.00\%}   & \multicolumn{4}{c}{26.00\%}   & \multicolumn{4}{c}{38.00\%}   & \multicolumn{4}{c|}{2.00\%}    & \multicolumn{4}{c}{61.82\%}   & \multicolumn{3}{c}{0.63} & \multicolumn{3}{c}{1.59} & \multicolumn{3}{c}{0.35} \\
    \multicolumn{3}{c}{Cauchy} & \multicolumn{4}{c}{38.00\%}   & \multicolumn{4}{c}{46.00\%}   & \multicolumn{4}{c}{10.00\%}   & \multicolumn{4}{c|}{8.00\%}    & \multicolumn{4}{c}{\textbf{81.82\%}} & \multicolumn{3}{c}{\textbf{0.81}} & \multicolumn{3}{c}{\textbf{4.63}} & \multicolumn{3}{c}{\textbf{0.65}} \\
    \bottomrule
    \end{tabular}}%
  \label{tab:wake}%
\end{table*}%

\bibliographystyle{IEEEtran}
\bibliography{Cauchy_SR}

\begin{IEEEbiography}[{\includegraphics[width=1in,height=1.25in,clip,keepaspectratio]{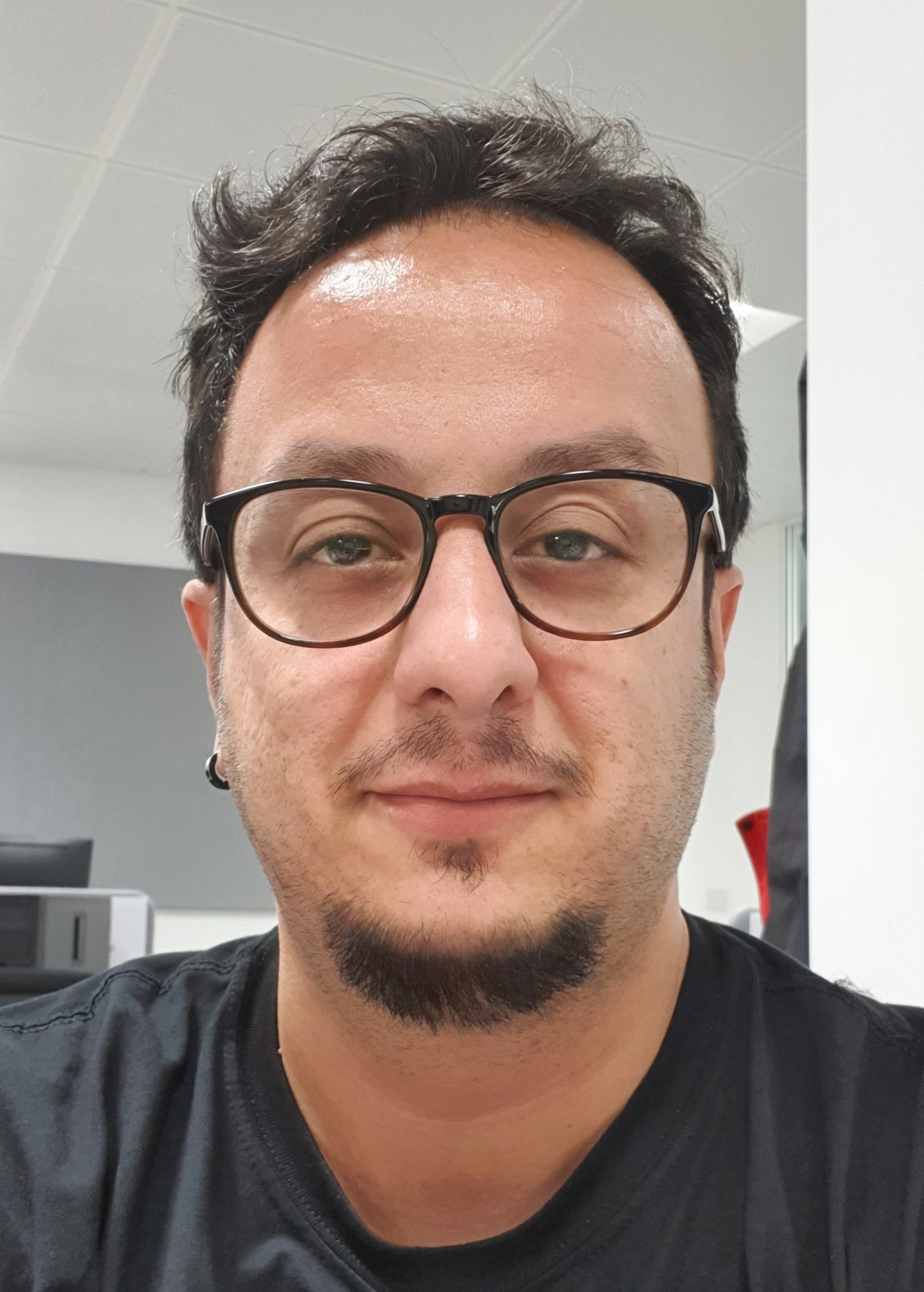}}]{Oktay Karakuş} (M'18) received his B.Sc. degree in electronics engineering (with Honours) from Istanbul Kültür University, Turkey in 2009. Then, he received his M.Sc. and Ph.D. degrees in electronics and communication engineering from İzmir Institute of Technology (IZTECH), Turkey, in 2012 and 2018, respectively. Between October 2009 to December 2017, he was associated with Department of Electrical and Electronics Engineering in Yaşar University, Turkey and Department of Electronics and Communication Engineering in İzmir Institute of Technology, Turkey as a research assistant. He was a Visiting Scholar with the Institute of Information Science and Technologies (ISTI-CNR), Pisa, Italy in 2017. From March 2018, He is working as a Research Associate in image processing at Visual Information Laboratory in University of Bristol, Department of Electrical \& Electronic Engineering. His research interests are mostly in statistical/Bayesian signal and image processing, inverse problems with applications on SAR and ultrasound imagery, heavy tailed data modeling, telecommunications and energy.
\end{IEEEbiography}

\begin{IEEEbiography}[{\includegraphics[width=1in,height=1.25in,clip,keepaspectratio]{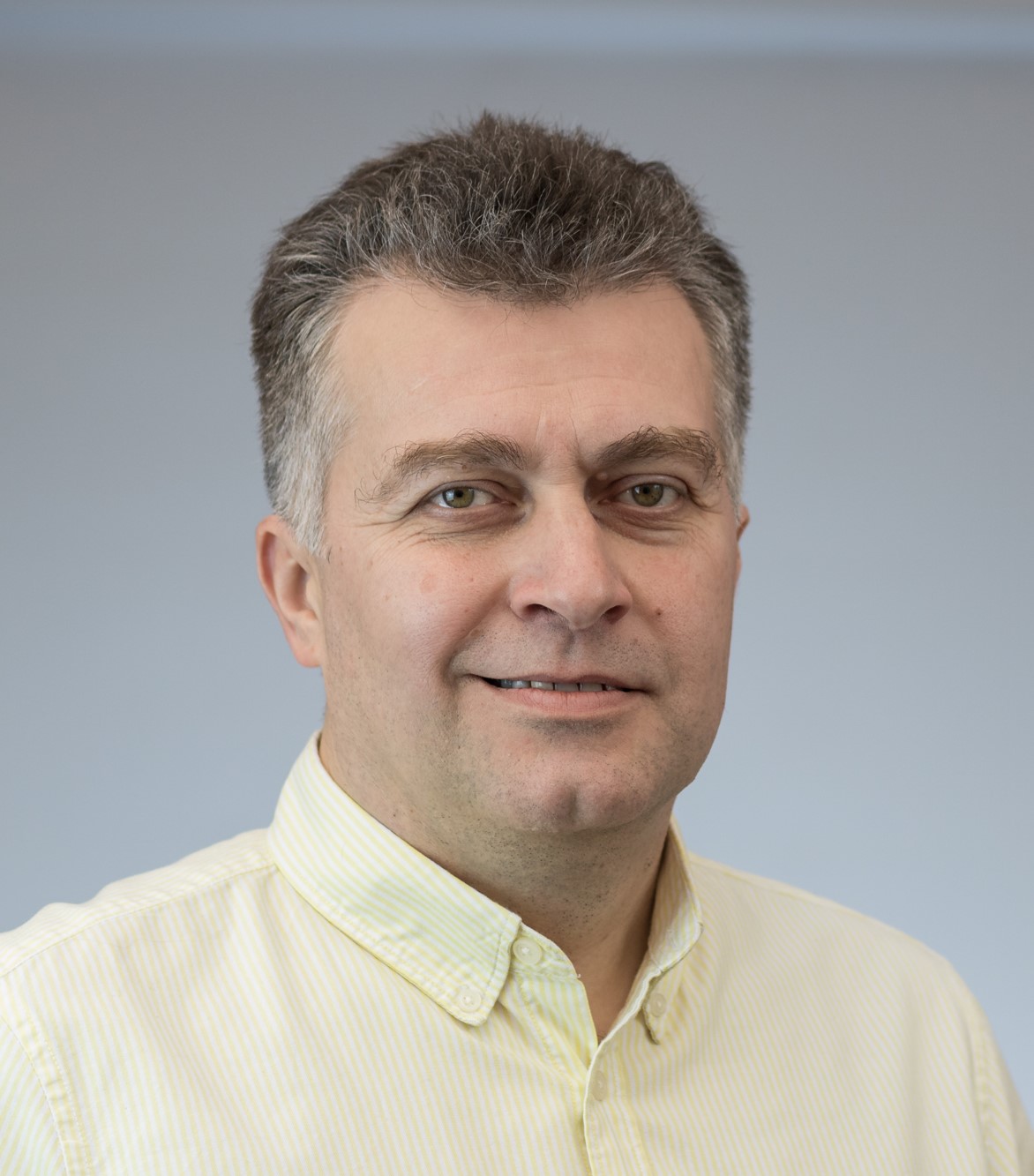}}]{Alin Achim} (S’99–M’03–SM’09) received the B.Sc. and M.Sc. degrees in electrical engineering from "Politehnica" University of Bucharest, Romania, in 1995 and 1996, respectively, and a Ph.D. in biomedical engineering from the University of Patras, Greece, in 2003. He then obtained an ERCIM (European Research Consortium for Informatics and Mathematics) postdoctoral fellowship which he spent with the Institute of Information Science and Technologies (ISTI-CNR), Pisa, Italy and with the French National Institute for Research in Computer Science and Control (INRIA), Sophia Antipolis, France. In October 2004 he joined the Department of Electrical \& Electronic Engineering at the University of Bristol, U.K., as a Lecturer, he became a Senior Lecturer (Associate Professor) in 2010, and a Reader in Biomedical Image Computing in 2015. From August 2018 he holds the Chair in Computational Imaging at the University of Bristol. He has co-authored over 140 scientific publications, including 40 journal papers. Alin's research interests include statistical signal, image and video processing with particular emphasis on the use of sparse distributions within sparse domains and with applications in both biomedical imaging and remote sensing. He is senior area editor of the IEEE Transactions on Image Processing, an associate editor of the IEEE Transactions on Computational Imaging, and an editorial board member of MDPI's Remote Sensing. He was/is an elected member of the Bio Imaging and Signal Processing Technical Committee of the IEEE Signal Processing Society, an affiliated member (invited) of the same Society’s Signal Processing Theory \& Methods Technical Committee and a member of the IEEE Geoscience and Remote Sensing Society’s Image Analysis and Data Fusion Technical Committee.
\end{IEEEbiography}
\end{document}